\documentclass[]{aastex63}

\usepackage[caption=false]{subfig}
\usepackage{graphicx}

\newcommand{\p}{$^\prime$}
\newcommand{\pp}{${^\prime\prime}$}
\newcommand{\hi}{\ion{H}{1}}
\newcommand{\hii}{\ion{H}{2}}
\newcommand{\Lya}{Lyman $\alpha$}
\newcommand{\kms}{km~s$^{-1}$}
\newcommand{\ha}{H$\alpha$}
\newcommand{\mips}{24$\mu$m}
\newcommand{\sfr}{$\Sigma_{\rm{SFR}}$}
\newcommand{\sm}{$\Sigma_{\star}$}
\newcommand{\mhi}{$\Sigma_{\rm{H\,{\small I}}}$}
\newcommand{\kesd}{$\Sigma_{\rm{KE}}$}
\newcommand{\disp}{$ \sigma_{los}$}
\newcommand{\vlos}{$v_{los}$}

\accepted{September 29, 2021}
\submitjournal{ApJ}

\shorttitle{Star Formation in the XUV disk of NGC 3344}
\shortauthors{Padave et al. 2021}

\begin{document}

\title{\large DIISC-II: Unveiling the Connections between Star Formation and ISM in the Extended Ultraviolet Disk of NGC 3344}

\author[0000-0002-3472-0490]{Mansi Padave}
\affiliation{School of Earth \& Space Exploration, Arizona State University, Tempe, AZ 85287-1404, USA}

\correspondingauthor{Mansi Padave}
\email{mpadave@asu.edu}

\author[0000-0002-2724-8298]{Sanchayeeta Borthakur}
\affiliation{School of Earth \& Space Exploration, Arizona State University, Tempe, AZ 85287-1404, USA}

\author[0000-0003-1436-7658]{Hansung B. Gim}
\affiliation{School of Earth \& Space Exploration, Arizona State University, 781 E Terrace Mall, Tempe, AZ 85287-1404, USA}
\affiliation{Cosmology Initiatives, Arizona State University, 650 E Tyler Mall, Tempe, AZ 85287-1404, USA}
\affiliation{Department of Physics, Montana State University, P.O. Box 173840, Bozeman, MT 59717, USA}

\author[0000-0003-1268-5230]{Rolf A. Jansen}
\affiliation{School of Earth \& Space Exploration, Arizona State University, Tempe, AZ 85287-1404, USA}

\author[0000-0002-8528-7340]{David Thilker}
\affiliation{Department of Physics \& Astronomy, Johns Hopkins University, Baltimore, MD, 21218, USA}

\author[0000-0001-6670-6370]{Timothy Heckman}
\affil{Department of Physics \& Astronomy, Johns Hopkins University, Baltimore, MD, 21218, USA}

\author[0000-0001-5448-1821]{Robert C.  Kennicutt}
\affil{Department of Astronomy and Steward Observatory, University of Arizona
Tucson, AZ 85721, USA}
\affil{Department of Physics and Astronomy, Texas A\&M University
College Station, TX 77843, USA}

\author[0000-0003-3168-5922]{Emmanuel Momjian}
\affiliation{National Radio Astronomy Observatory, 1003 Lopezville Rd, Socorro, NM 87801, USA}

\author[0000-0003-0724-4115]{Andrew J. Fox}
\affiliation{AURA for ESA, Space Telescope Science Institute, 3700 San Martin Drive, Baltimore, MD 21218, USA}



\begin{abstract}

We present our investigation of the Extended Ultraviolet (XUV) disk galaxy, NGC~3344, conducted as part of Deciphering the Interplay between the Interstellar medium, Stars, and the Circumgalactic medium (DIISC) survey. We use surface and aperture photometry of individual young stellar complexes to study star formation and its effect on the physical properties of the interstellar medium. We measure the specific star-formation rate (sSFR) and find it to increase from $\rm10^{-10}~yr^{-1}$ in the inner disk to $\rm>10^{-8}~yr^{-1}$ in the extended disk. This provides evidence for inside-out disk growth. If these sSFRs are maintained, the XUV disk stellar mass can double in $\sim$0.5~Gyr, suggesting a burst of star formation. The XUV disk will continue forming stars for a long time due to the high gas depletion times ($\tau_{dep}$). The stellar complexes in the XUV disk have high-\mhi\ and low-\sfr\ with $\tau_{dep}\sim$10~Gyrs, marking the onset of a deviation from the traditional Kennicutt-Schmidt law. We find that both far ultraviolet (FUV) and a combination of FUV and \mips\ effectively trace star formation in the XUV disk. \ha\ is weaker in general and prone to stochasticities in the formation of massive stars. Investigation of the circumgalactic medium at 29.5~kpc resulted in the detection of two absorbing systems with metal-line species: the stronger absorption component is consistent with gas flows around the disk, most likely tracing inflow, while the weaker component is likely tracing corotating circumgalactic gas.
\end{abstract}



\section{Introduction} \label{sec:intro}

The connections between star formation 
and cold interstellar gas are paramount to the understanding of 
galaxy evolution. The interstellar medium (ISM) provides the essential matter and hospitable environment that facilitate the conversion of cold atomic gas (\hi) to its molecular (H$_2$) form, which further collapses to form stars. 
In general, this relation between stars and gas 
follows the empirical Kennicutt-Schmidt law \citep[K-S law; ][]{kenn98b, schm59}. 
However, when extended to the outskirts of galaxies where gas mass surface density, $\Sigma_{\rm{gas}}\lesssim9$~M$_\odot$~pc$^{-2}$,
a considerable drop in the star formation rate (SFR) per unit gas is observed \citep{kenn89, bigi10}. Discerning the origin of this downturn calls for a thorough investigation of the causal nexus between the low surface mass density ISM and star formation. 

The discovery of extended ultra-violet (XUV) 
disks in the nearby Universe \citep{thilk05, thilk07, gild05} enabled by 
Galaxy Evolution Explorer  \citep[GALEX;][]{mart05, morr07} strengthened the notion of star formation beyond the optical disk ($\gtrsim R_{25}$). These galaxies showed UV bright extended disks at large
galactocentric distances (Type-I XUVs) or exhibited large blue (\emph{UV} $-$ \emph{K}) color low surface
brightness zones outside the optical disks (Type-II XUVs). Deep H$\alpha$ imaging and spectroscopy \citep{ferg98,leli00}, also uncovered young
stellar populations in the outer disk of spiral galaxies where $\Sigma_{\rm{gas}}$ falls below 10 M$_\odot$~pc$^{-2}$. These regions can be distinguished from the inner metal-rich disks by 
their sub-critical conditions for star formation, such as low densities, low dust abundances, and an \hi\ dominated ISM. 
Nevertheless, these environments do support star formation. 
The outer, predominantly high angular momentum zones in a galaxy begin star-forming activity later than the inner disk, causing the disk to grow \citep[theory of inside-out growth;][]{bard05, muno07, rosk08, goga10, gonz14}. 
Therefore, understanding the physics of star formation in these confounding regions would be pivotal for understanding galaxy growth and evolution. 

One approach to studying the growth of galaxy disks is to probe channels of 
gas from the circumgalactic medium (CGM) to the star-forming disks 
of galaxies. The condensation of gas into the \hi\ disk is expected to be active in the disk-CGM interface \citep{sancisi08}. We designed the Deciphering the Interplay between the Interstellar medium,
Stars, and the Circumgalactic medium (DIISC) survey that aims to do so by 
tracing the cycle of gas from the CGM to \hi\ disks, finally to regions where
young stars are forming. 
The COS-GASS survey using QSO absorption lines tracing the CGM \citep{bort15, bort16} discovered a significant correlation between the strength of Lyman-$\alpha$ absorbers and the \hi-21 cm gas mass of the galaxies, thus suggesting a connection between gas content of the CGM and the disk of galaxies.
This correlation is believed to be a consequence of the CGM feeding the \hi\ disks, which will then fuel star formation. 
On the other hand, outflowing gas would enrich the CGM, although outflows are believed to not be the primary origin of cool CGM in low-z galaxies.

In this paper, we present the results of a pilot study investigating the interplay between star formation, ISM, and CGM in the low-density, \hi\ dominated regions of one special case within the DIISC sample, NGC 3344. This galaxy is known to exhibit a Type-I extended UV disk showing UV-bright structures at large
galactocentric distances ($>\rm R_{25}$), indicating recent star formation in the outer disk \citep{thilk07}. NGC~3344 is ideal to probe the connections between gas flows and young stars, as -- 
(1) it is undergoing active star formation in the outer disk, where stellar densities are low, thus making it easy to identify individual star-forming regions, 
(2) it possesses a huge gas reservoir in the form of an extended \hi\ disk of M(\hi)=$\rm 2.2 \times 10^9~M_{\odot}$ that is sustaining star formation, and 
(3) there is a UV-bright Quasi-stellar object (QSO) at an impact parameter of 29.5~kpc that allows us to probe the CGM close to the disk (Figure \ref{fig:ngc3344}). We summarize some of the key properties of the galaxy in Table \ref{tb:ngcprop}. For this study, we adopt a distance of 8.28$\pm$0.7 Mpc for NGC~3344 from \cite{sabb18}. This corresponds to a linear scale of 40.14~pc/\arcsec. An updated distance of 9.83$\pm$1.7 is provided by \cite{anand21}, for which the linear scale is 47.65~pc/\arcsec 
(or 1.18 times our adopted scale).

\begin{figure}[!t]
\setcounter{figure}{0}
\subfloat[]{\includegraphics[width=0.6\textwidth, trim=0cm 0cm 2cm 2cm, clip]{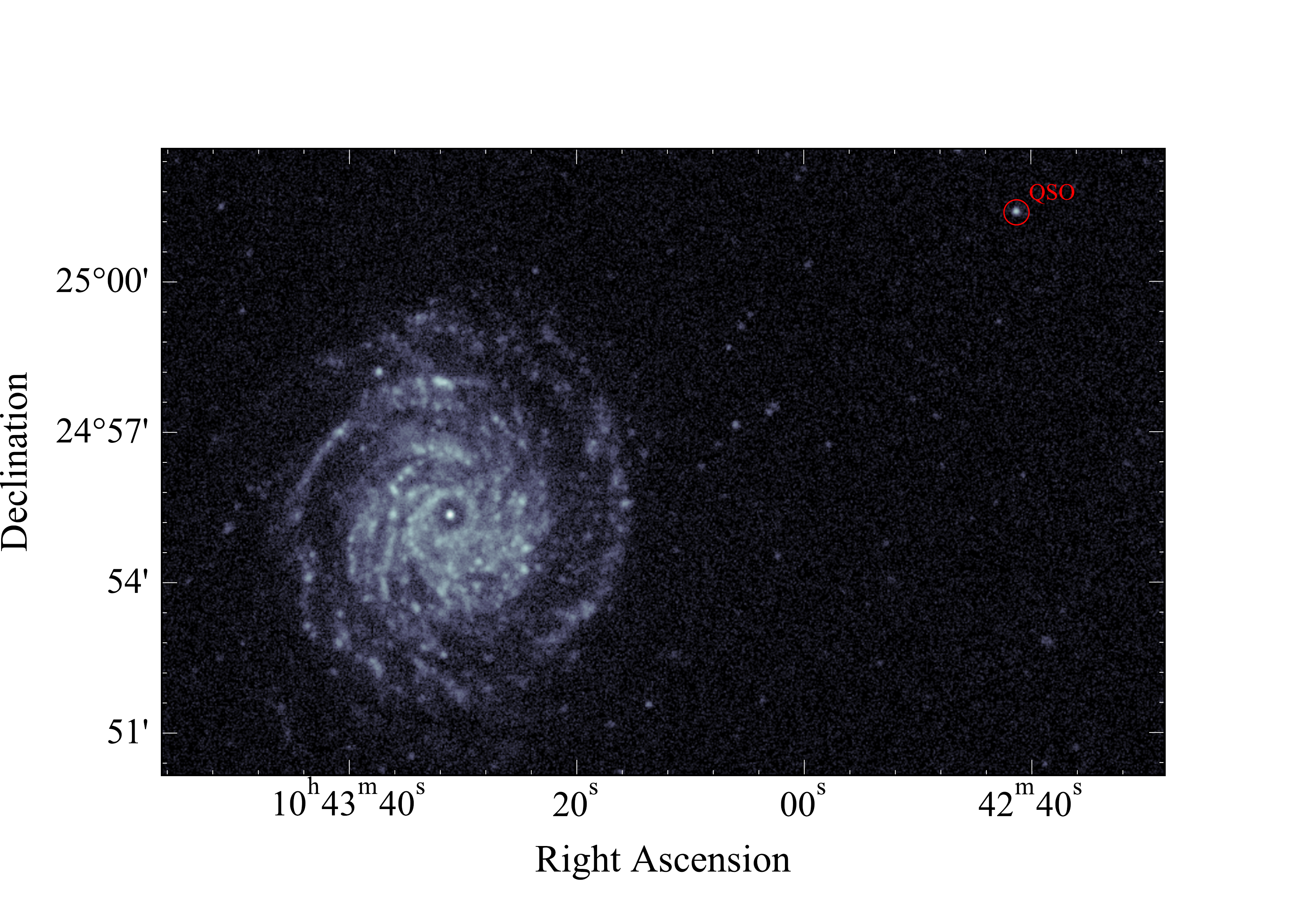} \label{fig:qso}}\hfill
\subfloat[]{\includegraphics[width=0.45\textwidth, trim=3cm 0cm 0cm 0cm, clip]{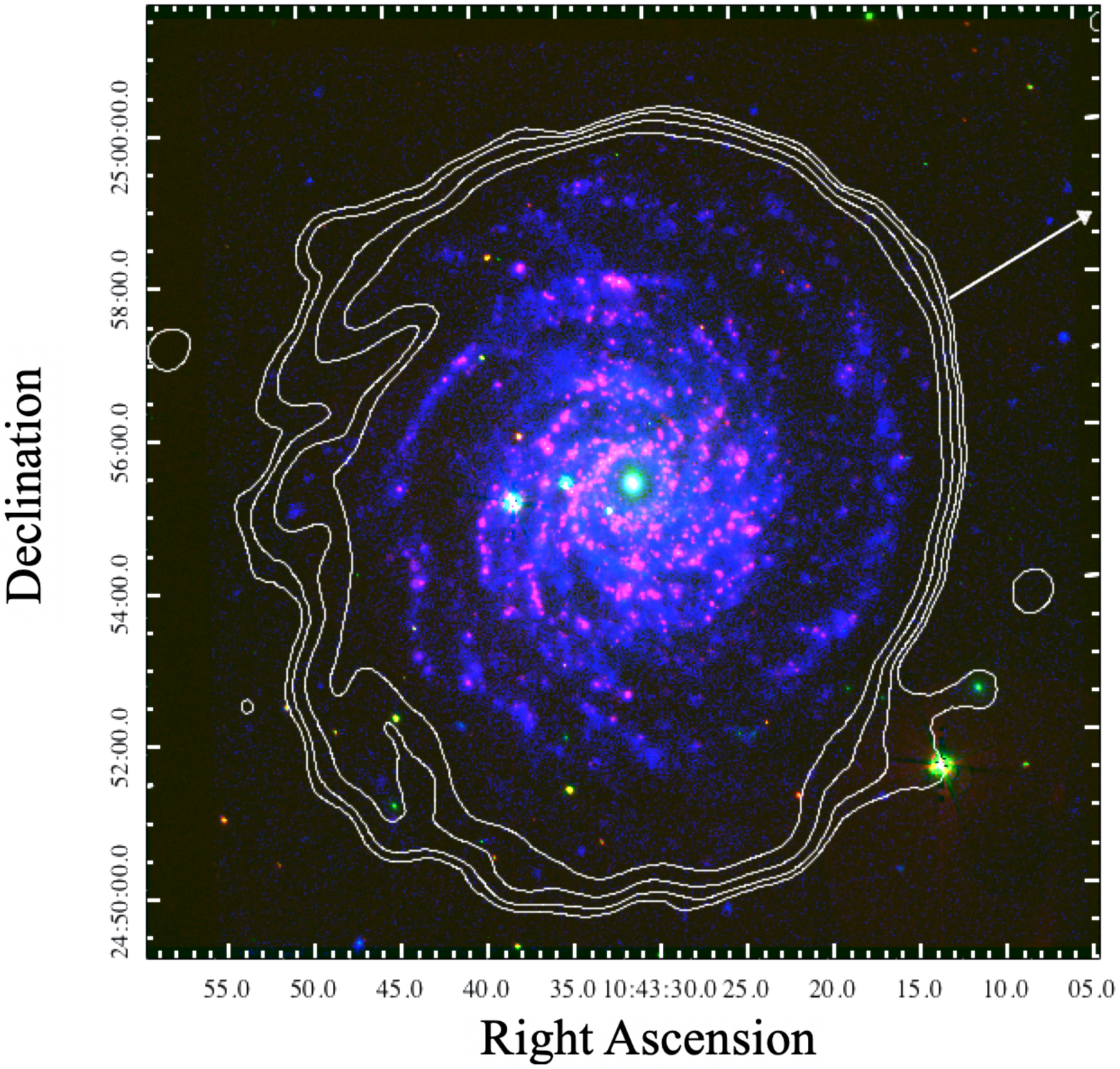} \label{fig:ngc3344col}} 
\setcounter{figure}{0}
\caption{(a) GALEX FUV image of NGC 3344 and the background QSO, SDSS J104241.27+250122.8 at an impact parameter of 29.5~kpc from the center of the disk. (b) False-color composite image of NGC 3344 with GALEX FUV in blue, VATT H$\alpha$ in red, VATT SDSS {\em g}-band in green showing the XUV disk
and the \hii\ regions. The white contours represent \hi-21~cm column densities of 3, 5, 10, 20$\times 10^{19}{\rm ~cm}^{-2}$.
The arrow indicates the direction to the quasar sightline. }
\label{fig:ngc3344}

\end{figure} 

\begin{table}
\begin{flushleft}
\centering
\vspace{-10pt}
\caption{Properties of NGC~3344}\label{tb:ngcprop}
\vspace{-0.2cm}
\begin{tabular}{lc}
\hline
\noalign{\smallskip} 
parameter & value \\
\noalign{\smallskip} 
\hline
\noalign{\smallskip} 
Alternate name & UGC 5840  \\
Right ascension, $\alpha_{2000}^a$  & 10$^h\;$43$^m\;31\fs1$\\
Declination, $\delta_{2000}^a$  & $+24^\circ\;55^\prime\;19\farcs9$  \\
Morphological Type$^b$ &  (R)SAB(r)bc \\
Redshift & 0.00194\\
Heliocentric systemic velocity$^c$ & 580 km s$^{-1}$  \\
Distance$^d$ & 8.28 Mpc\\
R$_{25}^{e}$& 7.96 kpc  \\
E(\bv)$^{f}$& 0.0285  \\
SFR$^{eg}$ & 0.455 M$_\odot$ yr$^{-1}$ \\
Stellar mass$^e$ & 1.69$\times10^{10}$ M$_\odot$ \\
 \ion{H}{1} mass$^e$ & 2.2$\times10^{9}$ M$_\odot$  \\
 Molecular mass $^h$ &   3.1$\times10^{8}$ M$_\odot$   \\
 
\noalign{\smallskip} 
\hline
\end{tabular}
\flushleft
Source: $^{\rm a}$ \cite{2mass06}; $^{\rm b}$ \cite{rc391}; 
$^{\rm c}$ \cite{epinat08}; $^{\rm d}$ \cite{sabb18}; $^{\rm e}$ measurement presented in this paper; $^{\rm f}$ \cite{schl11};$^{\rm g}$ estimated using FUV+\mips\ tracer, see \S\ref{sec:smsfrpro}; $^{\rm h}$  Total molecular gas mass extrapolated from the central value \citep{Lisenfeld11}.
\end{flushleft}
\vspace{-15pt}
\end{table}

In this pilot work, we investigate the radial variations in the optical, UV, and \hi\ emission and various galaxy properties. We also identify bright stellar complexes to explore the connections between \sfr\ and gas properties and detect absorption tracing the inner CGM of the galaxy. 
The paper is organized as follows: in \S\ref{sec:data} we describe the UV, H$\alpha$, 24$\mu$m, \hi\ 21 cm imaging and QSO UV-absorption spectroscopy data used in this study. This is followed by mapping the surface densities of SFR (\sfr) using FUV, FUV+\mips , \ha , and \ha+\mips\ tracers, stellar mass (M$_\star$), and properties of \hi\ gas, such as mass, velocities, velocity dispersion, and kinetic energy in \S\ref{sec:analysis}. We present the radial profiles of emission from stars and dust, identification of stellar complexes, followed by the investigation of the CGM in NGC~3344 in \S\ref{sec:results}. In \S\ref{sec:res}, we study the dust-corrected FUV and \ha\ tracers and explore the implications of the observed relations 
between star formation and ISM kinematics, and the CGM. Finally, we summarize our findings and their implications in \S\ref{sec:summ}.

\section{The Data} \label{sec:data}

\subsection{GALEX UV data}

UV imaging data for our study comes from the archives of {\em Galaxy Evolution Explorer} \citep[GALEX][]{mart05, morr07} obtained from Milkuski Archive for Space Telescope (MAST). Figure \ref{fig:fuv} shows the FUV map of NGC~3344.
The FUV and NUV maps were produced using GALEX Nearby Galaxy Survey 
(NGS) and All-Sky Imaging Survey (AIS) data by stacking all exposure time weighted tiles from the
target field. Each resulting image stack was trimmed to the size of $800\times800$ pixels centered at the galaxy with a scale of 
1.5\arcsec/${\rm pixel}$. 
The FUV ($\lambda_{\text{eff}}\sim1538.6 {\text\AA}$) and NUV 
($\lambda_{\text{eff}}\sim2315.7 {\text\AA}$) observations have angular resolutions (FWHM) of
4\farcs2 and 5\farcs3, respectively. The background flux was estimated from flux-free regions away from the galaxy and then
removed.  Corrections for Milky Way 
extinction were calculated using \cite{schl11} dust maps and the Galactic extinction curve for a total-to-selective
extinction of R$_{\text V} = 3.1$ derived by \cite{card89}. FUV and NUV flux densities are hence corrected with  A$_{\text{FUV}}=7.29\times E(\bv)$ and A$_{\text{NUV}}=8.0\times 
E(\bv)$. 
The 1$\sigma$ sensitivity limit of the FUV and NUV maps are 1.75$\times10^{-19}$ erg s$^{-1}$ 
cm$^{-2}$\AA$^{-1}$ (28.38 mag/\arcsec$^2$) and 1.28$\times10^{-19}$ erg s$^{-1}$ 
cm$^{-2}$\AA$^{-1}$ (26.43 mag/\arcsec$^2$), respectively.

\begin{figure*}[!tb]
    \setcounter{figure}{1}
\centering
 \subfloat[GALEX FUV]{\includegraphics[width=0.5\textwidth, trim = 0cm 1.5cm 3cm 1cm, clip,scale=0.25]{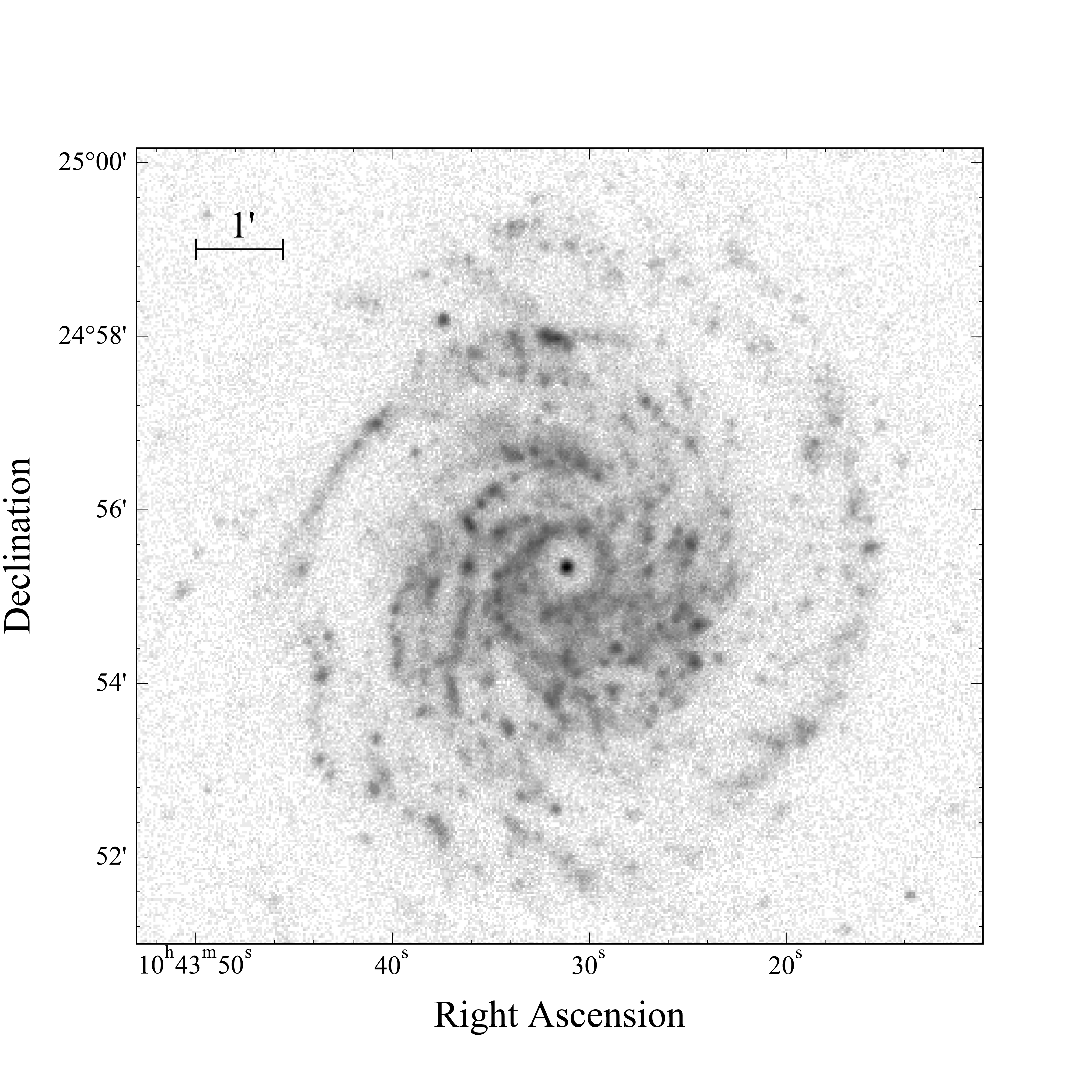}\label{fig:fuv}}\hfill
\subfloat[VATT \ha]{\includegraphics[width=0.49\textwidth, trim = 0cm 1.5cm 3cm 1cm, clip,scale=0.22]{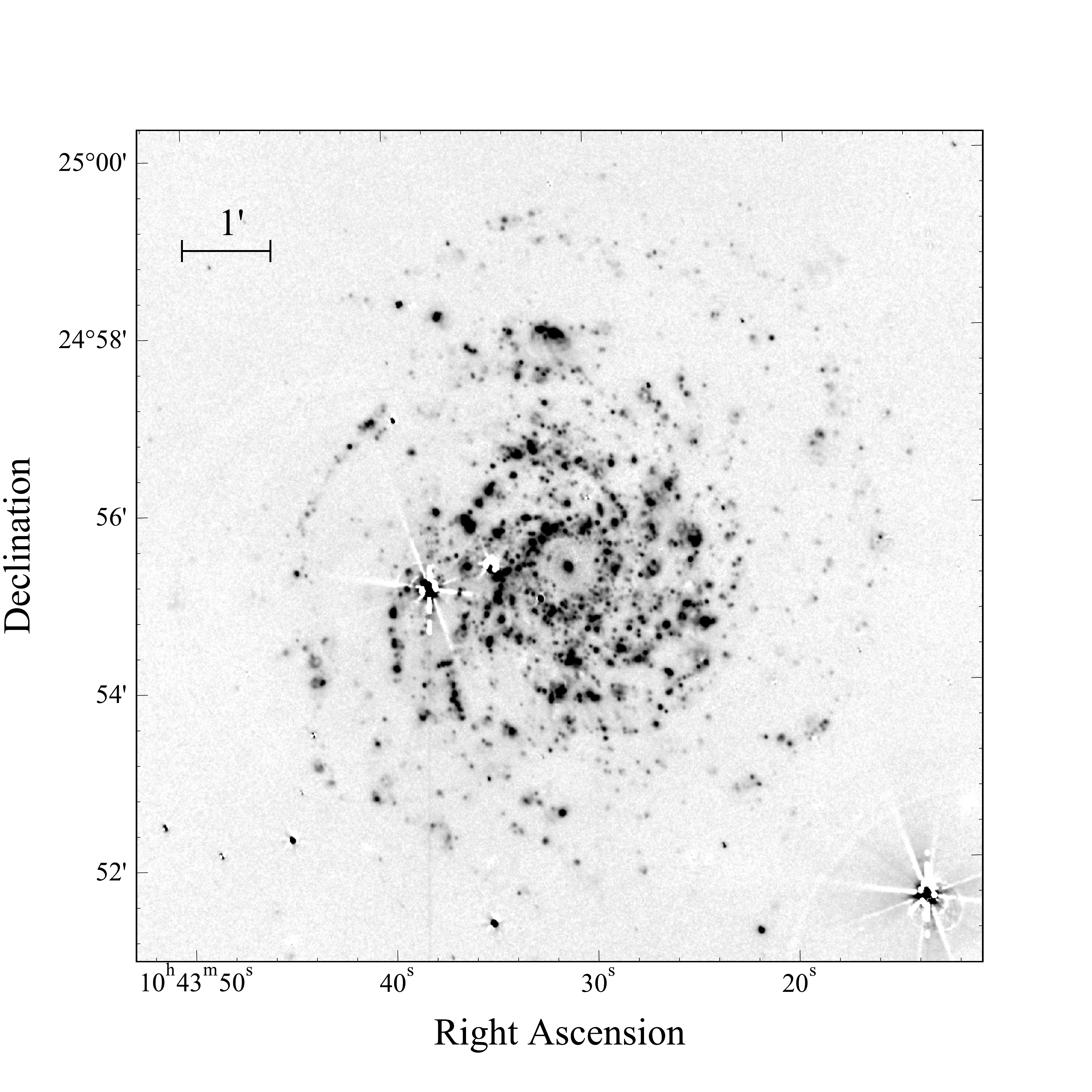}\label{fig:ha}}

 \subfloat[VATT r]{\includegraphics[width=0.49\textwidth, trim = 0cm 1.5cm 3cm 1cm, clip,scale=0.22]{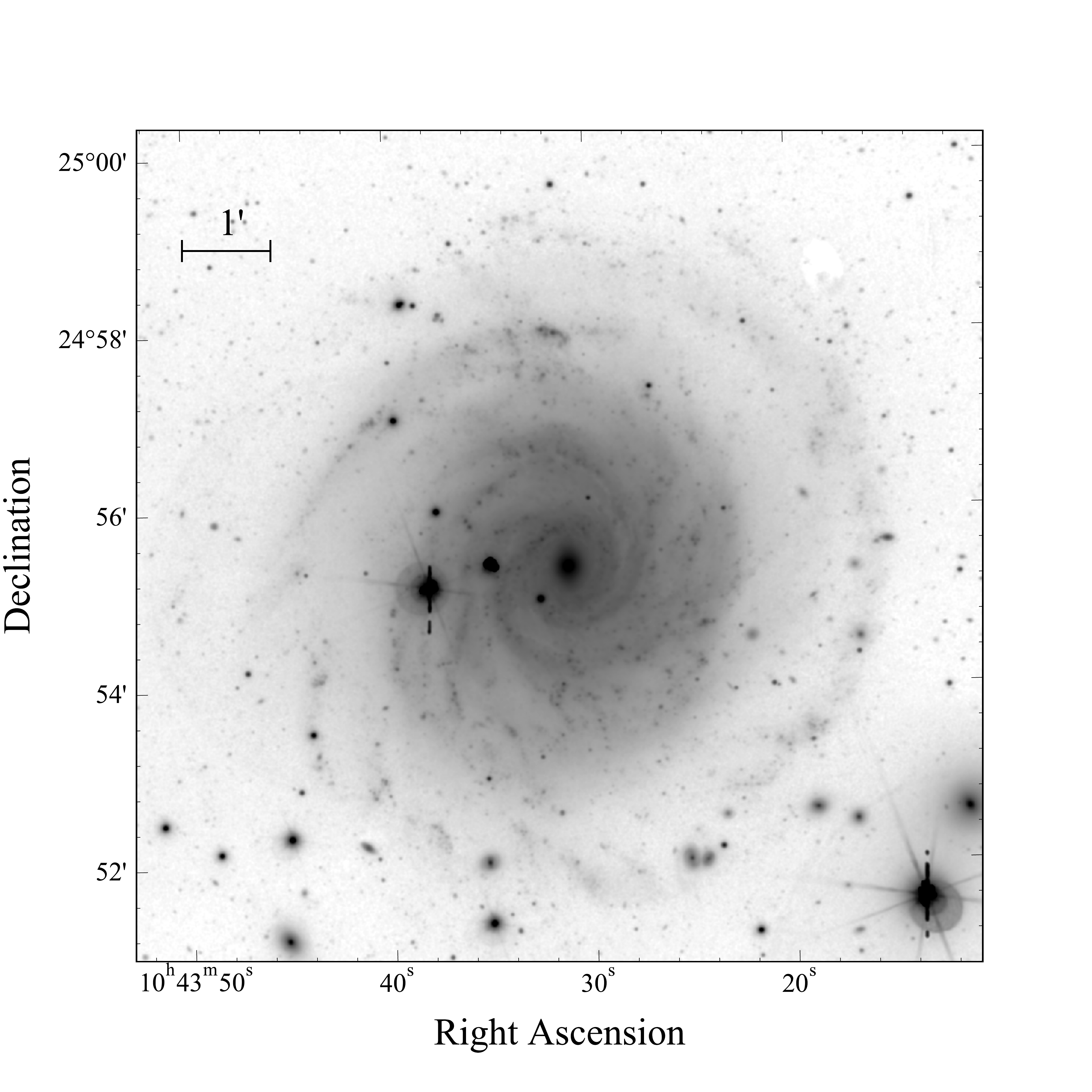}\label{fig:r}}\hfill
\subfloat[MIPS \mips]{\includegraphics[width=0.5\textwidth, trim = 0cm 1.5cm 3cm 1cm, clip,scale=0.22]{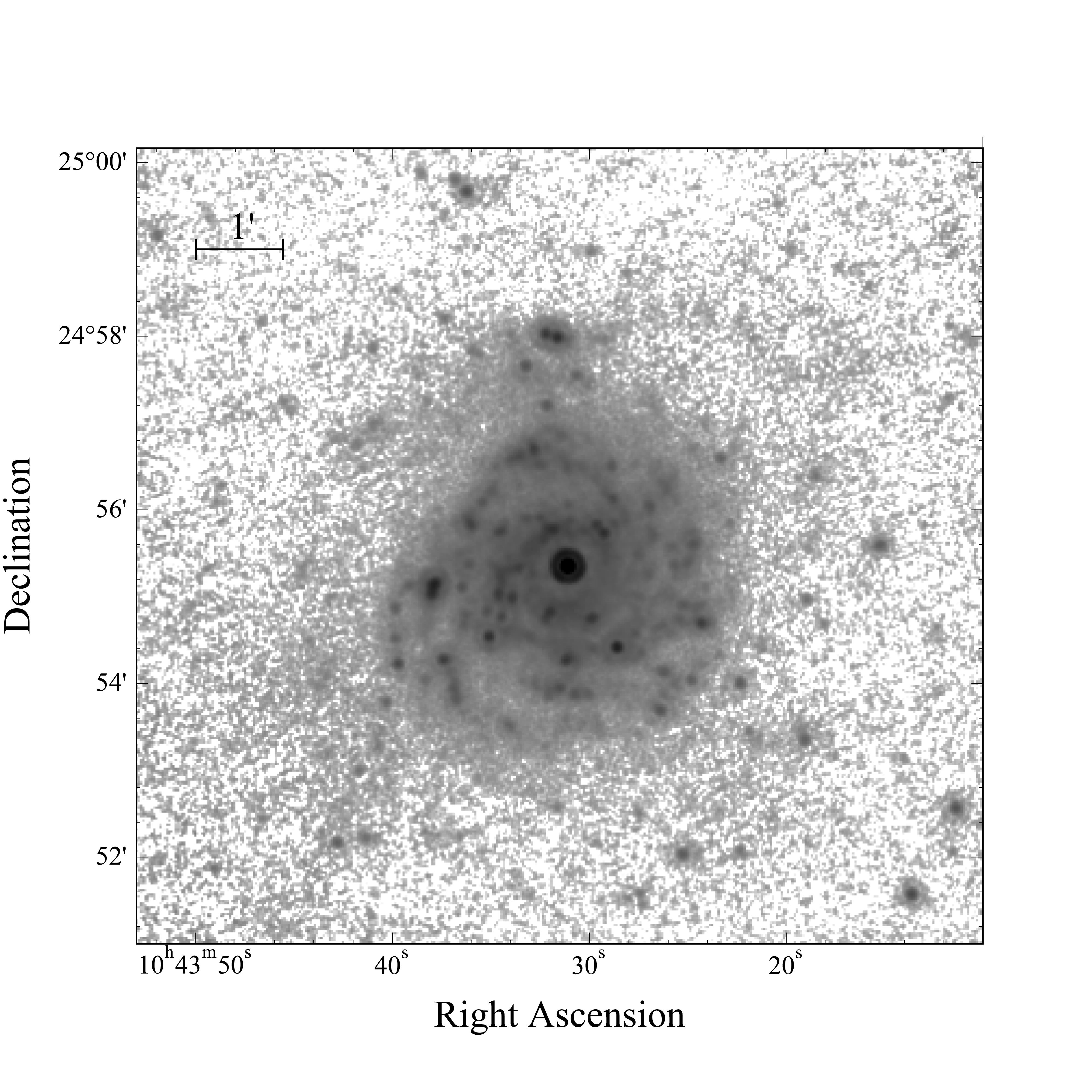}\label{fig:mips}}

\setcounter{figure}{1}
\caption{10\arcmin$\times$10\arcmin\,images of NGC 3344 as observed in (a) FUV, (b) H$\alpha$, (c) {\em r}, and (d) 24$\mu$m emission. 
The FUV image
trace stars $\lesssim$100 Myrs old, whereas the H$\alpha$
image is sensitive to the youngest population of massive stars of ages $\lesssim$10 Myrs. The {\em r} image traces old stellar populations while the 24$\mu$m map highlights dust. An angular scale of $1\arcmin$ corresponds to $\sim2.4$ kpc at the distance of NGC~3344 (8.28 Mpc) is shown on top left.\\ }
\label{fig:dat}

\end{figure*}

\subsection{VATT \ha\ data and reduction}
We obtained narrow-band H$\alpha$, continuum {\em r}-band and  {\em g}-band (in SDSS {\em r} and {\em g}) imaging of NGC~3344 on UT 2019 March 29, using the 
VATT4k CCD imager at the 1.8 m {\em Vatican Advanced Technology Telescope} (VATT) operated by the Mt. 
Graham Observatory. The VATT4k has a field of view of $\sim$12\farcm5$\times$12\farcm5, and a plate scale of 
0.375\pp/pixel, after $2\times2$ pixel
binning on read-out. The bandpass of the H$\alpha$ narrow-band interference filter is centered at 658 nm and has a nominal bandwidth of 5 nm. The observing conditions were 
photometric, measured with seeing between 0\farcs90, and 1\farcs1.  

The total integration times were 6000 s for narrow-band \ha\ image, 1200 s for continuum r-band image, and 120 s for the g-band image.  The data were reduced using
standard image processing routines in the Image Reduction and Analysis Facility (IRAF). The science images were 
bias subtracted and flat fielded using both dome and twilight sky flats. The sky 
level was measured by calculating the average intensity in the source-free regions of each exposure and subtracted from
the images. Cosmic rays were removed using the \texttt{L.A.COSMIC} 
routine \citep{vand01}. Field stars were used as a reference to shift and 
align the narrow-band and continuum images. This was carried out using the IRAF routine, {\tt imregister} \footnote{http://www.public.asu.edu/$\sim~$rjansen/iraf/rjtools.html}. All images were smoothed to the resolution of the worst seeing image. Stacks were created for the narrow-band \ha, and 
continuum {\em r}-band 
images by appropriately weighting the images. The {\em r} image was then scaled to the H$\alpha$ image. 
This involved measuring the ratio of count rates of individual field stars in the {\em r}-band and H$\alpha$ image to 
compute the scale factor. The scale factor was estimated empirically to minimize over/under subtraction in the galaxy region. We note that this method implicitly assumes the galaxy disk spectral energy distribution matches the foreground stars and is invariant with location. An emission-line only image of NGC~3344 was obtained by subtracting the 
scaled {\em r}-band image from the H$\alpha$ image.

For flux calibration, we used SDSS DR7 r-band photometry of NGC~3344. We assumed a standard atmospheric extinction 
coefficient of 0.08 mag airmass$^{-1}$. Instrumental magnitudes 
of field stars were determined through aperture photometry from the {\em r} image. The photometric zero point of the {\em r}-band image 
was then calculated by comparing the instrumental magnitudes to the SDSS-r 
magnitudes for the same stars. The photometric zero point of the H$\alpha$ image was calculated using the scale factor and the photometric zero point of the {\em r}-band image.    

The narrow-band filter used for the observation also covers  
the neighbouring [\ion{N}{2}] $\lambda\lambda6548, 
6583$\AA\ forbidden lines along with H$\alpha$. 
\cite{kenn08} obtain disk-averaged [\ion{N}{2}]/H$\alpha$ = 0.52 for NGC~3344, which we use to scale the image to the net H$\alpha$ surface brightness in order to account for contamination by [\ion{N}{2}]. 
A foreground galactic extinction correction is also applied, with A$_{H\alpha}=2.5\times E(\bv)$. The
final reduced H$\alpha$ and {\em r}-band image is shown in Figure \ref{fig:ha} and \ref{fig:r}. The 1$\sigma$ sensitivity limit of the 
final H$\alpha$ image is 6.1 $\times10^{-21}$ erg s$^{-1}$ 
cm$^{-2}$\AA$^{-1}$. 

\subsection{MIPS \mips\ Data} 
NGC~3344 was observed at \mips\ as part of the
{\em Spitzer} Local Volume Legacy survey \citep{dale09}, using the
Multiband Imaging Photometer for Spitzer (MIPS) instrument
\citep{mips} with the effective exposure time of 147 s. The \mips\ image used in this work has a resolution of 6\farcs0 FWHM, a 1$\sigma$ sensitivity limit of 6.81 $\times10^{-2}$ MJy sr$^{-1}$
and is shown in Figure \ref{fig:mips}.

\subsection{VLA \hi~21cm Data}

We combined archival and newly observed data for the atomic \hi\ 21 cm emission from NGC~3344. The archival data were obtained from the program AB365 observed with the NSF’s Karl G. Jansky Very Large Array (VLA)\footnote{The National Radio Astronomy Observatory is a facility of the National Science Foundation operated
under cooperative agreement by Associated Universities, Inc.} in its D-configuration on UT 1985, December 6, with a total integration time of 11h 52 m. These observations were made with a channel spacing of 48.8~kHz (10.36~\kms) in 31 channels. We used the Common Astronomy Software Application version 5.4.0 \citep[CASA; ][]{mcmullin07} to reduce the data following the standard \hi\ data reduction. We used 3C286 as a flux calibrator and 1108+201 and 1040+123 as phase calibrators. These observations were observed in the B1950 coordinate frame and were converted to J2000 using CASA task \textit{fixvis}. 

New observations were performed with VLA in C- and B-configurations (project id: 20A-125). The C-configuration observations were carried out for a total of 12 hours from UT 2020, April 13 to UT 2020, April 21, and the B-configuration observations were for a total of 36 hours from UT 2020, UT July 7 to UT 2020, August 28. These observations had a channel spacing of 7.812~kHz (1.65~\kms). The data reduction was performed with CASA 5.6.1 using 3C286 as a flux calibrator and J1021+2159 as a phase calibrator. Hanning smoothing was applied to the data in order to reduce the Gibbs ringing, which made the effective velocity resolution to be 3.3~\kms. The target field in C- and B-configuration observations were also hampered by side lobes of two strong sources outside the primary beam. These were removed by modeling the point sources with the CASA task \textit{tclean} and removing the modeled point sources with the task \textit{uvsub}. The amplitude errors of the target image in C-configuration were observed and corrected by multiple self-calibrations with a solution interval of 40 seconds. 
The self-calibration was stopped when the range of gain phases in the self-calibration solution was less than 1 degree.

After concatenating the data in each configuration with the task \textit{concat}, we calculated the weight of each configuration data based on the scatters of visibilities in the line-free channels using the task \textit{statwt}.
The weights were estimated with channels outside the \hi\ spectra to adjust relative weights among observations with different scales, i.e., D-, C-, and B-configurations. The final image cube was made using the task \textit{tclean} with a cell size of 1\farcs5 and 1280$\times$1280 pixels. The channel width of the final image cube is 48.8~kHz, the same as the D-configuration data. The final image cube was smoothed to a synthesized beam size of 6\farcs7$\times$5\farcs4. The RMS noise was $\sim$185 $\mu$Jy beam$^{-1}$ per channel in the line-free channels. Figure \ref{fig:ngc3344} shows the \hi\ 21 cm column
density contours at 3, 5, 10, 20 $\times10^{19}$ cm$^{-2}$.

\subsection{QSO-absorption Spectroscopy with HST}

The inner CGM of NGC~3344 was probed via UV absorption spectroscopy of the background QSO, SDSS~J104241.27+250122.8, at an impact parameter of 29.5~kpc ($\sim$12\farcm29). 
The sightline is 19.26~kpc 
from the closest point on the \hi\ disk as mapped by VLA, down to an \hi\ column density of $\rm 3 \times  10^{19}~cm^{-2}$.

\begin{figure}[!t]
  \setcounter{figure}{2}
    \centering
    \includegraphics[trim = 0cm 0mm 10cm 0cm, clip,scale=0.65]{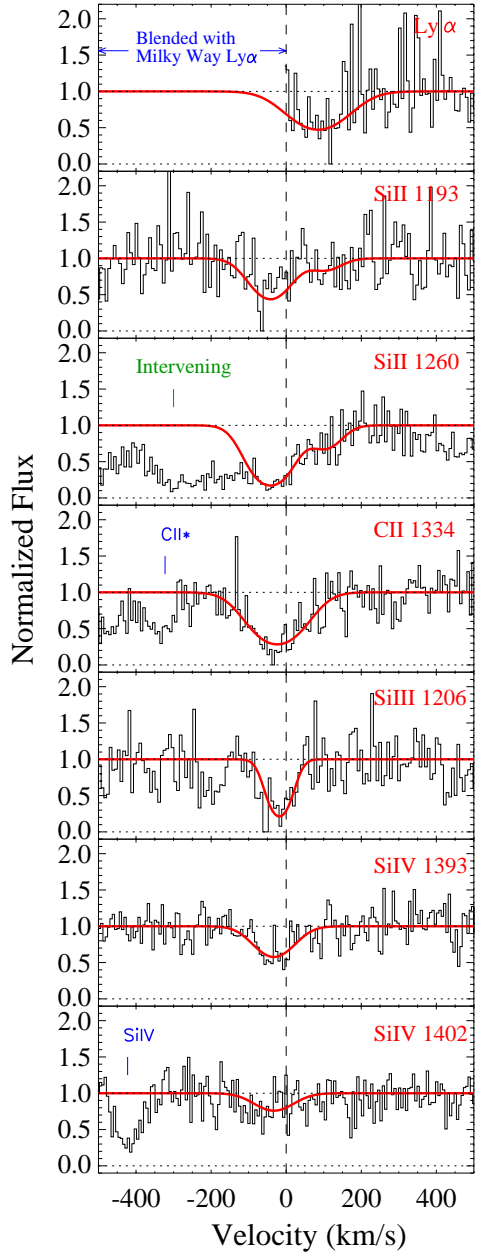}
    \caption{ COS FUV spectra of QSO, SDSS J104241.27+250122.8, tracing the circumgalactic medium 
  of NGC~3344 at an impact parameter of 29.5 kpc from the center of the disk. We detected \ion{H}{1} $\lambda$ 1215 (\Lya), \ion{Si}{2} $\lambda\lambda$ 1193, 1260, 
    \ion{C}{2} $\lambda$1334, \ion{Si}{3} $\lambda$1206, \ion{Si}{4} 
    $\lambda\lambda$1393, 1402. We detected two components in the absorption profile. The weighted average puts the strongest metal-line component at -28.5~~km~s$^{-1}$ and the weaker component at 
    88.9~km~s$^{-1}$, respectively. The second component was only detected in
   Ly$\alpha$ and \ion{Si}{2} indicating its low-column density and low-ionization state. }
    \label{fig:cos_spec}
        \setcounter{figure}{3}
\end{figure}

The observations were carried out with G310M grating of the Cosmic Origins Spectrograph \citep[(COS);][]{greenc12} aboard the Hubble Space Telescope (HST) for a total exposure of 10069~s under the COS-DIISC Survey\footnote{COS-DIISC is a large HST program (program ID: 14071) aimed at using UV bright QSO to trace the disk-CGM interface in 34 low-redshift with COS aboard the HST.} (Borthakur et al. in prep). The spectra covered a wavelength range from 1140-1430~$\rm \AA$ in the observed frame. 
The data were calibrated and reduced using the standard COS pipeline \citep[procedure described in the COS Data Handbook;][]{rafel18}.
The spectral resolution of the data was $\approx$ 20~\kms ~(R$\sim$15000). 
Owing to the low redshift of the galaxy ($z$=0.00194), the rest-frame wavelength coverage is almost identical to that of the observed frame. The data covered multiple line transitions including \ion{H}{1} $\lambda$1215 (\Lya), \ion{Si}{2} $\lambda$ 1190, 1193, 1260, \ion{Si}{3} $\lambda$1206,  \ion{Si}{4} $\lambda\lambda$1393, 1402, and \ion{C}{2} $\lambda$1334. The COS FUV spectra is shown in Figure \ref{fig:cos_spec}.

\begin{deluxetable*}{llc crr  ccc   }
\tabletypesize{\scriptsize}
\tablecaption{Sightline toward J1042+2501 at z=0.0020}\label{tb:qso}
\tablewidth{0pt}
\tablehead{
\\ \colhead{Species} & \colhead{$\lambda_{rest}$}  & \colhead{$W_{rest}^a$} &\colhead{Centroid}  & \colhead{b} &  \colhead{log N}\\
\colhead{} & \colhead{($\rm \AA$)} &  \colhead{($\rm m\AA$)} &  
\colhead{(km~s$^{-1}$)}  & \colhead{(km~s$^{-1}$)}  &\colhead{(log cm$^{-2}$)} 
&\colhead{} } 
\startdata
 HI$^b$ & 1215.67  &  $>$ 432$^b$ &   85.04 $\pm$ ~9.36  & 102$\rm ^{+17}_{-15}$ &$>$ 14.00$^b$~~~~~~~\\
 SiII$^c$ &       1260.42  &   ~~  668  &  -41.77 $\pm$ 24.37  &  69$\rm ^{+50}_{-29}$ & $>$ 13.81$\pm$0.19 \\
      &                &          &  102.37 $\pm$ 33.05  &  54$\rm ^{+101}_{-35}$ & 13.05$\pm$0.33 \\
 SiII$^c$ &       1193.29  &   ~~  375  &   -41.77 $\pm$ 24.37  &  69$\rm ^{+50}_{-29}$ & $>$ 13.81$\pm$0.19 \\
      &                &          &  102.37 $\pm$ 33.05  &  54$\rm ^{+101}_{-35}$ & 13.05$\pm$0.33 \\
 SiII &       1190.42  &     $\le$ 370  &  $-$ & $-$ ~~~~~ & $\le $ 14.07 ~~~~~~~~\\
 SiII$^d$ &       1304.270  &     $-$   &  $-$ & $-$ ~~~~~ & $-$  ~~~~~\\
 OI$^d$   &       1302.168  &     $-$   &  $-$ & $-$ ~~~~~ & $-$  ~~~~~\\
 CII$^e$ &        1334.53  &   ~~  608  &  -23.87 $\pm$ 13.75  &  90$\rm ^{+20}_{-17}$ & $>$ 14.69$\pm$0.09 \\
 SIII$^e$ &       1206.50  &   ~~  272  & -18.24 $\pm$ 20.36 &  39$\rm ^{+30}_{-17}$ & $>$ 13.3 $\pm$0.26  \\
 SiIV$^f$ &       1393.78  &   ~~  228  & -32.67  $\pm$   13.77   &  69$\rm ^{+21}_{-16}$ & 13.6 $\pm$0.11  \\
 SiIV$^f$ &       1402.77  &   ~~  145  & -32.67  $\pm$   13.77   &  69$\rm ^{+21}_{-16}$ & 13.6 $\pm$0.11  \\
\enddata
\tablenotetext{a}{Due to low signal-to-noise ratio in our data, equivalent widths were estimated based on the Voigt profile fits.}
\tablenotetext{b}{ \Lya\ profile arising from the process of de-blending it from the Milky Way's \Lya\ profile. Uncertainties from the de-blending process add to the uncertainty in the H\,{\tiny I} column density and were not quantified.}
\tablenotetext{c}{The fit was produced by simultaneously fitting the Si~II 1260 and the 1193 transitions. Profile is likely saturated as the ratio of the two equivalent widths is 1.8 as opposed to the predicted value of 2.}
\tablenotetext{d}{No measurement could be made due to overlap with geocoronal OI, Milky Way Si~II 1304, and a large set of intervening and QSO host galaxy line transitions.}
\tablenotetext{e}{The fits are likely saturated, although the level of saturation could not be determined.}
\tablenotetext{f}{The fit was produced by simultaneously fitting the Si~IV 1393 and 1402 transitions. The profile is likely saturated as the ratio of the two equivalent widths is 1.8 as opposed to the predicted value of 2.}
\end{deluxetable*}

The spectra were normalized by identifying a continuum. 
 Absorption-free regions within $\pm1500$~\kms~from the line center were identified visually to fit a continuum except for the \Lya\ transition where we chose a region of $\pm4000$~\kms\ in order to 
cover the Milky Way damped \Lya\ profile (DLA). 
The continuum was estimated by fitting a Legendre polynomial of order between 1 and 5, similar to the procedure used by \cite{semb04}. This corrected for any low variations in the QSO flux near the position of the lines.
Then we proceeded with fitting Voigt profile to the absorption features to estimate velocity centroid, column density, and the Doppler b-parameter which defines the width of the observed spectral line of the profile.
 The fitting applied the appropriate line-spread functions \citep[LSFs; ][]{oste11} for the aperture of the spectrograph from the COS Instrument Handbook \citep{dash19}. 
The associated uncertainties were estimated using the error analysis method published by \cite{semb92}.
The prescription includes continuum placement uncertainties, Poisson noise fluctuations, and zero-point uncertainties. The estimated properties of the absorption features are shown in Table 2.

The \Lya\ transition associated with NGC~3344 was blended with the Milky Way's \Lya\ absorption feature. Milky Way's neutral hydrogen column produced a damped \Lya\ (DLA) profile at $\lambda~1215.67~\rm \AA$, which we modeled based on the contamination-free regions. The model was then subtracted out of the data to enhance the absorption feature associated with NGC~3344, which was located in the damping wings of the Milky Way's \Lya\ profile. In spite of this correction, we were not able to completely recover the blueward (low-velocity end) \Lya\ absorption profile for NGC~3344 owing to low signal strength at the base of the DLA. The redward part of the profile was retrieved with high confidence. Hence, a full estimate of \hi\ column density could not be made.

Fortunately,  none of the metal lines were subject to blending with corresponding Milky Way's lines. Therefore, we were able to use the metal-lines profiles to derive the structure of the CGM blueward of the galaxy's systemic velocity. It is worth noting that we may have missed low column density clouds at velocities blueward of the line-center that are too weak to produce metal lines.


\newpage
\section{Derived Galaxy properties} \label{sec:analysis}

\subsection{SFR Surface Density}\label{sec:sfrmap}

For the comparison of different tracers of star formation, we compute unobscured and internal-dust corrected SFR surface densities (\sfr).  We calculate \sfr\ for both FUV and \ha\ luminosities and use \mips\ to correct for dust. 

FUV flux is sensitive to {\em recent} star formation 
over timescales of 100 Myrs \citep[][and reference therein]{kenn98} as it stems from the 
photospheres of massive O and B stars. We use the prescription provided in \cite{kenn98, verl09, lee09} to estimate FUV \sfr.  For FUV luminosity, $I_{\rm{FUV}}$, in MJy sr$^{-1}$, \sfr\ is
\begin{equation}\label{eq:1}
\centering
     \Sigma_{\rm SFR}\rm{(FUV)}~[\rm{M}_\odot \rm{~yr}^{-1} \rm{~kpc}^{-2} ] = 0.17~{\rm I}_{\rm{FUV}}
\end{equation}

However, high energy UV photons originating from the photospheres of these massive stars are absorbed by
small dust grains, which in turn produce thermal emission at \mips. As a result, a combination of FUV and \mips\ is required to recover both 
unobscured SFR via FUV \citep{sali07} and dust-embedded SFR
via \mips\ \citep{calz07}. 
To compute FUV+\mips\ \sfr s, we follow the method prescribed in \cite{lero08}. For FUV and \mips\ luminosity,  $I_{\rm{FUV}}$ and $I_{24}$, respectively, in MJy sr$^{-1}$, the \sfr\ is estimated as:

\begin{equation}\label{eq:2}
\centering
    \Sigma_{\rm SFR} \rm{(FUV} + 24 \mu \rm{m)} ~[\rm{M}_\odot \rm{~yr}^{-1} \rm{~kpc}^{-2} ] = 0.081~I_{\rm{FUV}} + 0.0032~I_{24}
\end{equation}
For the UV luminosity to SFR calibration, equation \ref{eq:2} assumes a \cite{kroupa01} initial mass function (IMF) with maximum
mass of 120~M$_\odot$ while equation \ref{eq:1} assumes a 0.1-100~M$_\odot$ \cite{salp55} IMF. A correction from Salpeter to Kroupa IMF is done by dividing the calibration constant by 1.59 \citep{lero08} in equation \ref{eq:1} .

Additionally, to trace the {\em current} star formation, we estimate \sfr\ from \ha\ and \mips. Independent of previous star formation 
history, \ha\ emission from \hii\ regions is sensitive to star formation over timescales of $\lesssim 10$ Myrs. Further, the combination of \ha\ and \mips\ gives the total star formation \citep{kenn07} from unobscured and dust-obscured components of star formation. For \ha\ luminosity, $I_{{\text H}\alpha}$, in MJy sr$^{-1}$, \sfr\ is:
\begin{equation}\label{eq:3}
   \centering 
     \Sigma_{\rm{SFR}} \rm{(H}\alpha\rm{)}~[\rm{M}_\odot \rm{~yr}^{-1} \rm{~kpc}^{-2} ]= 4.32~{\rm I}_{\rm{H}\alpha} 
\end{equation}
The non-dust corrected \ha\ \sfr\ calculated using the calibration provided in \cite{kenn98, blanc09, kenn12} assumes solar abundance and 0.1-100~M$_\odot$ Salpeter IMF.
\cite{gall18} provide a prescription for the 
H$\alpha$+24$\mu$m
\sfr\ with a 0.1-100~M$_\odot$ Kroupa IMF. For I$_{{\text H}\alpha}$ and I$_{24}$ in units of erg~s$^{-1}$~cm$^{-2}$, and MJy~sr$^{-1}$  respectively, \ha+\mips\ \sfr\ is:
\begin{center}
$ \Sigma_{\rm SFR} {\rm(H}\alpha+24\mu{\rm m)}~[\rm{M}_\odot \rm{~yr}^{-1} \rm{~kpc}^{-2} ]= 634~I_{{\rm H}\alpha} + 0.0025~{\rm I}_{24}$
\end{center}
Converting $I_{{\text H}\alpha}$ to MJy sr$^{-1}$, we obtain:
\begin{equation}\label{eq:4}
\Sigma_{\rm SFR} {\rm(H}\alpha+24\mu{\rm m)}~[\rm{M}_\odot \rm{~yr}^{-1} \rm{~kpc}^{-2} ]= 2.9~I_{{\rm H}\alpha} + 0.0025~{\rm I}_{24}
\end{equation}
Once again, correction from Salpeter to Kroupa IMF is obtained by dividing the calibration constant by 1.59.

\begin{figure*}[t] 
  \setcounter{figure}{3}
  \centering
 
 \subfloat[FUV+\mips\ SFR Surface Density]{\includegraphics[width=0.49\textwidth, trim = 0cm 3.5cm 0cm 3cm, clip,scale=0.26]{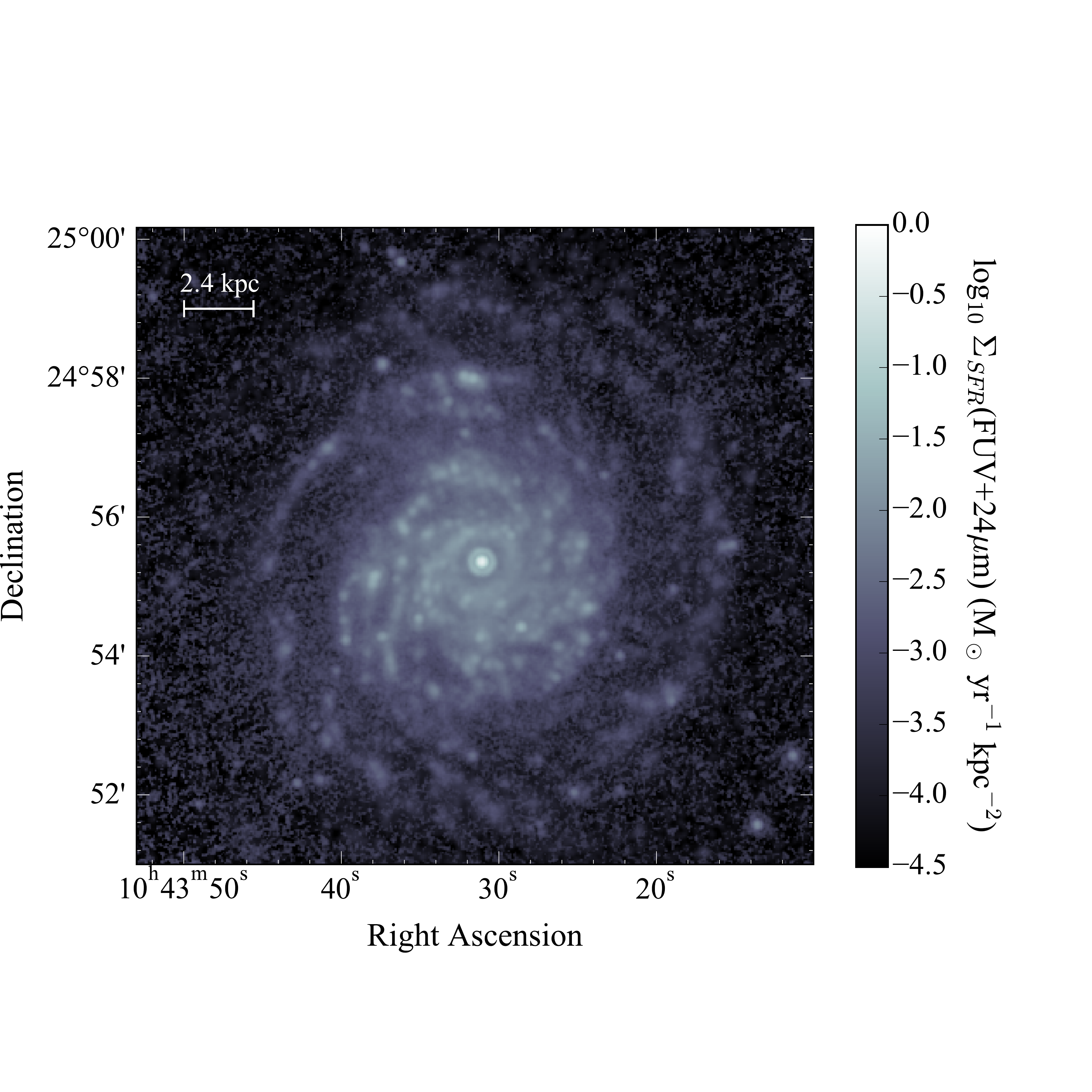} \label{fig:fuvsfrd}} 
   \hfill
 \subfloat[\ha +\mips\ SFR Surface Density]{\includegraphics[width=0.49\textwidth, trim = 0cm 3.5cm 0cm 3cm, clip,scale=0.26]{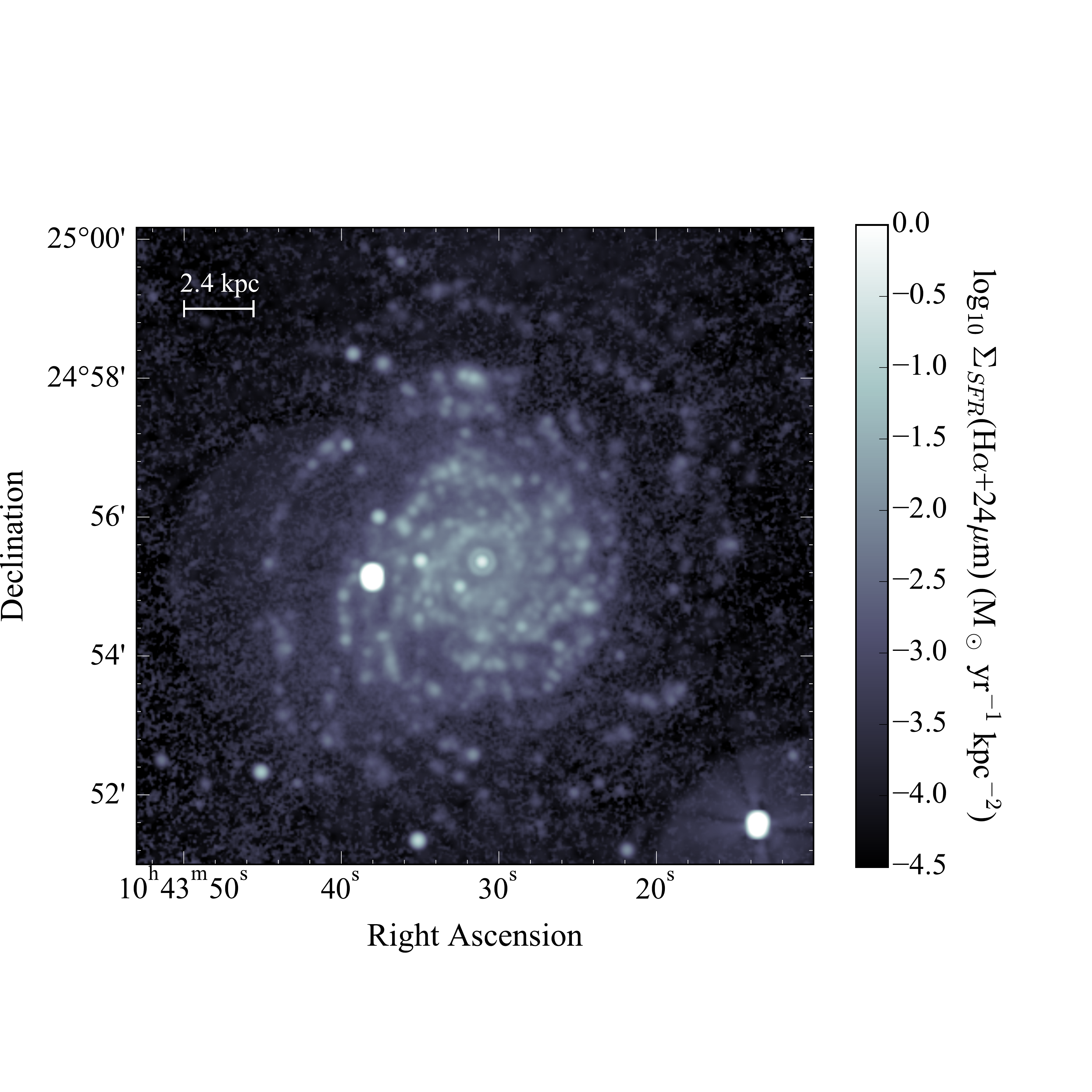} \label{fig:hasfrd}} 

\setcounter{figure}{3}
\caption{Star formation rate surface density, \sfr , in NGC 3344 obtained by combining  {(\textit a)} FUV and \mips\ maps and {(\textit b)} \ha\ and \mips\ emission following the prescriptions shown in equation \ref{eq:2} and \ref{eq:4} and color coded in M$_\odot$~yr$^{-1}$~kpc$^{-2}$. The maps have a resolution of 6\arcsec\,FWHM and a scale of 1.5 \arcsec/pixel. Foreground stars visible in the \ha+\mips\ map were masked. A physical scale of 2.4 kpc (1\arcmin) is shown on top left corner. }
    \label{fig:sfrmap}

\end{figure*}

Figure \ref{fig:fuvsfrd} and \ref{fig:hasfrd} show the FUV+\mips\ and \ha+\mips\ \sfr\ maps for NGC~3344, respectively. 
The resolution of these maps is 6\arcsec. These were generated by matching the PSFs of the GALEX FUV and VATT \ha\ data
to the PSF of MIPS \mips\ data.
The FUV and \ha\ images were gridded to the pixel scale of the 
\mips\ image. Gaussian smoothing was performed to match the PSFs of FUV and \ha\ images 
with widths $\sigma$ to the PSF of MIPS \mips\ image with width $\sigma_{24}$. 
The width of the convolution kernel, $\sigma_{kern}$ in pixels 
was then calculated using $\sigma_{24}^2=\sigma_{kern}^2+\sigma^2$ 
for the FUV and \ha\ images. The images were individually smoothed using
Python {\texttt {astropy}} functions {\it Gaussian2DKernel} and {\it convolve}. 

\begin{figure}[!t]
\setcounter{figure}{4}
    \centering
    \includegraphics[trim = 0cm 3.5cm 0cm 3cm, clip,scale=0.25]{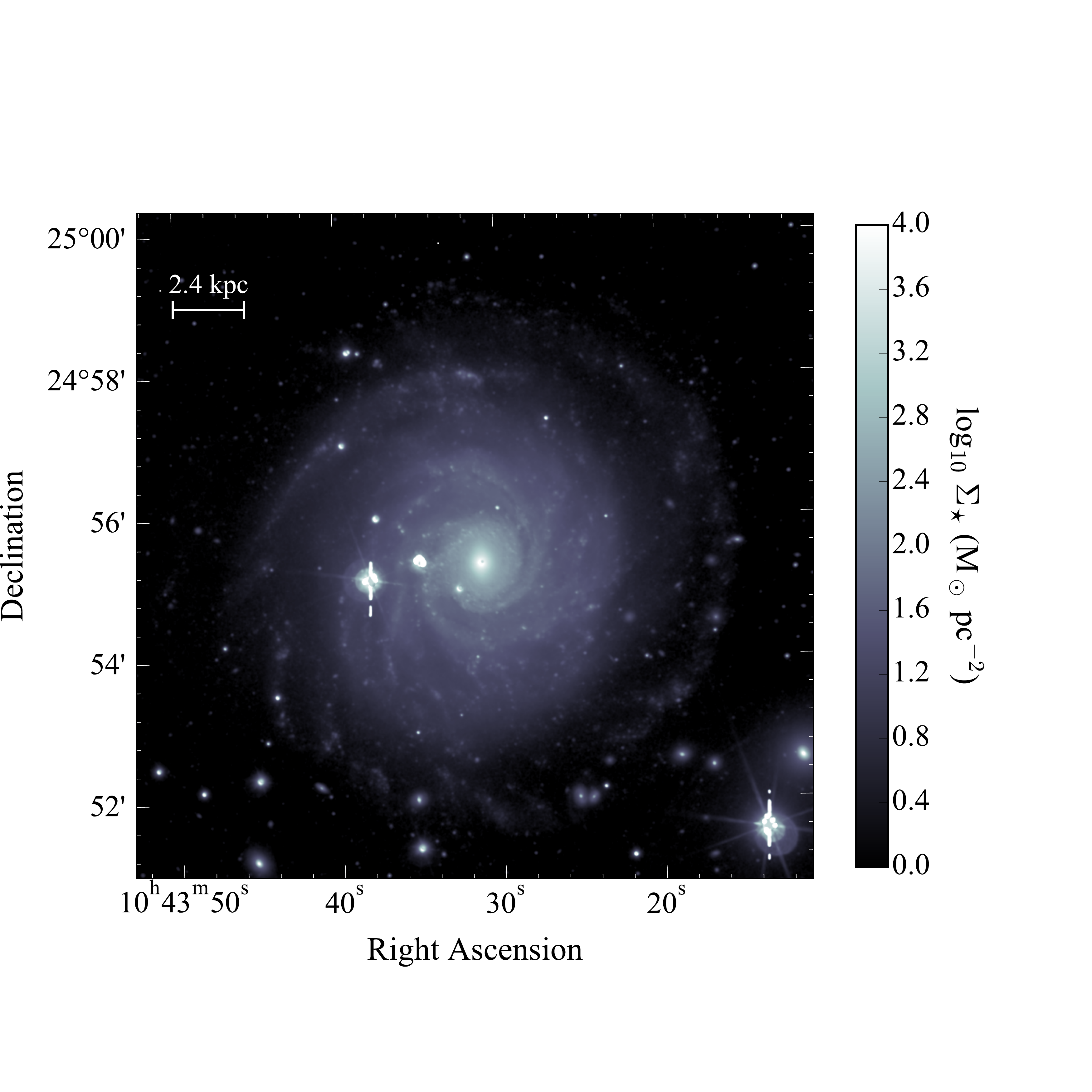}
    \caption{Stellar mass surface density, \sm , in NGC 3344 derived using ({\em g-r}) colors and the prescription shown in equation \ref{eq:5}, color coded in M$_\odot$~pc$^{-2}$. The map has a resolution of 1\farcs68 FWHM and a scale of 1.5\arcsec /pixel. Foreground stars in the field of view of the galaxy were masked. A physical scale of 2.4 kpc (1\arcmin) is shown on top left corner. }
    \label{fig:smpc}
    \setcounter{figure}{4}
\end{figure}

\subsection{Stellar Mass}\label{sec:sm}

To calculate the stellar mass surface density, we follow the prescription presented in \cite{yang07}. They use the relation between
the stellar mass-to-light ratio and color from \cite{bell03} and compute
\begin{equation}\label{eq:5}
    \log\bigg[\frac{{\rm M}_{\star}}{h^{-2} {\rm M}_\odot}\bigg] = -0.406 + 1.097(g-r) - 0.4({\rm M}_r-5\log h - 4.64)
\end{equation}
\\where ({\em g$-$r}) and M$_r$ refer to the color derived from SDSS {\em g}- and {\em r}- band colors and absolute magnitude in SDSS {\em r}-band. We adopt $h=H_0/(100~{\rm km~s}^{-1}{\rm Mpc}^{-1})=0.73$ to derive stellar mass. To create a spatial stellar mass map, we estimate ({\em g$-$r})  at each pixel. The stellar masses are then converted to stellar mass surface density (\sm) using a scale value of 40.14 pc/\arcsec\ for NGC~3344 at a distance of 8.28 Mpc. Figure \ref{fig:smpc} shows the derived \sm\ map of NGC~3344. 

\subsection{Properties of the Interstellar Medium}\label{sub:ism}

\begin{figure*}[!t] 
\setcounter{figure}{5}
\centering

 \subfloat[\hi\ mass surface density,  \mhi]{\includegraphics[width=0.49\textwidth , trim = 0mm 5cm 3cm 5cm, clip,scale=0.25]{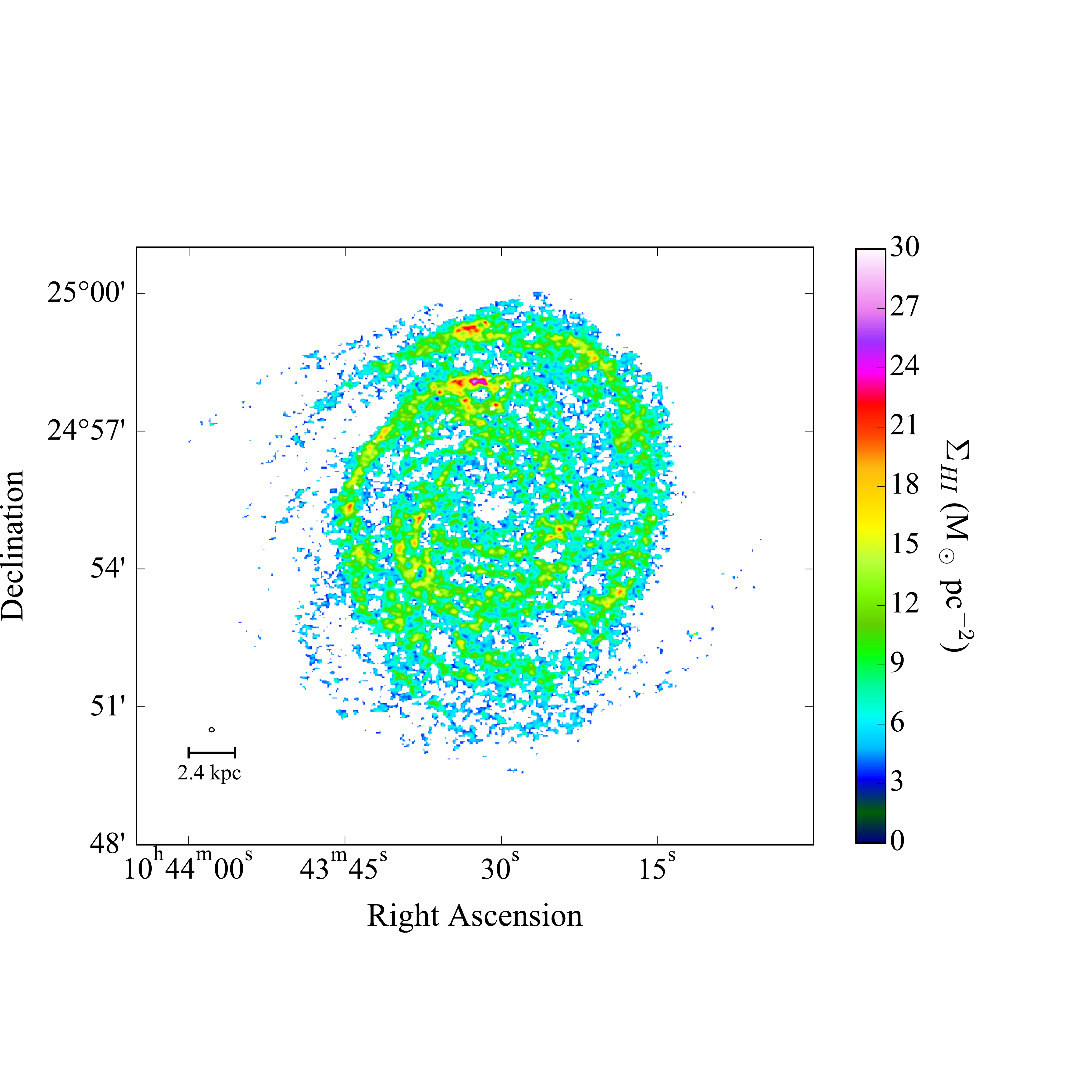}}\hfill
 \subfloat[\hi\ velocity, \vlos]{\includegraphics[width=0.49\textwidth , trim = 0mm 5cm 3cm 5cm, clip,scale=0.25]{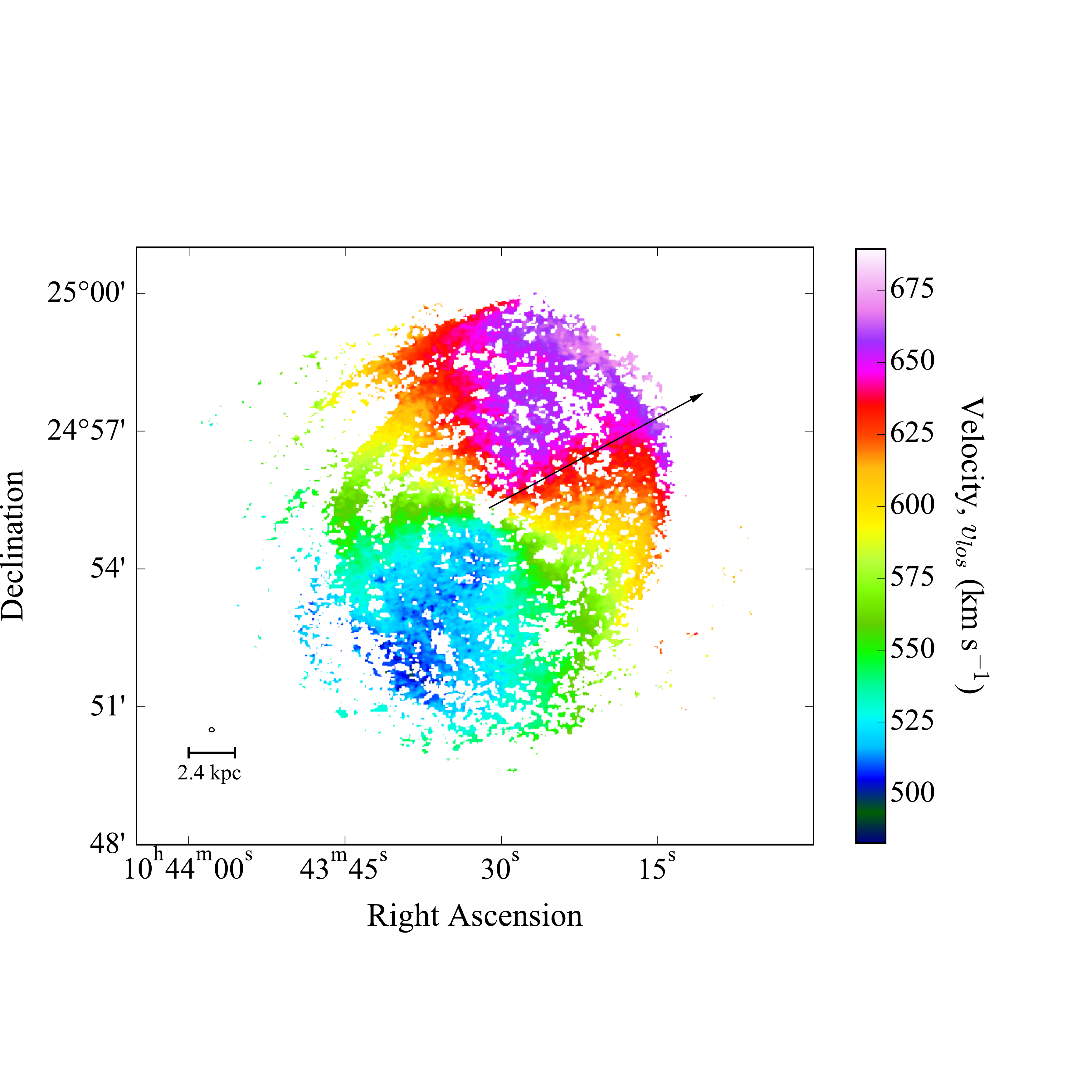}\label{fig:velarrow}} 
 \vfill
\subfloat[ \hi\ velocity dispersion, \disp]{ \includegraphics[width=0.49\textwidth , trim = 0mm 5cm 3cm 5cm, clip,scale=0.25]{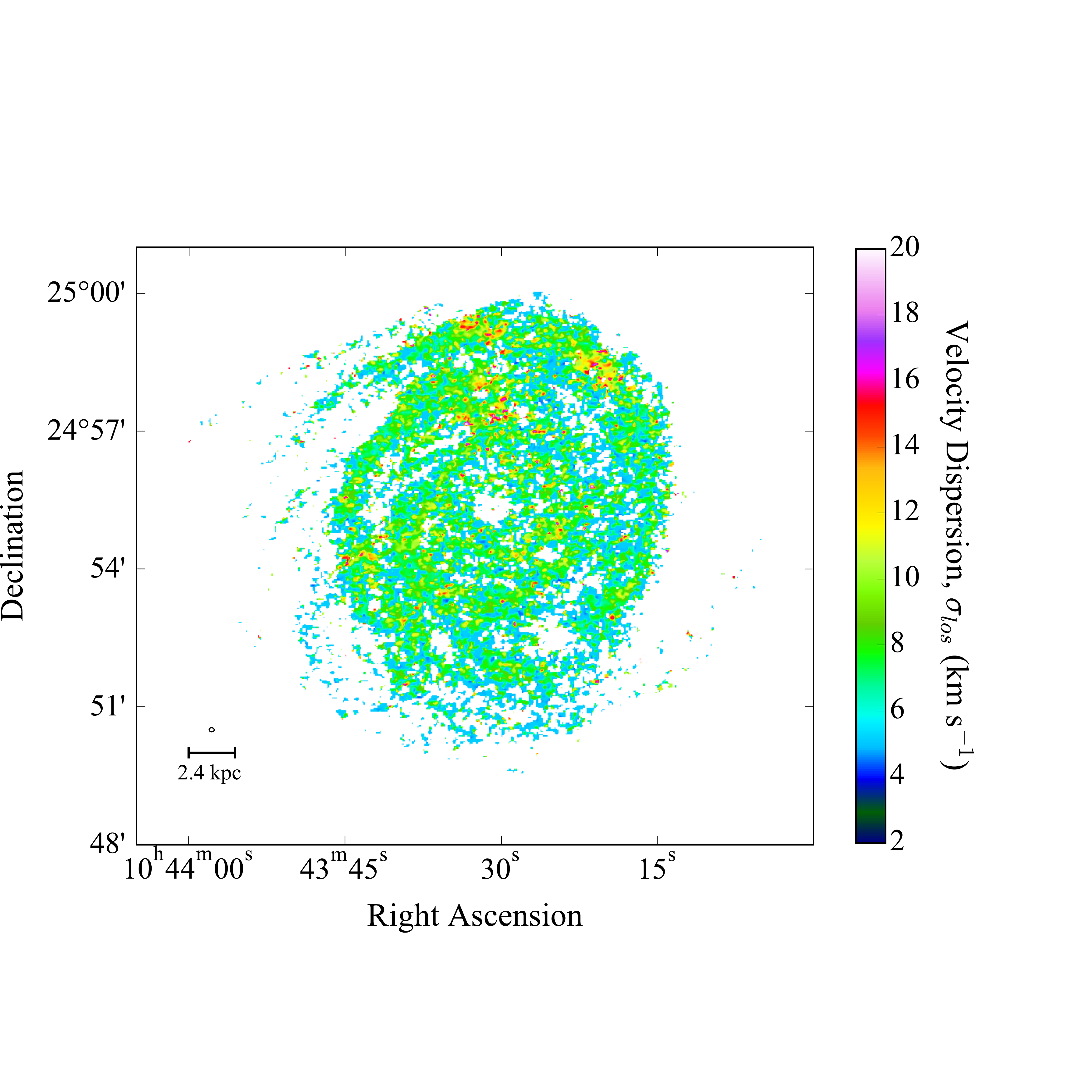}} \hfill
\subfloat[\hi\ kinetic energy surface density, \kesd]{ \includegraphics[width=0.49\textwidth , trim = 0mm 5cm 3cm 5cm, clip,scale=0.25]{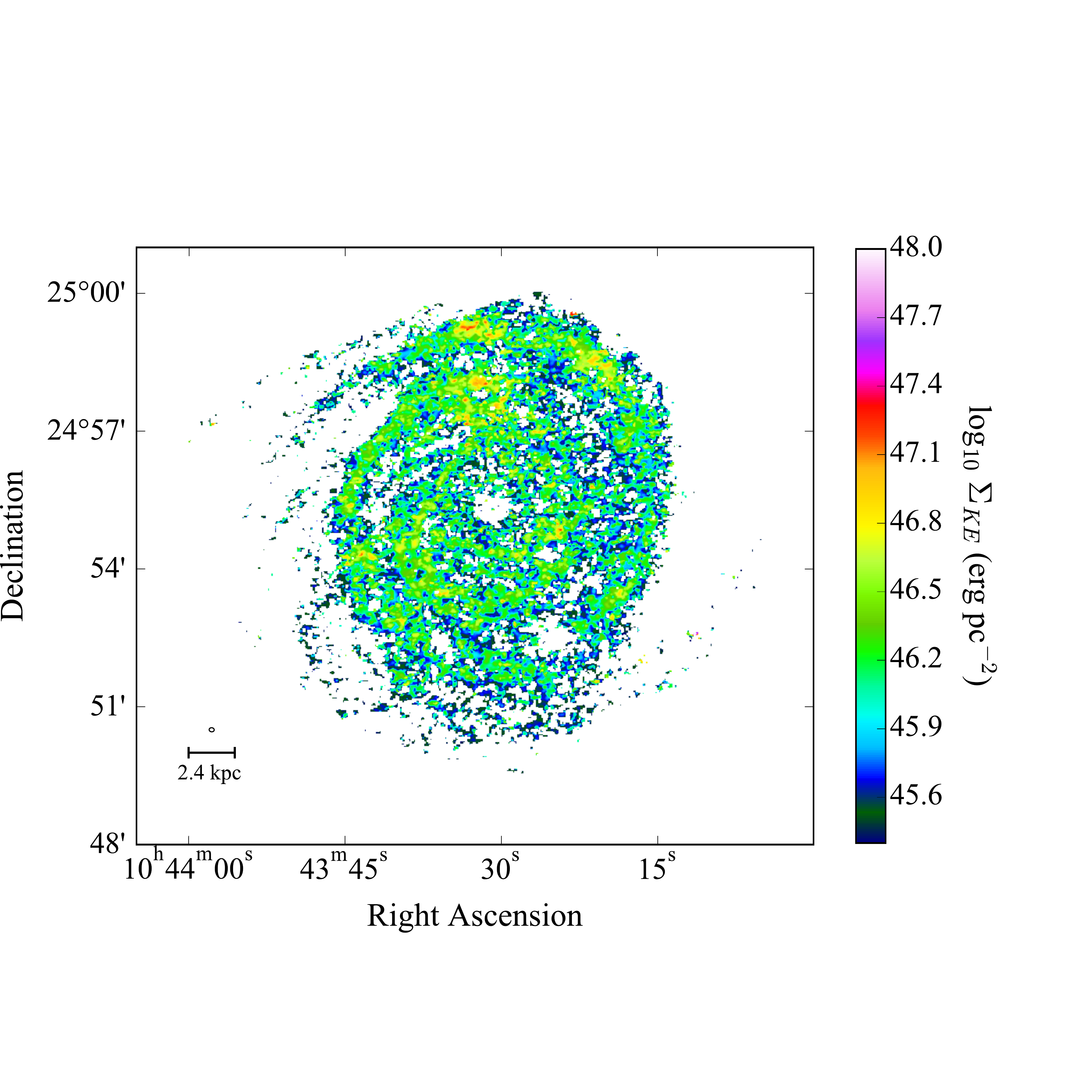}}
 \setcounter{figure}{5}
 \caption{Properties of the ISM -- (a) \ion{H}{1} mass surface density (\mhi), (b) velocity ($v_{los}$), (c) velocity dispersion (\disp), and (d) kinetic energy 
    surface density ($\Sigma_{KE}=1.5{\rm{M}}_{\rm{HI}}\sigma_{los}^2$) derived using VLA H\,{\small I} 21 cm imaging. The color-bars indicate the observed strength of the maps. The black arrow in panel (b) indicates the direction to the quasar sightline. A physical scale of 2.4 kpc (1\arcmin) is shown at bottom left. The ellipse on the top of the scale shows the beam size of 6\farcs7$\times$ 5\farcs4  with a position angle of 83.8$^\circ$. }
    \label{fig:himaps}
\end{figure*}

We trace the ISM in NGC 3344 using VLA \hi\ 21 cm imaging. The velocity-integrated flux density ($I_{tot}$), flux density-weighted velocity ($v_{los}$), and the 
velocity dispersion (\disp) maps above 3$\sigma$ were computed using the \hi\ Source Finding Application-2 \citep[SoFiA-2;][]{serra15}. 

We estimate the \hi\ mass surface density, \mhi , and the kinetic energy surface density, \kesd , in units of M$_\odot$ pc$^{-2}$ and erg pc$^{-2}$, respectively, at each pixel using the relations presented in \cite{mullan13}:
\begin{equation}\label{eq:6}
    \Sigma_{\rm {H\,\small{I}}} = 1.0 \times 10^{4}\; \frac{I_{tot}}{\rm{A}_{\rm beam}}
\end{equation}
\begin{equation}\label{eq:7}
    \Sigma_{KE} = 1.5 \times 10^{4} \; \frac{I_{tot}\; \sigma_{los}^{2}}{A_{beam}}
\end{equation}

where $I_{tot}$ and \disp\ have units of Jy~beam$^{-1}$ km~s$^{-1}$ and km~s$^{-1}$ with the beam area, $A_{beam}$, expressed in squared arcsecond. The resulting, \mhi, $v_{los}$, \disp, and \kesd, maps are shown in Figure \ref{fig:himaps}. 
The \mhi\ map informs that the \hi\ disk extends to 12.64~kpc at \mhi\ = 1~M$_{\odot}$ pc$^{-2}$.

\section{Results}\label{sec:results}

\subsection{Radial Profiles}\label{sec:radpro}

\begin{figure}[!t] 
\centering
\setcounter{figure}{6} 

    \includegraphics[trim = 5cm 0mm 4cm 0cm, clip,scale=0.6]{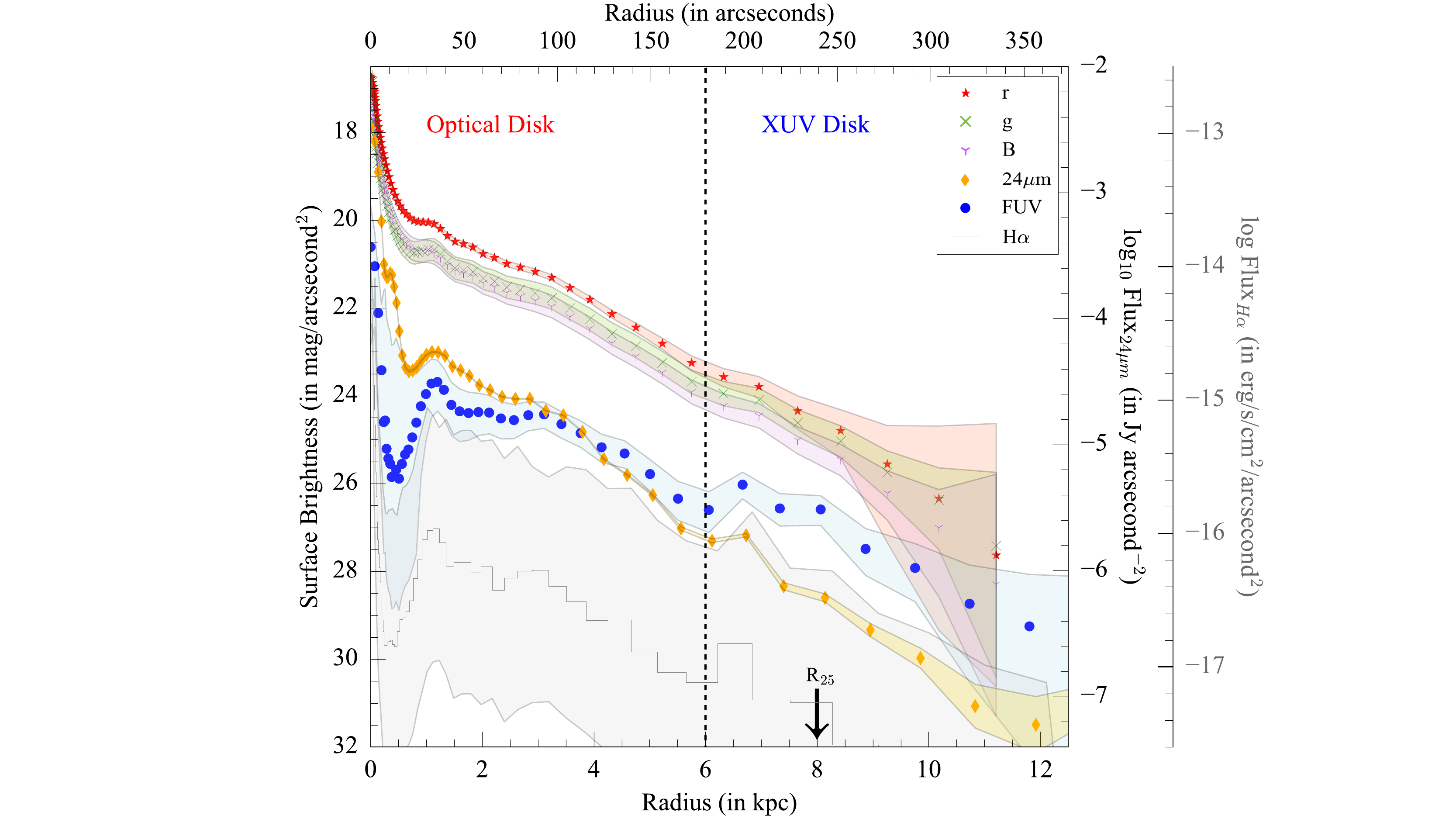}     

     \setcounter{figure}{6}
\caption{ {\small Radial profiles of {\em r-}, {\em g}-, B-bands, \mips, FUV, and \ha\ surface brightness. The shaded area shows the uncertainty in the respective quantities. The arrow marks the position of R$_{25}$ estimated from the B-band profile and the vertical {\em black} dashed line marks the break radius. }}
    \label{fig:radpro}
\end{figure}

Surface brightness ($\mu$) profiles of NGC 3344 are shown in Figure \ref{fig:radpro}. 
We use IDL routine \texttt{galprof} \footnote{http://www.public.asu.edu/$\sim$rjansen/idl/galprof1.0/galprof.pro}, which fits elliptical isophotes with fixed center positions, taking into account the ellipticity and position angle of a galaxy, to perform surface photometry. 
We show the surface brightness profiles in FUV, {\em g }, B, {\em r}, \ha , and \mips . The \mips\ surface brightnesses are estimated in Jy/arcsecond$^2$ while all others are in units of mag/arcsecond$^2$. The B-band profile is created using the transformation, ${\rm B}=g+0.33(g-r)+0.20$ from \cite{jest05}. We estimate R$_{25}$, i.e., the radius at 25 mag/arcsecond$^2$, of 7.9~kpc from the B-band radial profile, which is indicated by the black arrow. 

Using the method to derive the break radius prescribed in \cite{pohl06}, we find that the {\em r}-band profile shows multiple break locations. Breaks are observed at 1.6, 3.3, 5.9, and 7.6~kpc, likely caused due to asymmetries in spiral arms. The FUV, \ha, and \mips\ profiles also show breaks at similar locations. However, at 6.0~kpc, the FUV emission in the disk show a sudden increase along with an increase in \ha. We characterize this discontinuity at 6.0~kpc as the break radius, where a transition from the optical to the XUV disk occurs. This criteria differs from \cite{thilk07}'s definition of the starting point of an XUV disk beyond a single surface brightness contour corresponding to an FUV \sfr\ of 3.0$\times$10$^{-4}$~M$_\odot$~yr$^{-1}$~kpc$^{-2}$. At 6.0~kpc, for NGC~3344, we find an FUV \sfr\ of 3.8$\times$10$^{-4}$~M$_\odot$~yr$^{-1}$~kpc$^{-2}$. The XUV disk extends to a radius of 10.0 kpc at 3$\sigma$ FUV surface brightness, covering an annular area of $\sim$202~kpc$^2$. Following this, our XUV disk region is 1.3$\times$ larger in area than that
 of \cite{thilk07}.   

We also determine scale lengths (h) from the surface brightness profiles by fitting the exponential function: $\mu(R)=\mu_0+1.086 (R/h)$. Scale lengths estimated from the radial profiles in the optical and XUV disks are summarized in Table \ref{tab:scale}. These measurements exclude the central part of the surface brightness profiles which are dominated by the presence of a bulge. We find that the galaxy disk flattens from longer to shorter wavelengths. Scale lengths in the FUV are at least 1.4 times larger than those in the {\em r}-band.
This implies that young stars show an extended distribution in the XUV disk while older stellar populations are concentrated towards the center of the disk. Dust traced by \mips\ shows an even more centrally-concentrated distribution. 

\begin{table}
\caption{Exponential Scale lengths measured from the surface brightness profiles of NGC 3344}\label{tab:scale}
\vspace{0.2cm}
\centering
\begin{tabular}{cccc}
\hline
\noalign{\smallskip} 
Band & & Scale length &  \\
 & Optical Disk  & XUV Disk & Entire Disk \\
 & (kpc) & (kpc) & (kpc) \\
\noalign{\smallskip} 
\hline
\noalign{\smallskip} 
 FUV &  2.38 & 2.44 & 2.40 \\
{\em g} & 1.82 & 1.68 & 1.79 \\
 B & 1.70 & 1.53 & 1.66\\
{\em r} & 1.70 & 1.54 & 1.66\\
 \ha\ &1.88 & 1.78 & 1.98\\
 \mips\ & 0.51 & 0.55 & 0.51\\
\noalign{\smallskip} 
\hline
\end{tabular}
\vspace{-5pt}
\end{table}

\subsubsection{SFR and Stellar Mass profile}\label{sec:smsfrpro}

\begin{figure*}[!t] 

\setcounter{figure}{7}
 \centering
   \subfloat[]{\includegraphics[trim = 5mm 0mm 0cm 0.5cm, clip,scale=0.095]{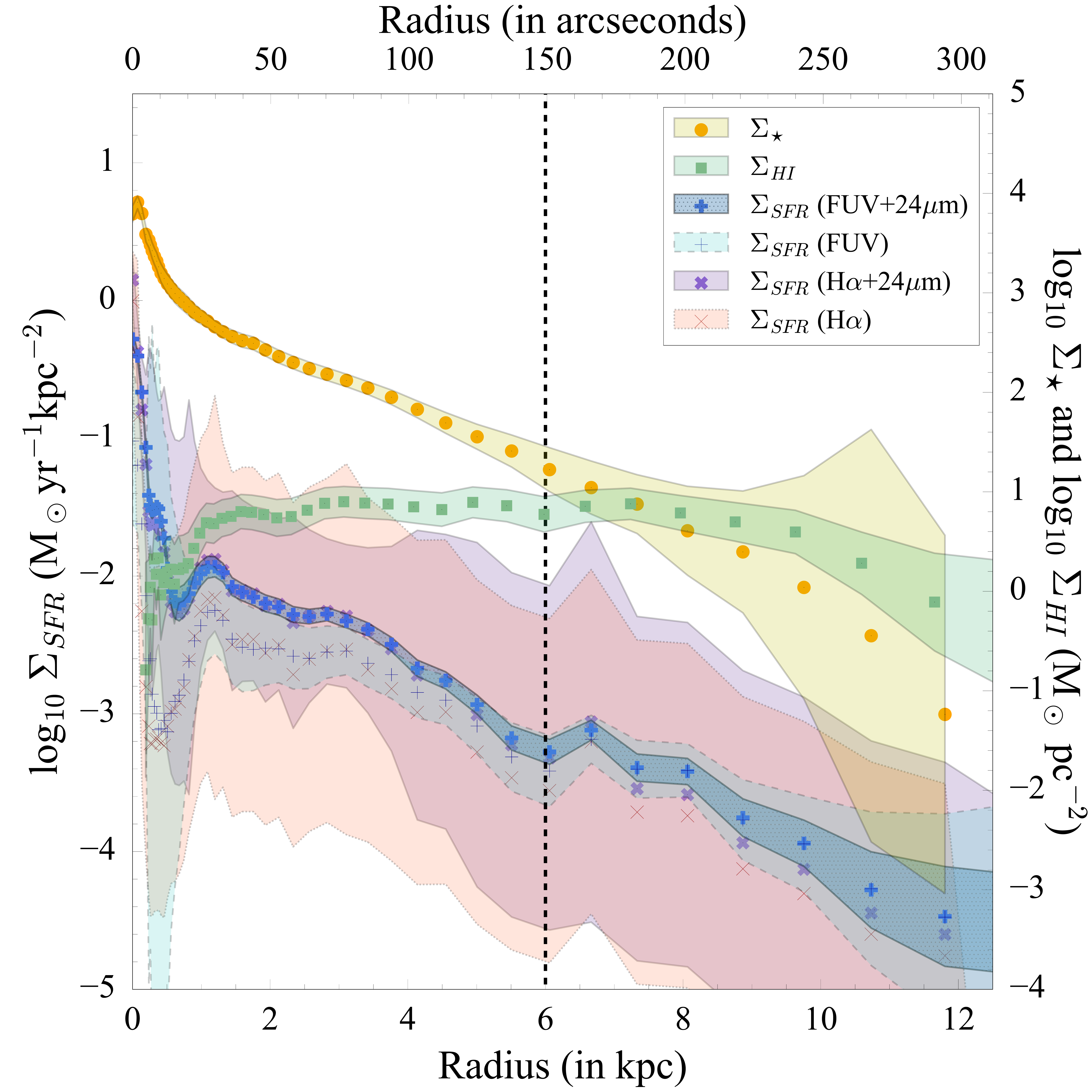} \label{fig:sigpro}}
   \vfill
\subfloat[]{\includegraphics[width=0.49\textwidth , trim = 5mm 0mm 0cm 0.5cm, clip,scale=0.08]{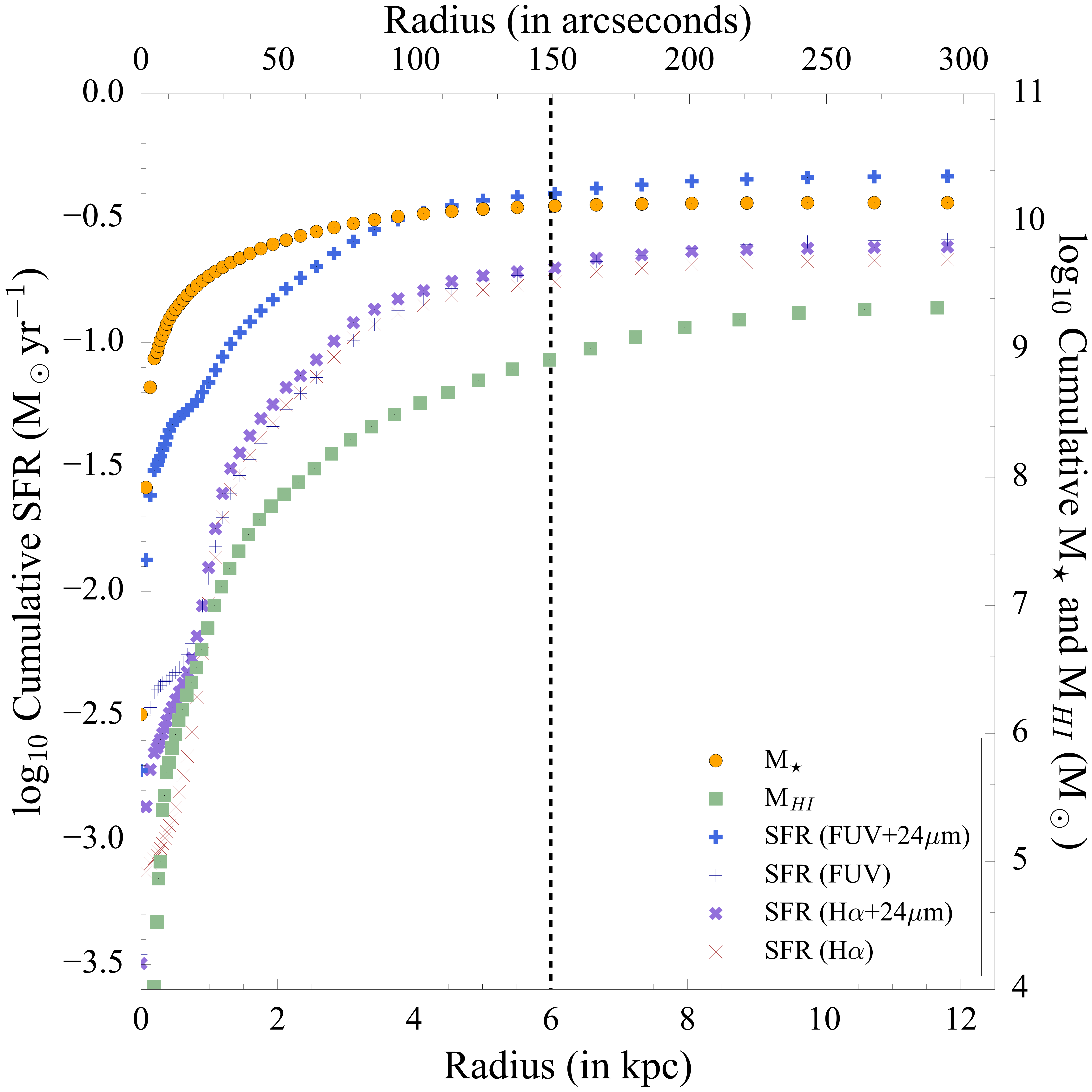}  \label{fig:cumupro} } \hfill
\subfloat[]{ \includegraphics[width=0.49\textwidth , trim = 5mm 0mm 0cm 0.5cm, clip,scale=0.08]{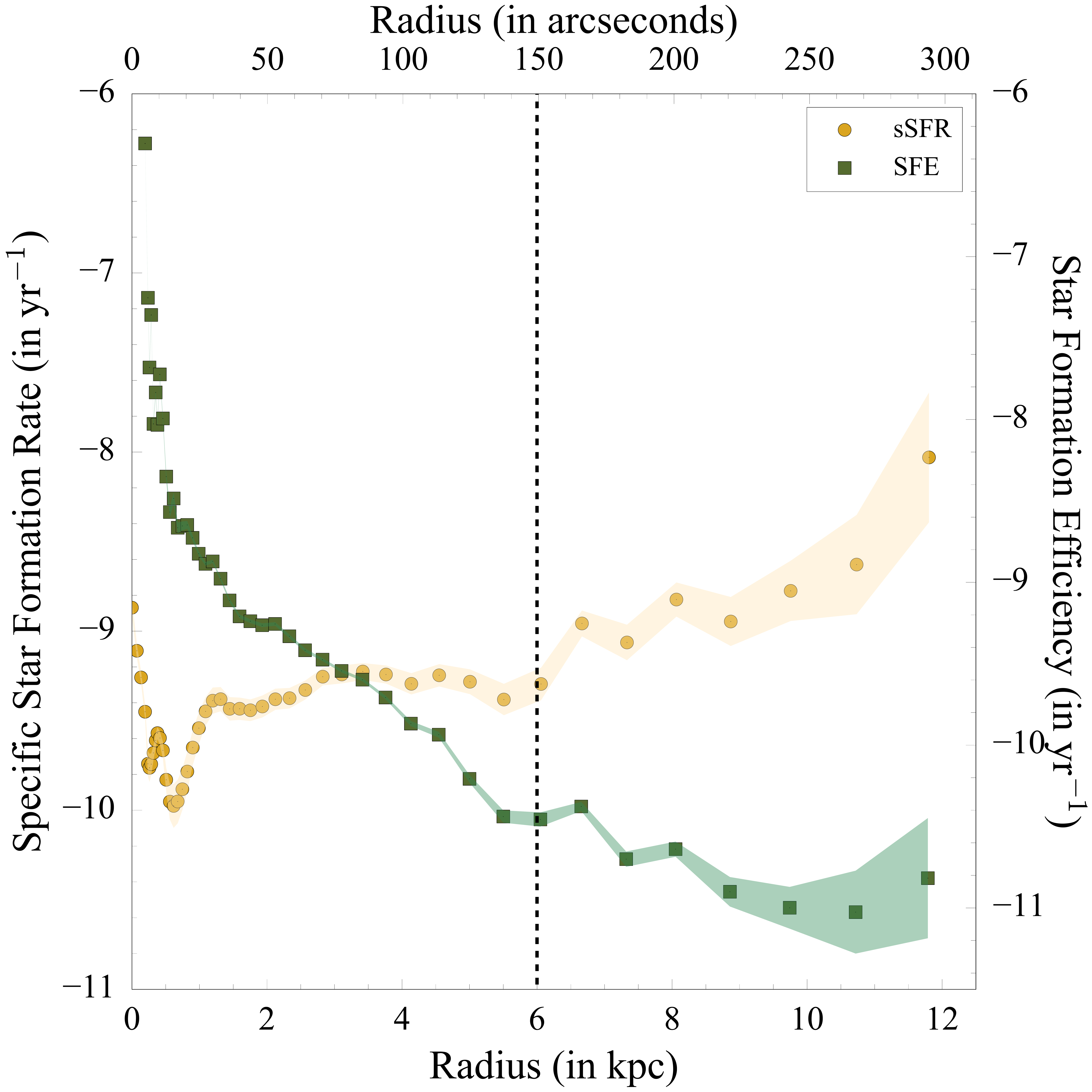} \label{fig:ssfrpro} }

 \setcounter{figure}{7}
\caption{ {\footnotesize Radial profiles of (a) \sm, \mhi, and \sfr, (b) cumulative M$_\star$, M$_{HI}$ and SFR, and (c) specific SFR and H\,{\tiny I} star formation efficiency of NGC 3344. The SFR and \sfr\ are estimated using FUV ({\em green}), FUV+\mips\ ({\em blue}), \ha\ ({\em pink}), and \ha+\mips\ ({\em purple}). The shaded area shows the uncertainty in the respective quantities. The vertical {\em black} dashed line at 6 kpc marks the break radius.}}
    \label{fig:profs}
\end{figure*}

We use the surface brightness profiles and transform them into star formation rates, stellar masses and atomic gas mass using the conversions provided in \S\ref{sec:analysis}. The radial profiles of \sm\, \mhi, and \sfr\ (estimated using FUV, \ha , FUV+\mips , and \ha+\mips), and the cumulative stellar and \hi\ mass, M$_\star$ and M$_{HI}$ (M$_\odot$) and SFRs (M$_\odot$~yr$^{-1}$) are shown in in Figure \ref{fig:sigpro} and \ref{fig:cumupro}. Figure \ref{fig:ssfrpro} shows the radial variation in specific SFR (sSFR), i.e., SFR per unit stellar mass, and \hi\ star formation efficiency (SFE), i.e., SFR per unit \hi\ mass.

The \sm\ profile shows a pure exponential profile and we observe no break at 6.0 kpc that is seen in the light profiles. This establishes that mass distribution has no impact on the location of the break and it rather originates from the radial variation in the age of stellar populations 
\citep{bakos08}. At 6.0 kpc, we find \sm\ equals 15.8~M$_\odot$~pc$^{-2}$ closer to a typical value of 13.6~M$_\odot$~pc$^{-2}$ at the break radius
of a typical galaxy exhibiting a truncated radial profile \citep{pohl06, bakos08}. We also find that only $\sim$5\% of the total stellar mass resides in the XUV disk with $\sim$14\% of the total SFR coming from this region.

The \sfr\ profiles also provide an opportunity to study the radial variation in the different tracers of star formation. We find the four tracers of star formation: FUV, \ha , FUV+\mips , and \ha+\mips\ to show different behavior in the optical and XUV disk. In the optical disk, dust plays an important role - both FUV+\mips\ and \ha+\mips\ show higher \sfr\ compared to FUV and \ha. However, as we move towards the XUV disk, FUV and FUV+\mips\ tracers pick up more of the star formation. Additionally, the \mips\ data is dominated by noise which is of the order of the low SFRs and hence, it is not sensitive to star formation in the XUV disk. 
Estimates from the cumulative profiles show that dust-corrected SFRs are higher, making them more effective in tracing star formation than non-dust corrected SFRs.  We note that SFR estimates from \ha\ and \ha+\mips\ are consistently lower compared to FUV and FUV+\mips\ SFRs. This is discussed in more detail in \S\ref{sec:hauv}. 

In the XUV disk, we also observe a rise in the sSFR after an almost flat trend in the optical disk while the \hi\ SFE declines as a function of galactocentric distance. The implications of this result are discussed in \S\ref{sub:iog}.

\subsection{Young stellar complexes in NGC 3344}\label{sec:regions}

We extracted 320 young stellar complexes in NGC 3344 
with typical physical sizes between 0.15--1~kpc. Figure \ref{fig:region} shows the individual 
star-forming complexes throughout the galaxy. 
Elliptical apertures above 3$\sigma$ threshold were identified by running SExtractor 
\citep{bert96} on cutouts of different regions of the FUV and NUV image. We applied a 5-pixel wide top-hat filter to detect regions of low surface brightness in the XUV
disk. Some apertures were redefined or added manually to optimally enclose faint regions. Any apertures belonging to background objects  
detected in the outer edge of the XUV disk were removed. This was done using the continuum-subtracted \ha\ image that enabled discerning objects present at the same redshift as the galaxy and utilizing SDSS DR 12 spectroscopy and imaging to verify if the objects were removed foreground stars or background galaxies. The aperture at the position of a bright foreground star near the center, clearly visible in the $r$-band image but not in the FUV image, was taken out. Care was taken to avoid contamination from neighboring apertures. These apertures were further divided into optical and XUV regions based on their galactocentric distance and the estimated break radius.
The red apertures are regions within the optical disk marked by the black dashed circle covering an area of $\sim$113~kpc$^2$. The blue apertures are regions in the XUV disk. 
Of the 320 young star-forming complexes, 132 are in the XUV disk.

\begin{figure}[!t]
    \centering
    \includegraphics[trim = 3cm 2cm 0cm 0cm, clip,scale=0.65]{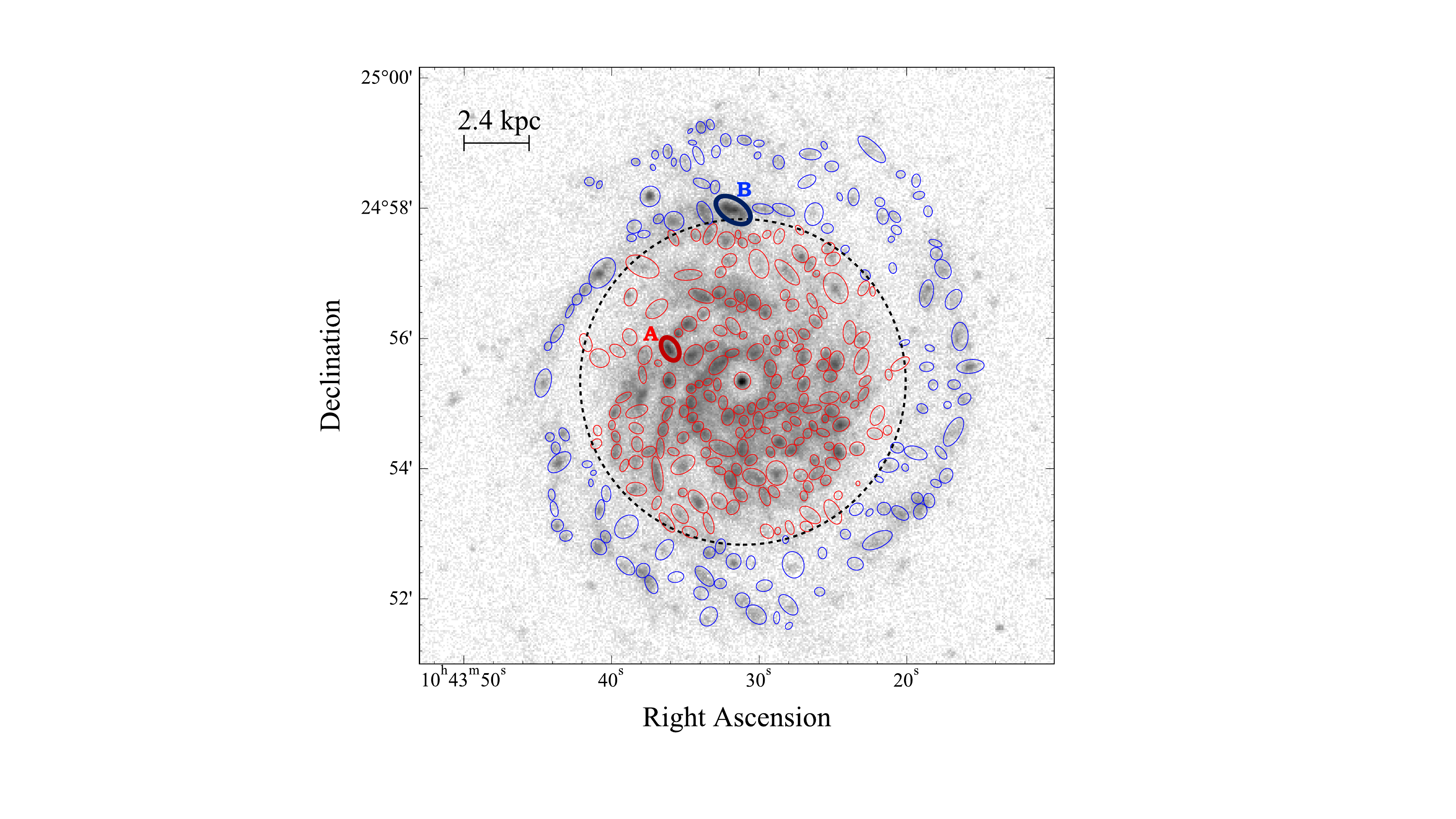}
    \caption{The greyscale
    FUV image of NGC 3344 with star-forming regions detected at 
    3$\sigma$ significance using SExtractor. The black dashed 
    circle represents the break radius of 6 kpc. The red 
    apertures are regions within the optical disk and blue apertures are regions
    in the XUV disk. 320 regions are extracted with 132 regions in the XUV disk.  A physical scale of 2.4 kpc (1\arcmin) is shown on top left corner. The highlighted regions, A and B (see Section \ref{sub:ha}) have SDSS DR12 spectra available. }
    \label{fig:region}
\end{figure}

\subsection{Circumgalactic Medium in NGC~3344}

We detected two distinct absorbing systems in the QSO sightline probing NGC~3344 at 
30~kpc. The strong component was seen in most of the metal-line transitions covering low-, intermediate-, and high-ionization states (\ion{Si}{2}, \ion{C}{2}, \ion{Si}{3}, and \ion{Si}{4}).
The weighted mean places the centroid of the strong component at $\sim -28.5$~\kms.
The weaker component was seen in \ion{H}{1} and \ion{Si}{2} at a mean centroid of $\sim$ 88.9~\kms. As noted before, the \hi\ line was blended with the 
damped \Lya\ profile of the Milky Way, so no measurements could be made for the 
atomic gas content of the first component. The measurements of the \Lya\ of the second component might 
 suffer from uncertainties pertaining to continuum identification issues. A single component fit to the profile yields a 
column density of Log N(\hi)=14.0. However, the column density log(N(Si~II))=13.06 corroborates our assertion of large uncertainties in the continuum fit of \Lya\ most likely leading to a lower column density measurement for \hi. 

In general, the QSO sightline probes a metal-rich circumgalactic medium in NGC~3344. The total \ion{Si}{2} content is $>\rm 1.3\times 10^{14}~cm^{-2}$. Assuming a solar metallicity, we expect the total \hi\ column density to be $\approx \rm 3.7\times 10^{18}~cm^{-2}$. Since we did not detect damping wings with the \Lya\ profile, we can also conclude that the metallicity 
of the gas in the CGM of NGC~3344 is no less than 0.1~Z$_{\odot}$ and is more likely $\approx$ 1~Z$_{\odot}$. This indicates that the circumgalactic gas at 30~kpc is well enriched by the products of stellar nucleosynthesis. The weaker component also showed \ion{Si}{2} indicative of a high metallicity and low-ionization state of the gas.

The kinematics of the two absorbers allows us to determine if they are consistent with corotation. The observed projected velocity of the \hi\ disk closest to the sightline is 665~$\pm$~5~\kms\ i.e. 85~\kms\ from the systemic velocity of 580~\kms\ (Figure~\ref{fig:velarrow}). The weaker component is consistent with rotation showing a deviation of 3.9~\kms, which is well within the uncertainties of our measurement. However, the stronger component observed at -28.5~\kms\ with respect to the systemic shows a deviation of -113.5~\kms\ from the part of the disk nearest to the sightline, thus indicating gas flow that is inconsistent with co-rotation.
The direction of the flow is unknown and could either be an inflowing or outflowing cloud with respect to the disk.
To put into perspective, this cloud is consistent with high-velocity clouds (HVCs) seen in the halo of the Milky Way showing velocity offsets of $\ge$~90~\kms\ \citep{wakker97} relative to the disk. 
The presence of similar clouds in the halos of other galaxies such as in M31 and M100 have been reported \citep{thilker04, westmeier05}.
The Milky Way HVCs are considered as one of the main pathways for bringing cold gas into the Milky Way disk. 
The same may be true for this cloud that is cold and metal-rich. 
We will discuss the likelihood of this cloud being an inflow or outflow in \S~\ref{sub:gas}.

\section{Discussion} \label{sec:res}

\subsection{Effectiveness of \ha\ and UV SFR Tracers}\label{sec:hauv}

\begin{figure*}
\setcounter{figure}{9}
\centering

\subfloat[]{\includegraphics[width=0.49\textwidth , trim = 1cm 0mm 0cm 0cm, clip,scale=0.08]{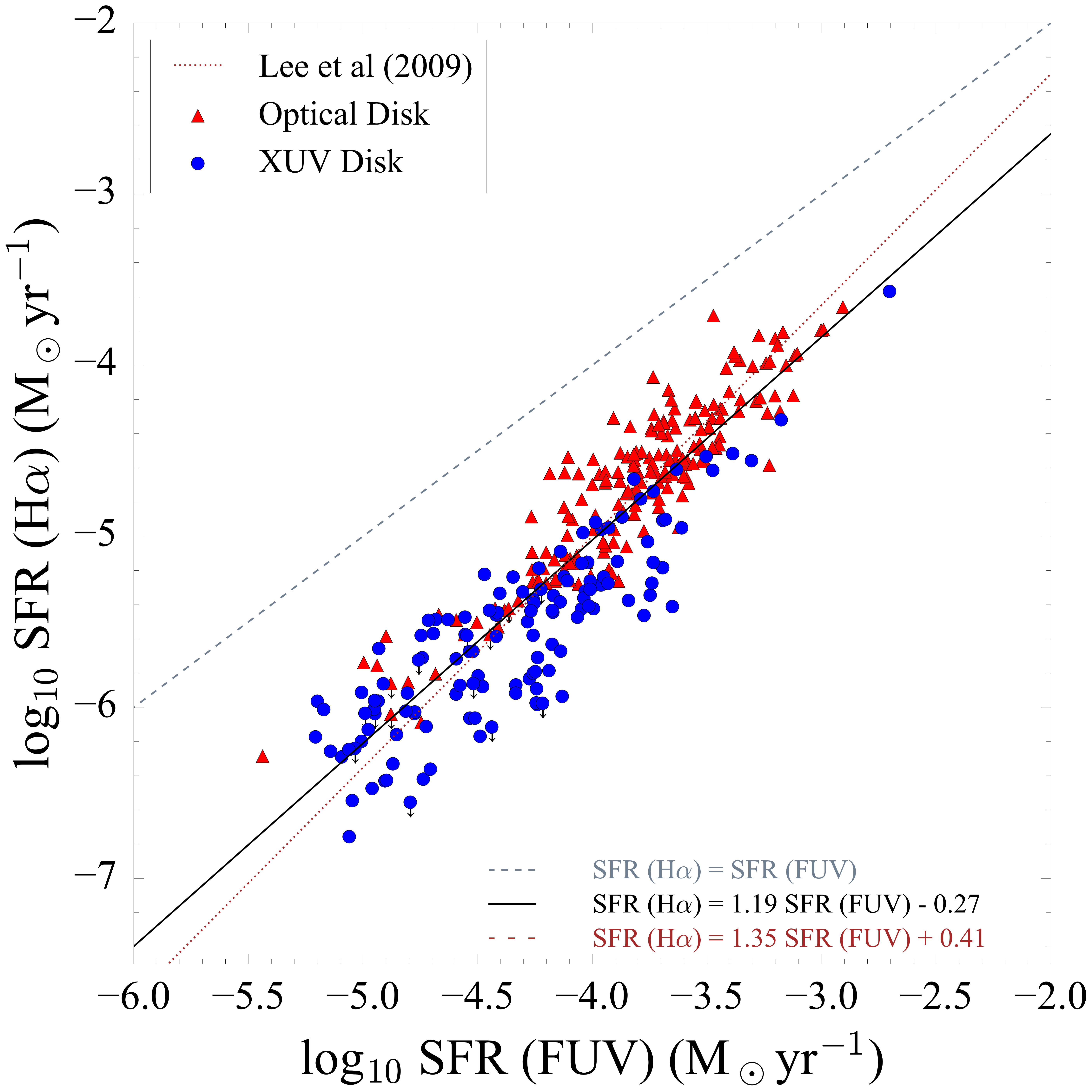} \label{fig:hatouv}}\hfill
\subfloat[]{\includegraphics[width=0.49\textwidth ,  trim = 1cm 0mm 0cm 0cm, clip,scale=0.078]{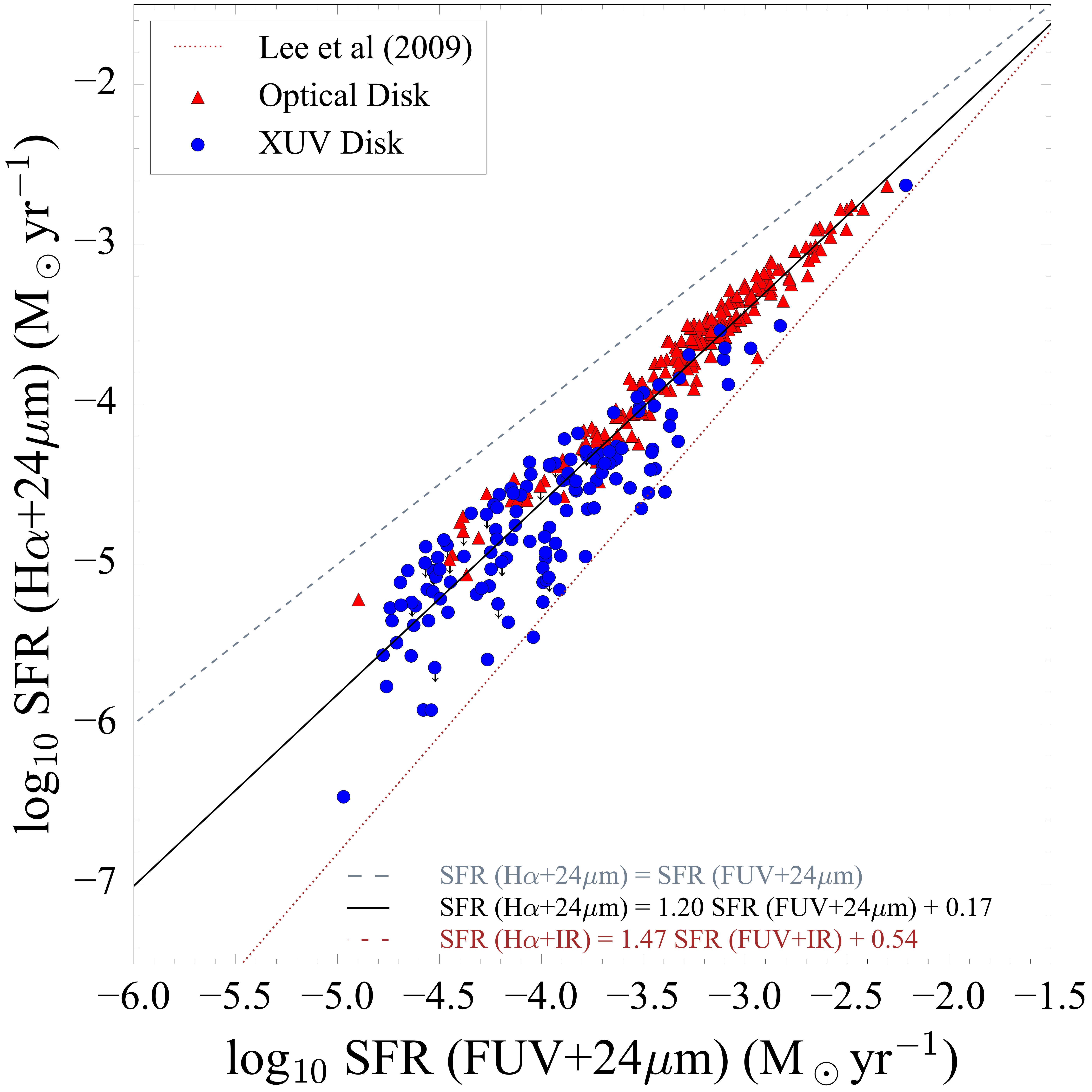}\label{fig:ha24touv24}}

\subfloat[]{\includegraphics[trim = 1cm 0mm 0cm 0cm, clip,scale=0.08]{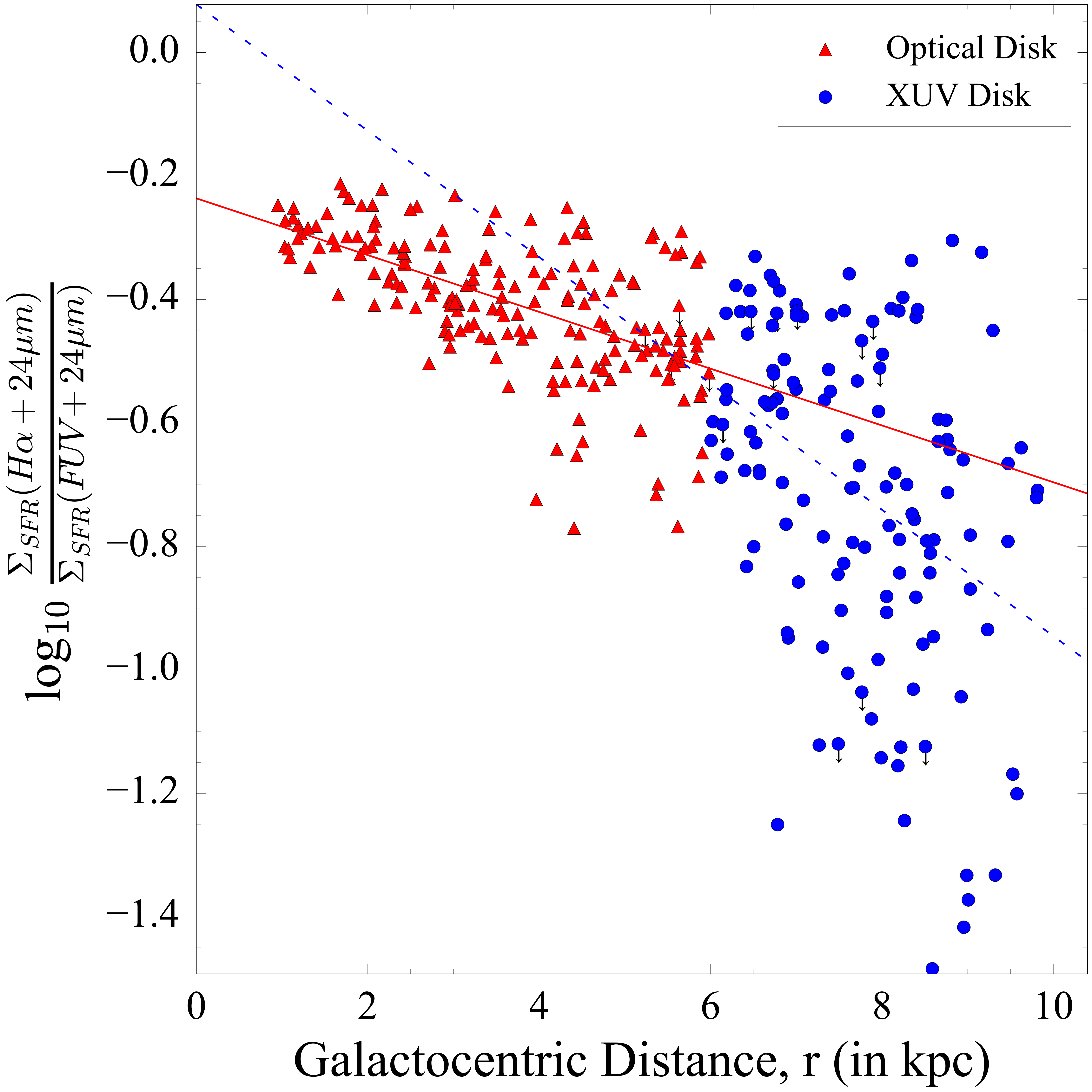} \label{fig:hauvtodist}}

    \setcounter{figure}{9}
    \caption{Comparison of the {\em a}) non-dust-corrected H$\alpha$
    SFR and FUV SFR and ({\em b}) dust-corrected \ha+\mips\
    SFR and FUV+\mips\ SFR, followed by ({\em c}) ratio of \ha-to-FUV \sfr\  as a function of galactocentric distance for the 
star forming regions in NGC 3344. The red triangles and blue circles are regions in the optical disk
and the XUV disk, respectively. In panel (a) and (b), the {\em solid} black line represents the best fit to the data. 
The {\em dashed} line shows the one-to-one correspondence between the two \sfr\ estimates, and the {\em dotted} line is the best-fit for non-dust-corrected and internal dust-corrected SFRs from \cite{lee09}, respectively. The parameters describing the line of best-fit are stated at the bottom right. The {\em solid} red and {\em dashed} blue lines in panel (c) represent the best-fit line to the optical disk and XUV disk points, respectively. In general, the \ha\ SFRs are found to be lower than the FUV SFRs.}
    \label{fig:hauvsfr}
\end{figure*}

We first tested how efficient the non-dust-corrected (\ha, and FUV) and dust-corrected (\ha+\mips\ and FUV+\mips) tracers are in probing star formation in the XUV disk.   
Both \ha\ and FUV tracers are known to give consistent estimates of SFRs in the optical disk. 
On the other hand, previous studies investigating the \ha\ and FUV emission in the outer disks of galaxies find the  
outer regions to be sparsely populated with \ion{H}{2} regions \citep{ferg98, gil05, meur09, godd10, werk10, barn11, 
watk17} and hence, show lower levels of \ha. In Figure \ref{fig:hatouv} and \ref{fig:ha24touv24}, we show the non-dust corrected and dust-corrected \ha\ SFRs against FUV SFRs for the young stellar complexes, respectively. 
The red and blue colors distinguish between the regions in the optical and the XUV disk, respectively, with the solid black line denoting the linear fit to the data. 
The one-to-one correlation between the \ha\ and FUV SFRs is represented by the dashed line. The dotted lines show the best-fit line for non-dust-corrected FUV and \ha\ SFRs and dust-corrected FUV and \ha\ SFRs for $\log$ SFR(\ha)$<-1.5$ from \cite{lee09} (Figure 2 and 5 therein, respectively). We note that the SFRs estimated using the prescriptions presented in \S\ref{sec:sfrmap}, assume fully-populated IMFs and a uniform distribution of ages of the regions. \cite{lee09} investigated $\sim$300 local star-forming galaxies and found a systematic offset between the non-dust corrected FUV and \ha\ SFRs where at SFRs~$\lesssim0.03$~M$_\odot$~yr$^{-1}$, \ha\ began to underestimate SFR. While we study $\sim$300 individual star forming regions within a single galaxy, what is striking is the similarity in the best fit line of our study (Figure \ref{fig:hatouv})  and the one from \cite{lee09} for the non-dust corrected SFRs. Our results extend the distribution to lower values $\log$ SFR(\ha)$<-2.0$ with \ha\ SFRs already lower than FUV SFRs. This not only corroborates \cite{lee09}'s result but also demonstrates that SFRs obtained on global scale and $\lesssim$1~kpc scales show a similar relation in their FUV and \ha\ emission.

However, our dust-corrected \ha\ SFRs are higher compared to the expected trend from \cite{lee09}, although still offset from the one-to-one line. In Figure \ref{fig:ha24touv24}, after accounting for internal dust correction via \mips, the optical disk points shift by $\lesssim$1\,dex and the XUV disk points move by $\lesssim$0.5\,dex. The addition of \mips\ does, however, add noise to the estimate. The upper envelope of the distribution in Figure \ref{fig:ha24touv24} has an almost constant offset from the one-to-one relation over a wide range in SFR that spans both the optical and the XUV portions of the
disk. So, a subset of star formation happens in the regions that behave similarly in the XUV disk and in the optical disk. The total \ha+\mips\ SFR of the stellar complexes in the optical disk is $\sim$0.09~M$_\odot$~yr$^{-1}$. This is $\sim$50~\% lower than the total FUV+\mips\ SFR of $\sim$0.18~M$_\odot$~yr$^{-1}$ in the optical disk. The offset further increases in the XUV disk. At SFRs $\lesssim$10$^{-4}$~M$_\odot$~yr$^{-1}$ in the XUV disk, \ha\ underpredicts the SFR by $\sim$75\%. 
This is better illustrated in Figure \ref{fig:hauvtodist}, which shows the ratio of \ha-to-FUV \sfr\ with galactocentric distance. The ratio of \sfr s range within factors of 0.03 to 0.88. In the XUV disk, the ratios show a larger scatter with a steeper correlation compared to the optical disk. Since galactocentric distance is representative of decreasing SFRs as \sfr\ drops with radius (Figure \ref{fig:sigpro}), the larger scatter then indicates that \ha\ SFRs drop faster at lower SFRs in the extreme outskirts of the galaxy. 

We note that a fixed value of the [\ion{N}{2}]/H$\alpha$ ratio of 0.52, adopted from \cite{kenn08}, is an average over the optical disk. However, the [\ion{N}{2}]/H$\alpha$ ratio varies with galactocentric radius and is likely low in the XUV disk. This would lead to an over-correction and a subsequent underestimation of \ha\ SFRs in the XUV disk. We calculated \ha\ SFRs without applying a [\ion{N}{2}] correction and found that the intercept in Figure \ref{fig:hatouv} changes from 0.27 to 0.36, i.e., the
 data would shift upward by $\sim$0.12\,dex.
So, even if a radially accurate [\ion{N}{2}] correction is applied, the deviations would be small and not greatly affect the results. In the following subsection, we explore other factors that cause lower \ha\ SFRs in the XUV disk.\\

\subsubsection{Possible Causes for Low \ha-to-FUV ratios}\label{sub:ha}

The drop in the \ha-to-FUV ratios in the XUV disk can be associated with (1) stochastic sampling of the stellar IMF, (2) truncation and/or steepening of the upper end of the IMF, (3) non-continuous star formation history, or (4) leakage of hydrogen-ionizing photons. Here we investigate these possibilities for lower levels of \ha\ seen in the XUV disk. 

The first possibility that can explain the discrepancy in the \ha\ and FUV emission is the stochastic sampling of the IMF. In the optical disk, high SFRs presume a large number of stars and hence, a nearly complete sampling of
the IMF. 
However, in the XUV disk, lower SFRs reduce the probability of finding massive O stars. 
As a result, the IMF may not be fully sampled (see \cite{bois07, godd10, fuma11, koda12}). Calculations done by \cite{lee09} show that above SFR~$\gtrsim$~1.4$\times10^{-3}$~M$_\odot$~yr$^{-1}$
or log~SFR~$\gtrsim-2.8$~, \ha\ flux should be robust against stochasticities. In NGC~3344, SFRs of the stellar complexes are below this value, with SFRs yielded by \ha\ mostly lower than the FUV SFRs. 
As a result, stochastic IMF sampling is likely to at least partially account for the observed \ha-to-FUV ratios.

The second possibility arises from the difference in the stellar mass ranges probed by \ha\ and FUV. 
\ha\ is sensitive to O-stars more massive than $\sim$17~M$_\odot$ while FUV traces both O, B stars that have masses $\gtrsim$4~M$_\odot$ \citep{meur09, lee09, godd10, koda12}. 
The observed \ha-to-FUV ratios could be a manifestation of steeper upper IMF slope, possibly combined with an upper truncation \citep{meur09, bruzz15, watts18, bruzz20}. Previous studies have, however, found that \ha\ and FUV ratios are consistent with a standard IMF \citep{godd10, koda12}. In the XUV disk of M83, \cite{koda12} found that O-stars may intermittently populate low mass clusters. Certainly, O-stars are forming in the XUV disk of NGC~3344 but the low \ha-to-FUV ratios may still be consistent with steeper upper IMF and/or truncation along with stochastic IMF sampling.   

Sensitivity to ages of stellar populations could also account for the decline in the \ha-to-FUV ratios.
The star formation history of a galaxy, however, need not be uniform and may comprise of multiple bursts of star formation over the past 10--100~Myrs. 
We find that the total SFR of $\sim$0.43~M$_\odot$~yr$^{-1}$ over a timescale of 10~Myrs probed by \ha\ is lower than the total SFR  of $\sim$0.46~M$_\odot$~yr$^{-1}$ over a timescale of $\sim$100~Myrs traced by FUV. The occurrence of a burst of star formation at a time $\gtrsim10$~Myrs but $\lesssim100$~Myrs in the past would leave B-type stars to emit FUV while the massive O-type stars are already dead  \citep{koda12}, accounting for the higher FUV SFRs in the XUV disk. Age-dating individual star-forming UV and \ion{H}{2} regions in the XUV disk of NGC~3344 and deriving their star formation history would provide more information on the contribution of the aging effect on the observed \ha\ and FUV fluxes.  Although this analysis is beyond the scope of this study, we do not completely rule out the effect of aging on the lower \ha-to-FUV ratios.

\begin{table}[h]
\caption{Emission Line Rations of the \hii\ regions}
\centering
\begin{tabular}{lcc}
\hline
\noalign{\smallskip} 
 & Region A & Region B \\
\noalign{\smallskip} 
\hline
\noalign{\smallskip} 
\ha/H$\beta$ & 3.69 & 3.69 \\
{[}\ion{O}{3}{]}~$\lambda 5007$/H$\beta$ & 0.68 & 4.28\\
{[}\ion{O}{1}{]}$~\lambda 6300$/\ha\ & 0.008 & 0.02\\
{[}\ion{N}{2}{]}$~\lambda 6548$/\ha\ & 0.26 & 0.07 \\
{[}\ion{S}{2}{]}$~\lambda 6716$/\ha\ & 0.14 & 0.068 \\
\noalign{\smallskip} 
\hline
\end{tabular}
\label{tab:hii}
\end{table} 

The fourth possibility is that some parts of the \hii\ regions may be optically thin causing the leakage of ionizing photons out of the star-forming regions. 
Higher forbidden-to-Balmer line ratios, such as, [\ion{S}{2}]$~\lambda6717$,6731/\ha, [\ion{N}{2}]$~\lambda6548$/\ha, [\ion{O}{1}]$~\lambda6300$/\ha, [\ion{O}{3}]~$\lambda 5007$/H$\beta$, indicate the leakage of Lyman continuum photons \citep{hoop03, mads06, voges06, pelle12, zast13, stas15, weil18, wang19} that are believed to ionize the surrounding diffused gas. 
To investigate if leaky \hii\ regions are responsible for lower \ha-to-FUV ratios, we use SDSS DR12 archival spectra of two \hii\ regions: Region A in the optical disk at a galactocentric distance of 2.95 kpc and Region B in the XUV disk at a galactocentric distance of 8.15 kpc. The regions are highlighted in Figure \ref{fig:region}. Both show high \ha\ SFR of $1.08\times 10^{-2}$ and $1.15\times10^{-3}$ ${\rm M}_\odot$~yr$^{-1}$, respectively and a Balmer decrement, \ha/H$\beta\approx3.69$.
A few emission-line ratios from archival SDSS DR12 optical spectra for the two regions are tabulated in Table \ref{tab:hii}. 
We find lower ratios of [\ion{S}{2}]/\ha\ and [\ion{N}{2}]/\ha\ ratios in Region B compared to Region A. Although, in this regard, the data is not sufficient to understand how significant the contribution of leakage of LyC photons is in producing the lower \ha-to-FUV ratios in the XUV disk. Future work involving highly multiplexed spectroscopy can shed more light on the same. 

\subsection{The Inside-Out Disk Growth \& XUV Disk Formation }\label{sub:iog}

Based on our analysis in \S\ref{sec:hauv}, we find FUV+\mips\ to be a more robust and effective tracer of star formation. Therefore, for the remainder of our analysis, we will use UV-based SFR. 
Figure \ref{fig:uvsfr} shows the FUV+\mips\ SFR for each of the 320 stellar complexes identified in the disk of NGC~3344 as a function of galactocentric distance. At large radii, the variation in the SFR between the stellar complexes can be up to 2 orders of magnitude versus a smaller scatter seen in the inner disk. 
As is evident by the radial profile of \sfr\ in Figure \ref{fig:sigpro}, star formation drops as a function of radius. This can be linked to the low SFRs of individual stellar complexes and not due a deficit of star-forming regions. The total FUV+\mips\ SFR in NGC 3344 is 0.458~M$_\odot$~yr$^{-1}$ -- only $\sim$10\% of which is in the XUV disk.
\begin{figure}[!t]
    \centering   
    \includegraphics[trim = 0mm 0mm 0cm 0cm, clip,scale=0.1]{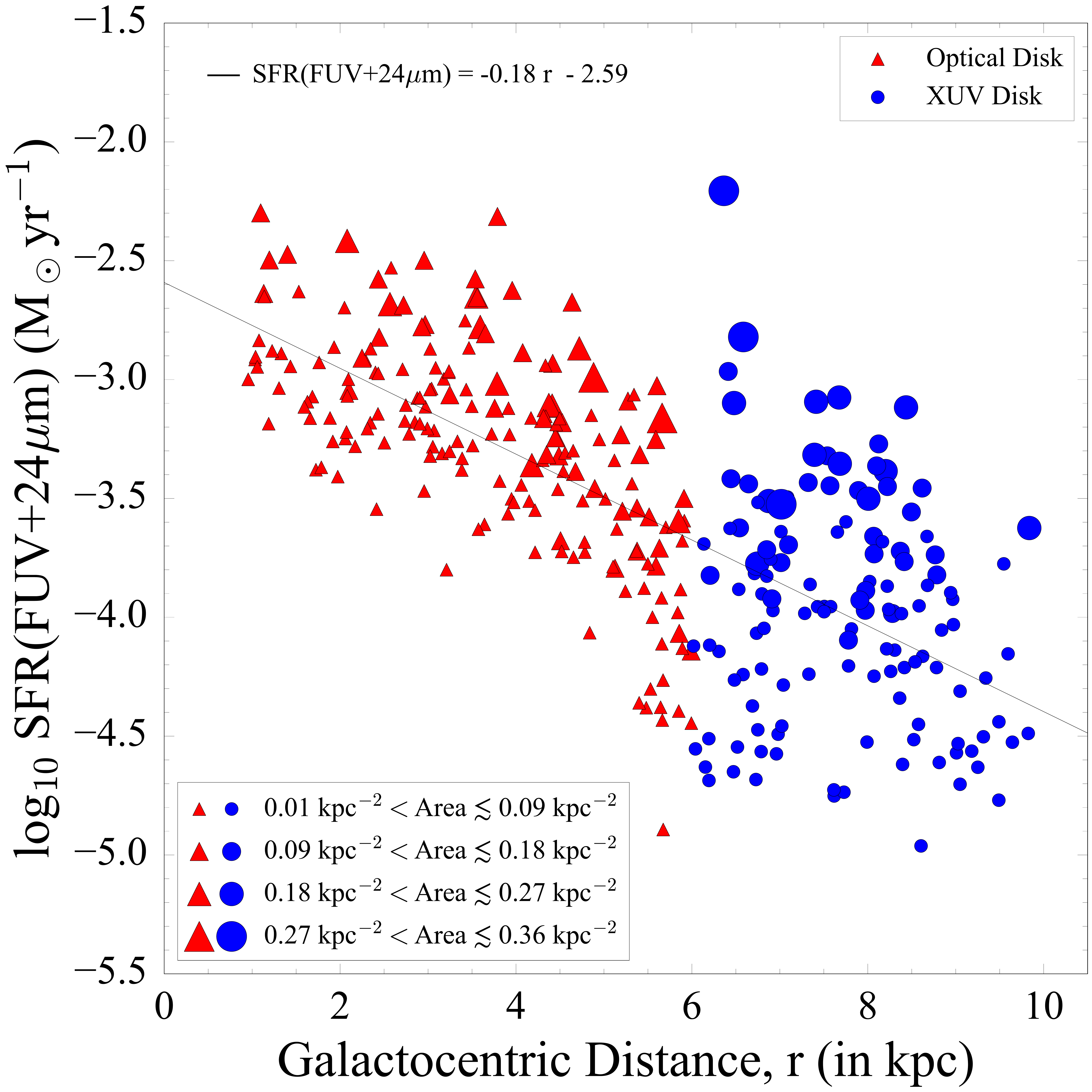}
    \caption{ FUV SFRs of the star forming regions as a function of galactocentric distance
    in NGC 3344. The red triangles and blue circles are regions in the optical  
and the XUV disk, respectively, with the {\em solid} black line representing the best fit to the data. The parameters describing the line of best-fit are stated at the top. Symbol sizes show the distribution of the area of the stellar complexes.}
    \label{fig:uvsfr}
\end{figure}

As discussed earlier in \S\ref{sec:radpro}, FUV shows a larger scale length compared to {\em r}-band illustrating an extended star formation. This also provides evidence for inside-out disk growth \citep{nels12}. The claim of inside-out disk growth is further supported by the rise in sSFR (Figure \ref{fig:ssfrpro}) beyond the break radius at 6~kpc. We observe an almost constant sSFR of $10^{-9.4}$~yr$^{-1}$ in the inner disk between $\sim$1--6~kpc with a sudden rise in the sSFRs from $10^{-9.3}$~yr$^{-1}$--$10^{-8.5}$~yr$^{-1}$ in the XUV disk. As a result, the inner disk is growing slowly while the XUV disk is actively forming stars suggesting an outward increase in the size of the galaxy. 

This leads us to infer that the entire XUV disk could essentially be a starburst. A starburst can be defined as a region in which the mass-doubling time for stars is $<<$ Hubble time, and a sudden infusion of gas has triggered the star formation. In the centers of the galaxies, this can happen rapidly due to the small size of the region ($\sim$kpc). On the contrary, the XUV disk is large and it is difficult to change things rapidly. The present stellar clumps in the XUV disk are a sprinkling of a recent activity on top of a longer-term-forming optical disk. The \hi\ map, however, does not show any strong dynamical perturbation in the XUV disk associated with a recent major accretion event. We find longer \hi\ gas depletion times (1/SFE) in the XUV disk (Figure \ref{fig:ssfrpro}) which suggest that the XUV disk has had a big \hi\ reservoir that only recently started forming stars and it could continue for a very long time ($\sim$10 Gyr). Additionally, the observed high sSFRs, if maintained, would double the disk mass in under one orbital period ($\sim$0.5 Gyr), indicating a burst of star formation.  

Simulations have shown spiral arms as being one possible mechanism for locally triggering the outer star formation and creating a Type-I XUV disk \citep{bush08, bush10, lemo11}. A preliminary investigation showed a spatial offset between the \ha\ and FUV star forming regions where the \ha\ regions in the XUV disk lead the UV along the direction of rotation of the galaxy. We believe that this observation confirms that spiral density wave propagation supports the formation scenario of XUV disks (Padave et al. in prep).

\subsection{Correlations between Star Formation and \ion{H}{1}}\label{sec:sfrism}

In this subsection, we explore the relations between star formation and atomic gas (\hi) and quantify the impact of stellar feedback on the ISM in the disk of NGC 3344. The \hi\ disk is a reservoir of cold gas 
that serves as the fuel for star formation. We correlate the FUV+\mips\ \sfr\ (hereafter, \sfr) with the \hi\ properties in the optical and XUV disk of NGC 3344 and investigate
(1) atomic gas mass surface density (\mhi), (2) SFR efficiency (\sfr/\mhi), and (3) star-formation-driven feedback on the ISM using \hi\ kinematics (\disp, \kesd).

\subsubsection{The Star formation Law in NGC 3344 }\label{sssec:ism1}

\begin{figure*}[!t]
\setcounter{figure}{11}
\centering
\subfloat[]{\includegraphics[width=0.49\textwidth , trim = 1cm 0mm 1cm 0cm, clip,scale=0.068]{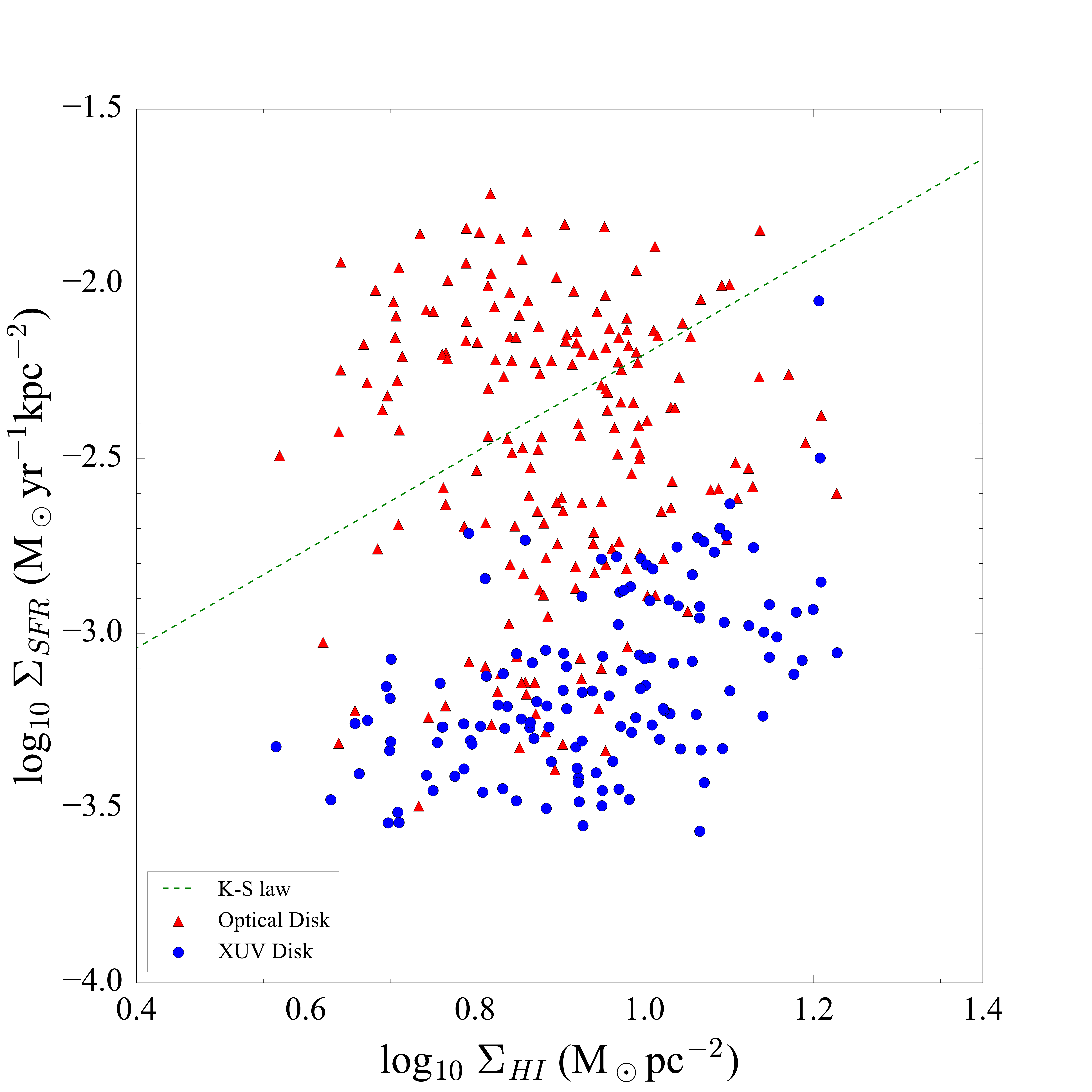} \label{fig:mhia}}\hfill
\subfloat[]{\includegraphics[width=0.49\textwidth , trim = 1cm 0mm 1cm 0cm, clip,scale=0.068]{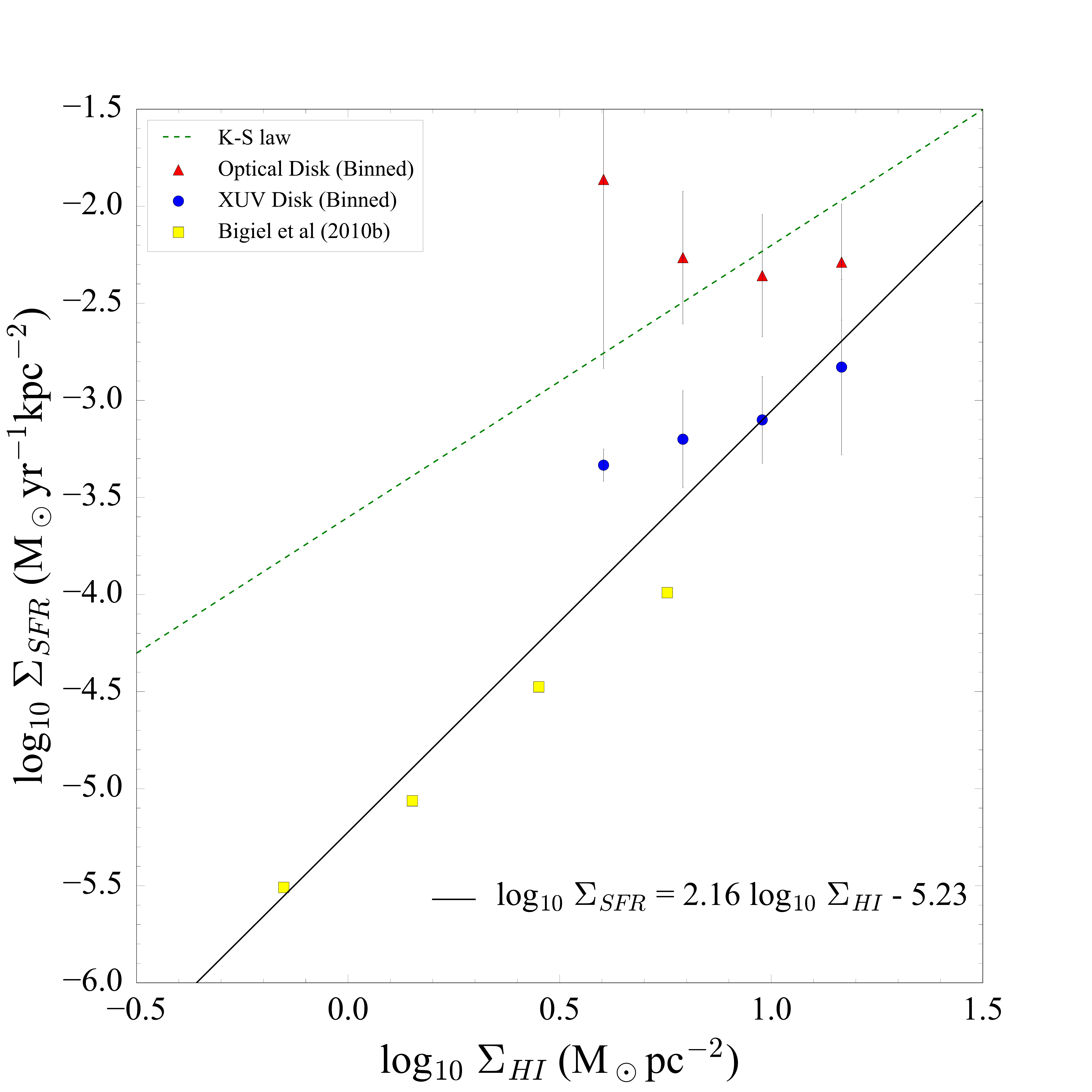} \label{fig:mhib}}
\setcounter{figure}{11}

\caption{\sfr\ and \mhi\ for the star forming 
    regions in NGC 3344. In panel (a), the red triangles and blue circles represent regions in the optical  
and the XUV disk, respectively. In panel (b), the red triangles and blue circles show binned optical and XUV data, respectively, from panel (a) along with data from \cite{bigi10} shown in yellow squares. The {\em solid} black line depicts the line of best fit to XUV data and data from \cite{bigi10}. The green ({\em dashed}) line shows the Kennicutt-Schmidt law with an index of 1.4 \citep{kenn98}. }
   \label{fig:mhi}
\end{figure*}

Figure \ref{fig:mhia} shows the relationship between surface densities of \hi\ and \sfr\ for the 320 stellar complexes in the optical ({\em red} triangles) and XUV ({\em blue} circles) disk.  The dashed green line marks the empirical K-S law that connects the total
 $\Sigma_{\rm SFR}$ to the total gas surface density, $\Sigma_{\rm gas}$,
 such that $\Sigma_{\rm SFR} = A \Sigma^{N}_{\rm gas}$ with $N = 1.4$ and
 $A = 2.5\times 10^{-4}$ yr$^{-1}$ \citep{kenn98}.
These plots only consider neutral gas and, hence, offsets from the K-S relationship are expected for regions where the contribution of molecular gas to $\Sigma_{\rm gas}$ is significant.
However, in that regime the points should lie systematically above the K-S relationship. This is seen in the star-forming regions in the optical disk, especially for regions at galactocentric distance $\lesssim 4$~kpc and \sfr$\gtrsim3.16\times10^{-3}$~M$_\odot$~yr$^{-1}$~kpc$^{-2}$.  

We find an overall lack of correlation between \sfr\ and \mhi\ which is consistent with what \cite{bigi10} and others have seen in the spatially resolved star formation law, or what \cite{reyes19} see in the disk integrated \hi\ star formation law. The distribution of \sfr\ has no relation with \mhi\ in the optical disk but shows a moderate correlation (Pearson {\it r}~=~0.51) in the XUV disk (Figure \ref{fig:mhia}). This result is the consequence of a predominantly molecular \citep{bigi08} ISM around the highly star forming inner parts of spiral galaxies and \mhi\ alone is not a good tracer of $\Sigma_{\text {gas}}$. Meanwhile, a correlation between \sfr\ and \mhi\ in the outer regions of galaxies points to the fact that high \hi\ column densities pave the way for star formation \citep{bigi10}. 

In Figure \ref{fig:mhib}, we show the median \sfr\ (and 1$\sigma$ range) in four bins of \mhi\ for both the optical ({\em red} triangles) and XUV ({\em blue} circles) data for NGC~3344, along with median outer disk data for the sample of spirals and dwarf galaxies from \cite{bigi10} ({\em yellow}  squares) that extends to much lower \hi\ surface densities. The optical and XUV disk of NGC~3344 cover a comparable range in \mhi\ but the optical disk is forming stars more {\em efficiently} by a factor of almost 10 on average. The XUV points fall in the
\mhi\ and \sfr\ regime where the deviation from the K-S law begins, connecting the H$_2$-dominated SF regions of inner disks of L$_{\star}$ galaxies to the \hi-dominated regions in the outskirts of spiral and dwarf galaxies studied by
\citep{bigi10}.

\cite{bigi10} discussed the lack of gas surface densities between $ \sim3-10 $~M$_\odot$ pc$ ^{-2} $ and low \sfr\ points conceived as a ``forbidden region" in the \mhi--\sfr\ relation. These points connect the star formation efficiency in H$_{2} $-dominated inner regions of L$_\star$ galaxies to the \hi-dominated regions in the outskirts of spiral and dwarf galaxies, thereby creating the ``S-shape" distribution seen in \cite[][their
Figure 13]{bigi10}. The drop in the \mhi--\sfr\ relation in the outer disk and the missing ``forbidden region" points found by \cite{bigi10} give rise to the question of whether the transition from the {\em efficiently} star-forming inner disk to the {\em inefficiently} star-forming outer disk is gradual or discontinuous?
The XUV disk data trace the missing parameter space of high-\hi\ low-\sfr\ points. And, the presence of the XUV points of NGC 3344 in the forbidden region in the \sfr-\mhi\ plot can be thought to connect the low-density outer disk to the H$_{2} $-dominated inner disk completing the ``S-shape" distribution. However, a larger sample of XUV disk data must be explored to investigate the origin of the break in the star-formation law.

\subsubsection{\hi\ Star Formation Efficiency in NGC 3344}

\begin{figure*}[!t]
    \centering
 \setcounter{figure}{12}

    \centering
\subfloat[]{\includegraphics[width=0.49\textwidth , trim = 0cm 0mm 1cm 0cm, clip,scale=0.068]{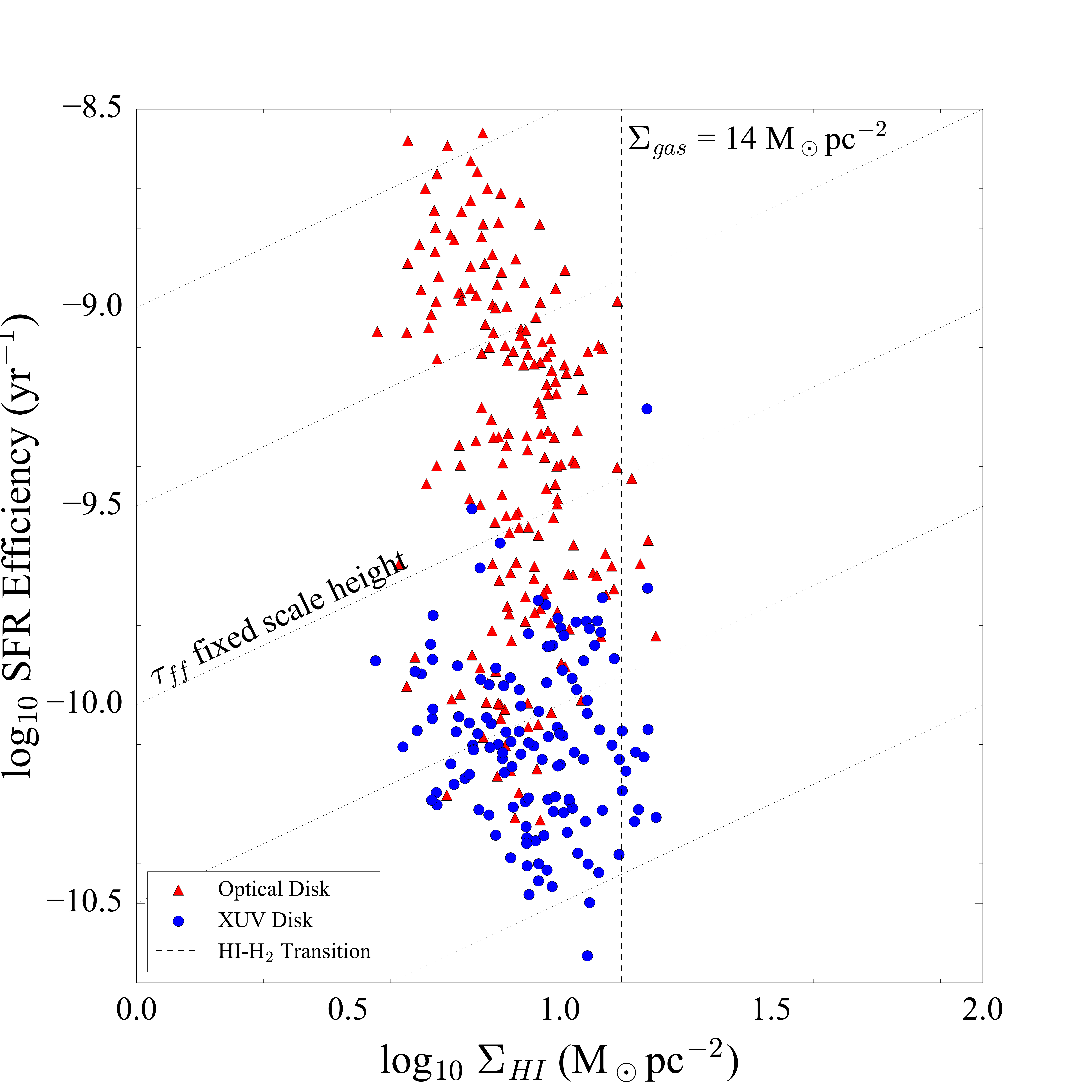} \label{fig:sfea}}\hfill
\subfloat[]{\includegraphics[width=0.49\textwidth , trim = 0cm 0mm 1cm 0cm, clip,scale=0.068]{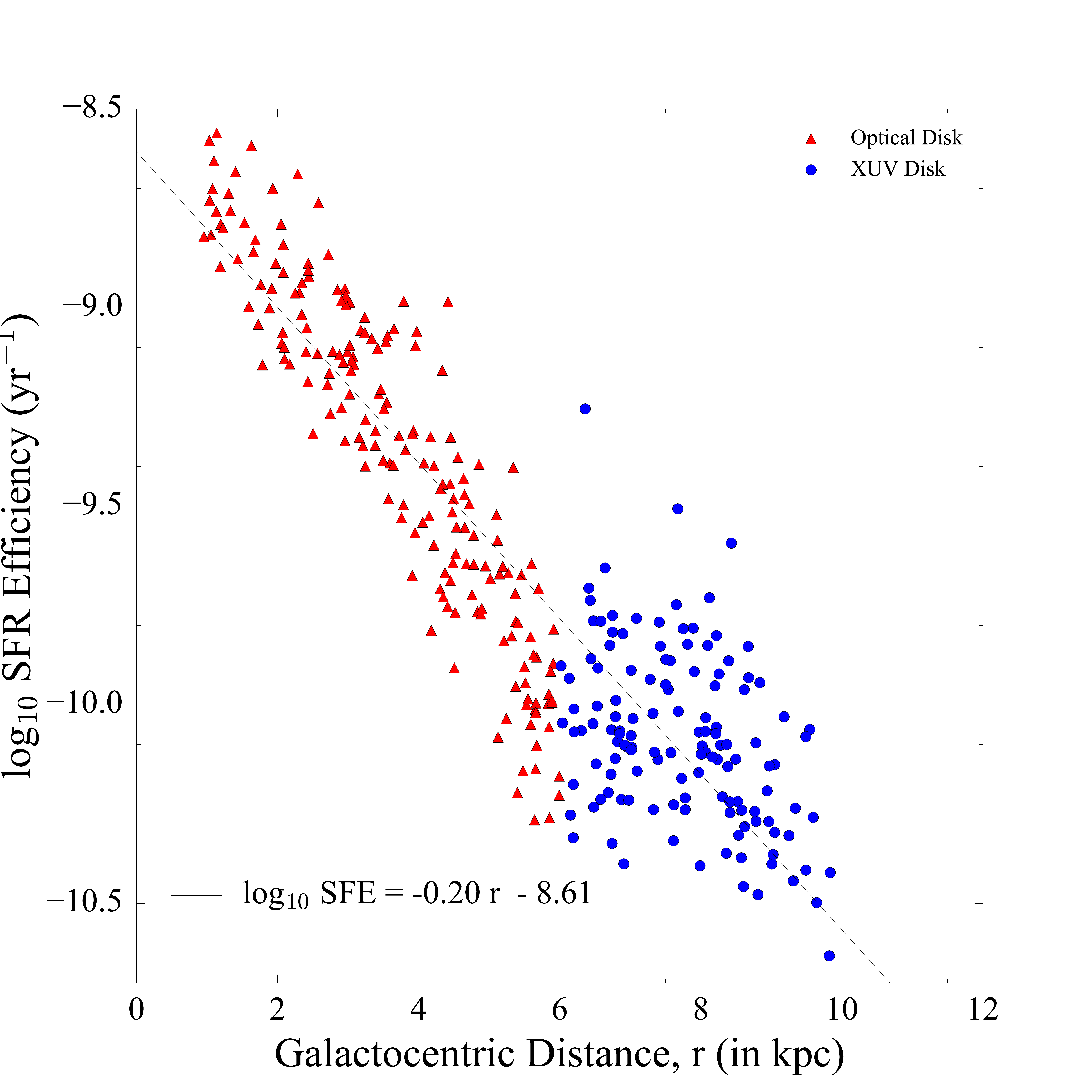} \label{fig:sfeb}}

   \setcounter{figure}{12}
\caption{\hi\ SFE as a function of (a) \mhi\ and
    (b)  galactocentric distance for the star forming 
    regions in NGC 3344. The red triangles and blue circles are regions in the optical  
and the XUV disk. The {\em dashed} line in panel (a) marks the transition from an H$_2$-dominated to an \hi -dominated environment (from \cite{lero08}). The {\em dotted} lines show the \hi\ SFE proportional to $\tau_{ff}$ in a fixed scale height gas disk for power-law coefficients of -11.0, 10.5, -10.0, -9.5, -9.0. The {\em solid} line in panel (b) represents the line of best fit to the points. In general, \hi\ plays no major role in regulating \hi\ SFEs.  The decrease in the \hi\ SFE as we move towards the outer parts of the galaxy, imply increase in the \hi\ depletion times.}
    \label{fig:sfe}
\end{figure*}

Star formation efficiency (SFE) or SFR per unit gas mass (here, \hi\ mass) indicates how good the available gas is at forming stars. It is the inverse of the gas depletion time, $ \tau_{dep} $, which is the time taken to exhaust the available supply of gas by the star-formation at the current rate. 
In Figure \ref{fig:sfe}, we plot the variation in the \hi\ SFE over the galaxy disk as a function of \mhi\ and galactocentric distance.  The {\em dashed} line marks the boundary between an \hi -dominated and H$ _{2} $-dominated ISM discussed by \cite{lero08}. The {\em dotted} lines in 
Figure \ref{fig:sfea} illustrate \hi\ SFE proportional to the free-fall time ($\tau_{ff}$) in a gas disk with a fixed scale height, , which is, $SFE\propto\Sigma_{gas}^{0.5}$\citep{kenn98b}. 
Irrespective of the power-law coefficient, the dotted lines cannot reproduce the observed relation between \mhi\ and \hi\ SFE due to the small range of \mhi\ ($ \sim 2.5-40$ M$_\odot$ pc$^{-2}$). This suggests that \mhi\ plays almost no role in governing the \hi\ SFEs. 

We find that the \hi\ SFEs drop with galactocentric distance for the star-forming complexes, also implying longer \hi\ depletion times (since SFE~$\propto\tau_{dep}^{-1}$) in the XUV disk similar to what was observed with the SFE radial profile (Figure \ref{fig:ssfrpro}). 
The correlation between \sfr\ and \hi\ and the longer \hi\ depletion times in the XUV disk infer the importance of \hi\ in regulating the star formation in the outer disk. The {\it in situ} star formation, however, would take at least 10~Gyrs to consume the current supply of \hi. Meanwhile, the longer-lived \hi\ gas can also act as a necessary source for the star-forming optical disk \citep{shlos89, blitz96, bauer10, bigi10}, in order for it to keep forming stars. 

Star formation presupposes that  
H$_2$/ giant molecular clouds (GMC) will form from \hi. Molecular gas has depletion times of $\sim2 $~Gyrs \citep{bigi08, bigi10, lero08} while the observed \hi\ depletion times in the XUV disk are much longer. Under the assumption that the depletion time of molecular gas stays the same in both the optical and XUV disk, longer \hi\ depletion times then illustrate that star formation in the outer disk is only limited by the formation of molecular clouds \citep{bigi10a, rafel16}.

\subsubsection{ISM Turbulence and Impact of Stellar Feedback in NGC 3344}\label{sub:kedisp}

\begin{figure*}[!t]
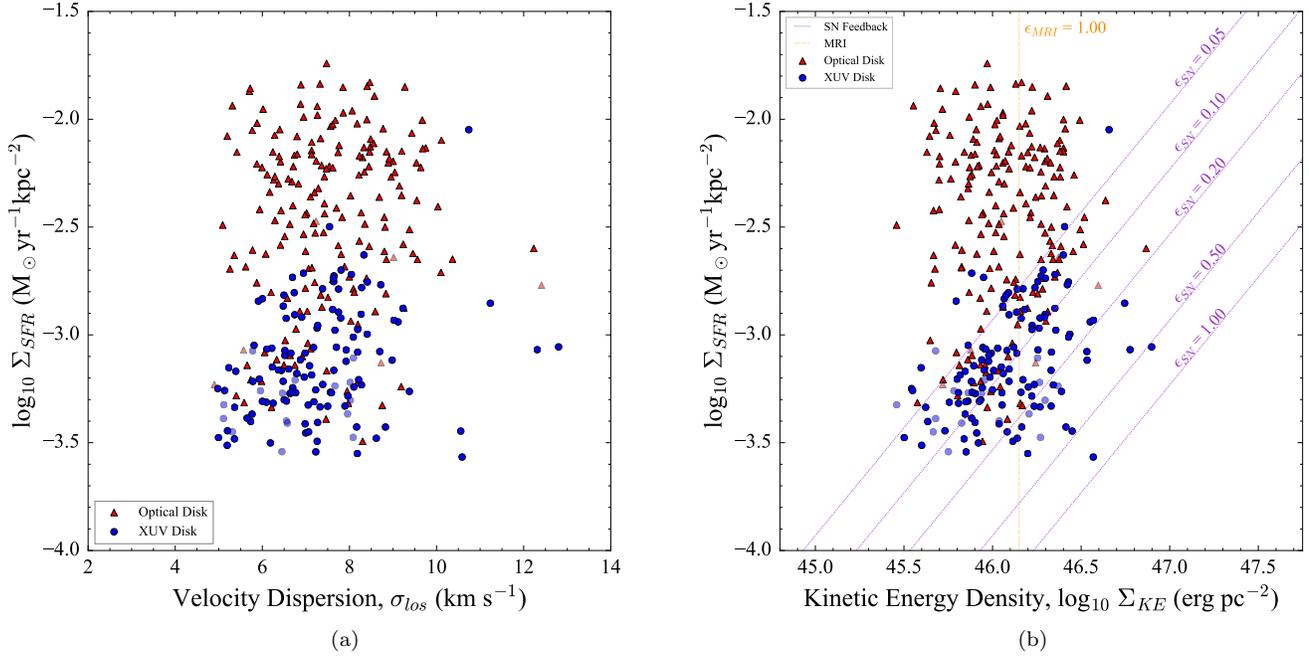

    \centering
 \setcounter{figure}{13}

\centering
\subfloat[]{\includegraphics[width=0.49\textwidth , trim = 1cm 0mm 1cm 0cm, clip,scale=0.068]{BCD_snr3_disp_vs_sfr.pdf}  \label{fig:turba}}\hfill
\subfloat[]{\includegraphics[width=0.49\textwidth , trim = 1cm 0mm 1cm 0cm, clip,scale=0.068]{BCD_snr3_ke_vs_sfr.pdf}\label{fig:turbb}}
  \setcounter{figure}{13}

\caption{Correlations between $\Sigma_{\text{SFR}}$ and (a) velocity dispersion and (b) \hi\ kinetic energy surface density $\Sigma_{{\rm KE}}$. The red triangles and blue circles are regions in the optical  
and the XUV disk, respectively. Regions with area smaller than the \hi\ beam are shown in different transparency. Models of SN energy input with different efficiency values ($\epsilon_{SN}=1$, 0.5, 0.2, 0.1, 0.05) and MRI input at maximum efficiency ($\epsilon_{MRI}=1$) from \cite{tamb09} are shown in purple and orange lines.  We find no correlations between $\Sigma_{\text SFR}$ and the gas kinematic in NGC~3344. Supernova explosions can maintain turbulence at high \sfr s but MRI, feedback mechanisms, spiral density wave may be important in the low \sfr\ - high \kesd\ regions. }
    \label{fig:turb}
\end{figure*}

To understand the connections between turbulence and star formation, we investigate the relationship between \sfr\ and \hi\ velocity dispersion (\disp) and \hi\ kinetic energy surface density (\kesd). In Figure \ref{fig:turba}, we observe \hi\ velocity dispersion between  4.8--12.8~km~s$^{-1}$ for the stellar complexes in NGC 3344. 
In both the optical disk and the XUV disk, we find a weak (Pearson {\it r}~$\sim$~0.15 and $\sim$~0.36, respectively) correlation, between \disp\ and \sfr. Also, on the galactic scale, \sfr\ varies by a few orders of magnitude, implying \sfr\ does not really depend on \disp. This observation suggests that triggering of star formation by turbulence does not dominate over large scales \citep{tamb09} as the timescales for supercritical density fluctuations that promote star formation are shorter than gas free-fall time ceasing cloud collapse \citep{kles00, elme02}. In fact, some simulations have shown that velocity dispersion can be emulated from self-gravity without any contribution from star formation \citep{hopkins11}.  
However, in the outer disk where self-gravity is weak \citep{elme06}, turbulence can be crucial in supporting the structure of the ISM \citep{lars81, elme04, macl04, mckee07} and cloud formation \citep{krum05}. This could explain the slightly higher Pearson {\it r} value in the XUV disk. 

We note that the \hi\ maps used here for the analysis were produced with masking imposed at 3$\sigma$ and a beam size of 6\farcs7$\times$5\farcs4. Stellar associations were extracted from the GALEX maps which has a resolution of 4\farcs2. As a result, some regions that are smaller than the \hi\ beam have poorly-resolved \disp\ measurements . These regions are shown in a different transparency in Figure \ref{fig:turb}. Excluding them from the analysis, we find that the Pearson {\it r} values change to $\sim$~0.11 and $\sim$~0.26 in the optical and XUV disk, respectively. Overall, this does not impact our result. In Appendix \ref{ap:hi}, we also discuss the effect of signal-to-noise ratio and resolution on the observed correlations.

In Figure \ref{fig:turbb}, we plot \sfr\ as a function of \kesd\ in the optical and XUV disk of NGC 3344. Energy inputs of different supernova (SN) efficiency values, $\epsilon_{\rm SN}=0.05, 0.1, 0.2, 0.5, 1.0$ and magnetorotational instabilities (MRI) energy input at maximum efficiency, $\epsilon_{\rm MRI}=1.0$ from \cite{tamb09} are plotted as purple and orange dashed lines. Since \kesd\ is a product of \mhi\ and \disp$^2$, the observed distribution of \kesd\ in NGC 3344 is mainly a result of variations in \mhi\ as \disp\ shows little variation with \sfr. 

We find almost no correlation (Pearson $r$ = 0.10) between \sfr\ and \kesd\ of the stellar complexes in the disk of NGC 3344.  \cite{hunter21} also found a poor correlation between \kesd\ and \sfr\ in the LITTLE THINGS sample of dwarf irregular galaxies. They found that both the \kesd\ and FUV maps of the dwarf irregular galaxies show clumps but the position of FUV clumps do not line up with those in the \kesd\ map. We also observe a similar disposition of the clumps in the maps of NGC 3344 which is possibly why we don't see a correlation. 

On sub-kpc scales, line widths of $\sim 8$~\kms\ are emitted by the warm neutral gas with temperatures of $\sim$~8$\times 10^3$~K \citep{wolf95}, whereas widths broader than 8~\kms\ are caused by turbulence stirring the ISM \citep{dick90,tamb09,koch18}. However, turbulence decays over a short timescale of $ \sim 10 $~Myrs \citep{macl99, ostr01}. 
In order to maintain a steady-state over the lifetime of a galaxy, turbulence is likely driven by various mechanisms that inject energy into the ISM, among which SN explosions and MRI are some of the most important. In the disk of NGC 3344, only $\lesssim 20$\% of the SN energy input would be required to maintain the observed \disp\ in most regions. Therefore, at high \sfr, SN feedback would be sufficient to support the ISM. Expected to play a major role in maintaining the velocity dispersion and magnetic field in regions with lower levels of star formation \citep{sell99, tamb09,utom19}, our results suggest that MRI may only be playing a small role in imparting energy to the ISM at low \sfr\ levels of $\lesssim10^{-3}$ M$_\odot$~yr$^{-1}$~kpc$^{-2}$. 

Additional sources of energy could be acting to drive the turbulence in the disk if the expected SN efficiency becomes larger than 50\% \citep{teno91}. In regions with $\epsilon_{\rm SN}\gtrsim50\%$ and high \kesd, MRI could be dominant. Other feedback mechanisms such as stellar winds, photoelectric heating by FUV radiation can inject energy into the ISM but their effect will always be sub-dominant compared to SNe \citep{korn00, macl04,bacch20}. 

\subsection{Connection Between Circumgalactic Gas Flows and Star Formation in the Disk }\label{sub:gas}

We discuss the two possibilities for the CGM gas flow detected in the stronger metal-line component of the QSO absorption system at an impact parameter of 29.5~kpc from the center of the galaxy and 19.26~kpc from the edge of the \hi\ disk. The blue-shifted absorption feature at 113.5~\kms\ with respect to the disk gas closest to the sightline is consistent with gas flowing into the disk from behind or out of the disk in front. We will consider the implications of both scenarios in an attempt to identify which is most likely.

First, we estimate how likely it is for us to detect outflowing material at the observed impact parameter and velocity offset. Assuming that the opening angle of the outflow cone is as large as some of those seen in starburst galaxies in the low-redshift Universe of $\theta \sim 60^{\circ}$ \citep{heckman90}, the ``true" 3-dimensional velocity of the cloud would be $\sim$131~\kms, and a transverse velocity of 66~\kms. At this rate, it would take the cloud about 500-600~Myrs to reach 29.5~kpc depending on its point of origin in the star-forming disk at a constant velocity. This would indicate the current star formation seen in the XUV disk is not responsible for generating the observed cloud as outflow. In addition, the cloud would encounter the gravitational pull of the galaxy as well as interaction with the hot gas (drag force) that would likely limit its survival times. 

Another line of argument would be to look at the velocity of ejection. It is expected that the velocity at the time of ejection was greater than the velocity inferred above as drag and ballistic nature of the motion would act to reduce the velocity as a function of distance from the point of ejection. A detailed analysis of the motion of starburst driven clouds in the CGM of starburst galaxies by \citet{afruni21} indicates that an initial kick velocity of $\approx$ 370 \kms\ would get as far as 40~kpc before returning back towards the galaxy. However, the kick velocity would imply that a much higher \sfr\ of the order of 1~$\rm M_{\odot}~yr^{-1}~kpc^{-2}$ \citep{heckman16} is needed, i.e., essentially requiring a starburst in the galaxy. Therefore, it is unlikely that the blue-shifted cloud seen as a QSO absorption feature is star-formation-driven outflowing material.

It is worth noting that the velocity kinematics of the blue-shifted cloud is consistent with HVCs seen in the Milky Way halo. 
They are predominantly tracing inflowing material \citep{fox19} that have a significant fraction of metals. These include returning material such as the Smith cloud and the complex C in the Milky Way halo \citep{lockman08,fox16,fraternali15}. Therefore, it is likely that the blue-shifted cloud could be infalling material that might have originated in the disk long back and is now returning gas from the CGM or the intergalactic medium that has undergone metal mixing. Similarly, kinematically anomalous clouds have been observed in nearby galaxies \citep{frat01,thilk04,west05,heald15,gim21}.

Additionally, the cloud might be tracing a faint but gradual inflow event. The atomic gas in NGC~3344 indicates no major merger signature in the gas dynamics, however, the entire outer disk covering an area of about 340~kpc$^2$ from a radius of 6 $-$ 12 kpc is undergoing a star formation event that is no more than 100~Myrs old. This suggests that there must have been a global event that triggered star formation throughout the outer disk. 
We can eliminate major mergers as we do not see a companion nor do we see the atomic gas disk showing any strong dynamical perturbations. Interestingly, the cold gas mass of this galaxy is dominated by atomic gas (87\% of total gas mass), and the depletion time for \hi\ is almost an order-of-magnitude that of molecular gas. This may suggest that perhaps this is an accretion event that is smooth and gradual, not to disrupt the disk but to build the immense gas reservoir. In this picture, the fact that most of the metals in the QSO sightline probing the CGM at 30~kpc is showing a gas flow velocity of about 113~\kms\ would support gradual gas accretion.

\section{Summary} \label{sec:summ}
In this paper, we study the star formation and its effect on the ISM and the inner CGM of the XUV disk galaxy, NGC 3344. We investigate radial variations in stars, dust, and gas using surface photometry of FUV, \ha, {\em r}-, {\em g}-band, \mips\ and \hi-21~cm emission. We identify 320 young stellar complexes in the disk of NGC 3344 with typical physical sizes between 150~pc--1~kpc. Further, we study the relationship between star formation and ISM properties for these stellar complexes using aperture photometry and investigate how these properties differ in the inner and outer disk. 
Our key results are summarized below:

\begin{itemize}
\item[1.] We find that FUV emission shows scale length 1.4 times larger than those from {\em g }, B, {\em r}, \ha, and \mips. This indicates that young stars have extended distribution while old stellar populations and dust are concentrated more towards the center. We identify a break at 6~kpc using the FUV and {\em r}-band surface brightness profiles, marking a transition from the {\em inner} optical disk to the {\em outer} XUV disk.  
\item[2.] Comparing the non-dust corrected (FUV, \ha) and dust-corrected (FUV+\mips, \ha+\mips) star formation tracers shows that both FUV and FUV+\mips\ tracers are more sensitive indicators of star formation in the XUV disk while dust-corrected tracers are more effective in the optical disk and on the galactic scale. 
\item[3.] We also find that the SFRs traced by \ha\ are consistently lower than FUV SFRs, especially in the XUV disk. Investigation of the \ha+\mips\ and FUV+\mips\ SFRs for the identified stellar complexes show that lower \ha-to-FUV ratios in the XUV disk are likely due to stochastic sampling of the IMF along with the effect of steeper upper IMF and/or truncation. 
\item[4.] We observe that sSFR increases from 10$^{-10}$~yr$^{-1}$ in the optical disk to 10$^{-8}$~yr$^{-1}$ the XUV disk, suggesting that the XUV disk is a ``starburst" actively forming stars with a slowly growing optical disk. This provides evidence for inside-out growth of the disk.
\item[5.] \sfr\ of the stellar complexes show no correlation with \mhi\ in the disk. In the XUV disk, however, we find a moderate correlation (Pearson $r$ = 0.51). This is a consequence of an ISM that is H$_2$-dominated in the optical disk and \hi-dominated in the XUV disk. The XUV disk points also show high-\mhi\ and low-\sfr\ and mark the onset of a deviation from the traditional Kennicutt-Schmidt law. 
\item[6.] The \hi\ star formation efficiency decreases as a function of galactocentric distance, and we find longer \hi\ depletion times. This suggests that the \hi\ reservoir, which has only recently started forming stars will continue to do so for $\sim$10~Gyrs in the XUV disk and also feed the optical disk. 
\item[7.] Correlations between \sfr\ and \kesd\ show that stellar feedback via supernova explosions can maintain the observed \hi\ velocity dispersion in the optical and XUV disk. MRI and additional mechanisms, such as, spiral density wave may be playing a sub-dominant role in the high \kesd\ - low \sfr\ regime in the XUV disk to maintain the ISM turbulence.
\item[8.] We detect two absorbing systems in the QSO-sightline probing the inner-CGM at 30~kpc from the center. The CGM is metal-rich and has a low-ionization potential as indicated by the presence of strong \ion{Si}{2}, \ion{Si}{3}, and \ion{C}{2} and a relatively weak \ion{Si}{4} associated with the strong component. The weak component was seen in \Lya\ and \ion{Si}{3} only. 
\item[9.] The CGM shows a component (the weaker component) that is consistent with rotation. The stronger metal-line component shows a velocity offset of 113~\kms, thus making it similar to high-velocity clouds in the Milky Way halo. While the velocity offset confirms gas flow, the cloud could either be inflowing or outflowing material. Based on the kinematics and the energies necessary to propel a gas cloud with the observed velocity out to 40~kpc, we conclude that it is unlikely that the cloud is tracing outflow. Instead, the observations are consistent with inflowing gas, which might perhaps be related to the triggers of star formation in the XUV disk. 
\end{itemize}

XUV disks are unique laboratories to test our theories of star formation and feedback. Future deep optical and UV imaging of XUV disks will further our understanding of the formation of XUV disks and galaxy growth.


\acknowledgments

MP, SB, RJ, and DT are supported by NASA ADAP grant 80NSSC21K0643, SB and HG are also supported by NSF Award Number 2009409, and SB, HG, and TH are supported by
 \emph{HST} grant HST-GO-14071 administrated by STScI which is operated by AURA
 under contract NAS\,5-26555 from NASA.

We thank the staff at the Space Telescope Science Institute, the National Radio Astronomy Observatory (NRAO) Array Operations center at Socorro,  the Steward Observatory and the Vatican Advanced Technology Telescope for their help and support on this project. We thank the referee for their constructive feedback. We thank the members of the ASU STARs lab (Jacqueline Monkiewicz, Lee Chiffelle, Chris Dupuis, Tyler McCabe, and Ed Buie II) for their extensive help and feedback during the course of this work. 
MP, SB, HG, and RJ acknowledge the land and the native people that Arizona State University's campuses are located in the Salt River Valley. The ancestral territories of Indigenous peoples, including the Akimel O’odham (Pima) and Pee Posh (Maricopa) Indian Communities, whose care and keeping of these lands allows us to be here today.

GALEX is a NASA Small Explorer, launched in 2003 April.
We gratefully acknowledge NASA’s support for the construction,
operation, and science analysis of the GALEX mission, developed in cooperation with the
sCentre National d'Etudes Spatiales (CNES) of France and the
Korean Ministry of Science and Technology.

This work is also partly based on observations with the VATT: the Alice P. Lennon Telescope and the Thomas J. Bannan Astrophysics Facility.

The National Radio Astronomy Observatory is a facility of the National Science Foundation operated under cooperative agreement by Associated Universities, Inc.

Based on observations made with the NASA/ESA Hubble Space Telescope, obtained from the data archive at the Space Telescope Science Institute. STScI is operated by the Association of Universities for Research in Astronomy, Inc. under NASA contract NAS 5-26555.

Funding for the SDSS and SDSS-II has been provided by the Alfred P. Sloan Foundation, the Participating Institutions, the National Science Foundation, the U.S. Department of Energy, the National Aeronautics and Space Administration, the Japanese Monbukagakusho, the Max Planck Society, and the Higher Education Funding Council for England. The SDSS Web Site is http://www.sdss.org/.

This work is based [in part] on observations made with the Spitzer Space Telescope, which was operated by the Jet Propulsion Laboratory, California Institute of Technology under a contract with NASA

The SDSS is managed by the Astrophysical Research Consortium for the Participating Institutions. The Participating Institutions are the American Museum of Natural History, Astrophysical Institute Potsdam, University of Basel, University of Cambridge, Case Western Reserve University, University of Chicago, Drexel University, Fermilab, the Institute for Advanced Study, the Japan Participation Group, Johns Hopkins University, the Joint Institute for Nuclear Astrophysics, the Kavli Institute for Particle Astrophysics and Cosmology, the Korean Scientist Group, the Chinese Academy of Sciences (LAMOST), Los Alamos National Laboratory, the Max-Planck-Institute for Astronomy (MPIA), the Max-Planck-Institute for Astrophysics (MPA), New Mexico State University, Ohio State University, University of Pittsburgh, University of Portsmouth, Princeton University, the United States Naval Observatory, and the University of Washington.

\facilities{GALEX, HST, Sloan, Spitzer, VATT, VLA}

%



\appendix
\section{Effect of spatial resolution and S/N on relations of \mhi, \disp, \kesd\ with \sfr}\label{ap:hi}

We investigated the influence of resolution and signal-to-noise ratio (S/N) of the various \hi\ maps on the relations observed in \S\ref{sec:sfrism}. We first study the effects of S/N on the ISM-\sfr\ relations. For this, we produced the \mhi, \disp, and \kesd\ maps above 2$\sigma$ and 4$\sigma$ using the method prescribed in \S\ref{sub:ism}. Pixels with extremely small values of velocity dispersion were masked in all maps. For the 4$\sigma$ maps, the number of pixels masked was the highest which in turn reduced the number of regions with valid pixel values. Figure \ref{fig:sd2} and Figure \ref{fig:sd4} show the ISM-\sfr\ relations for the 2$\sigma$ and 4$\sigma$ maps plotted along with 3$\sigma$ (Figure \ref{fig:sd3})  for comparison. We do not find much variation in the \mhi-\sfr\ distribution in the three plots except the regions with all pixels masked are missing in the 4$\sigma$ plot. The variations in the \kesd-\sfr\ plots are, hence, caused by the variations in the \disp\ values at different S/N levels. At a higher S/N cut, we get lower \disp\ values compares to the lower S/N cuts as we lose out on the high velocity wing of the profile.

\begin{figure*}[!htb]
    \centering
 \setcounter{figure}{14}
\includegraphics[width=0.325\textwidth ,  trim = 0cm 0mm 0cm 1cm, clip,scale=0.05]{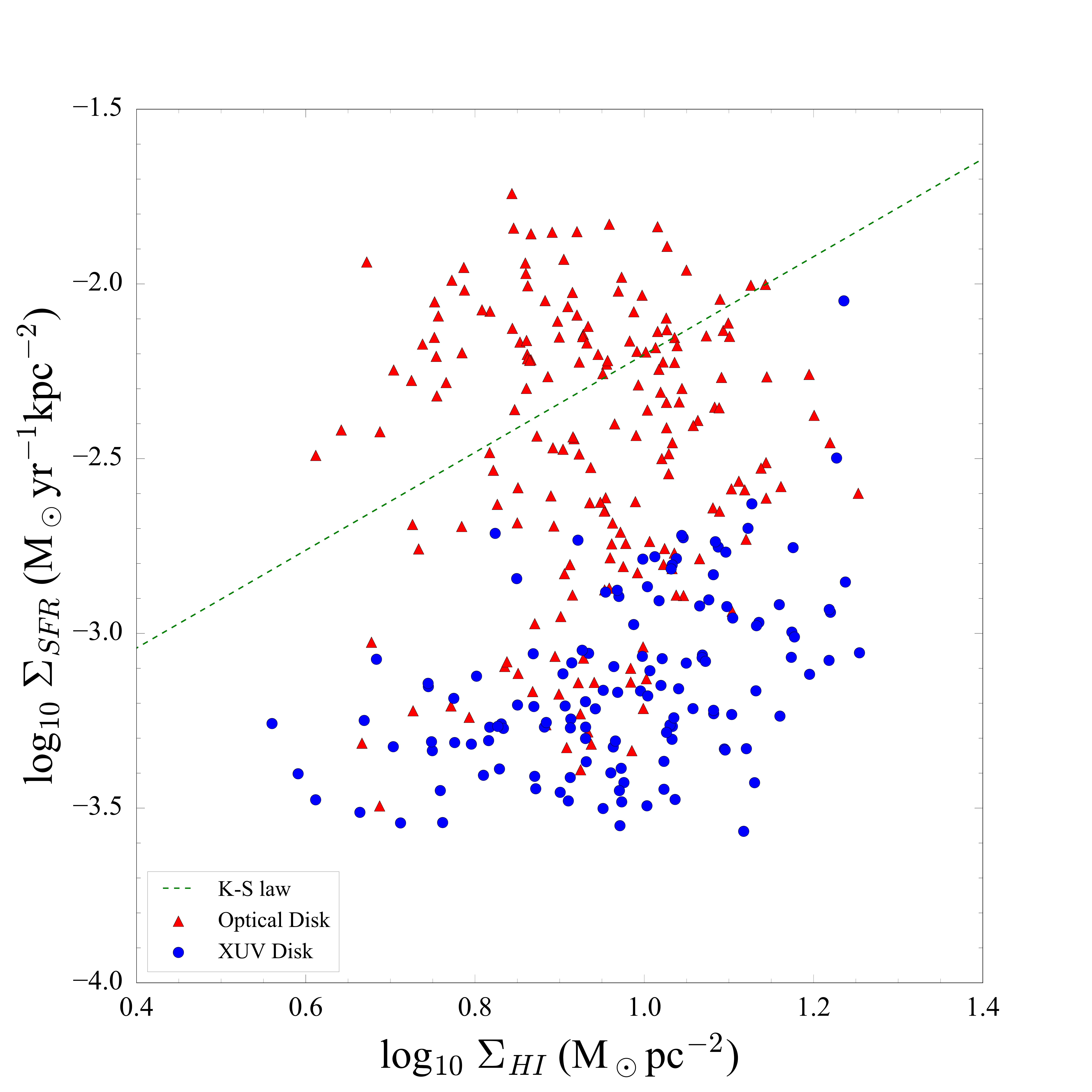}
\includegraphics[width=0.325\textwidth ,  trim = 0cm 0mm 0cm 1cm, clip,scale=0.05]{BCD_snr3_mhi_vs_sfr.pdf}
\includegraphics[width=0.325\textwidth ,  trim = 0cm 0mm 0cm 1cm, clip,scale=0.05]{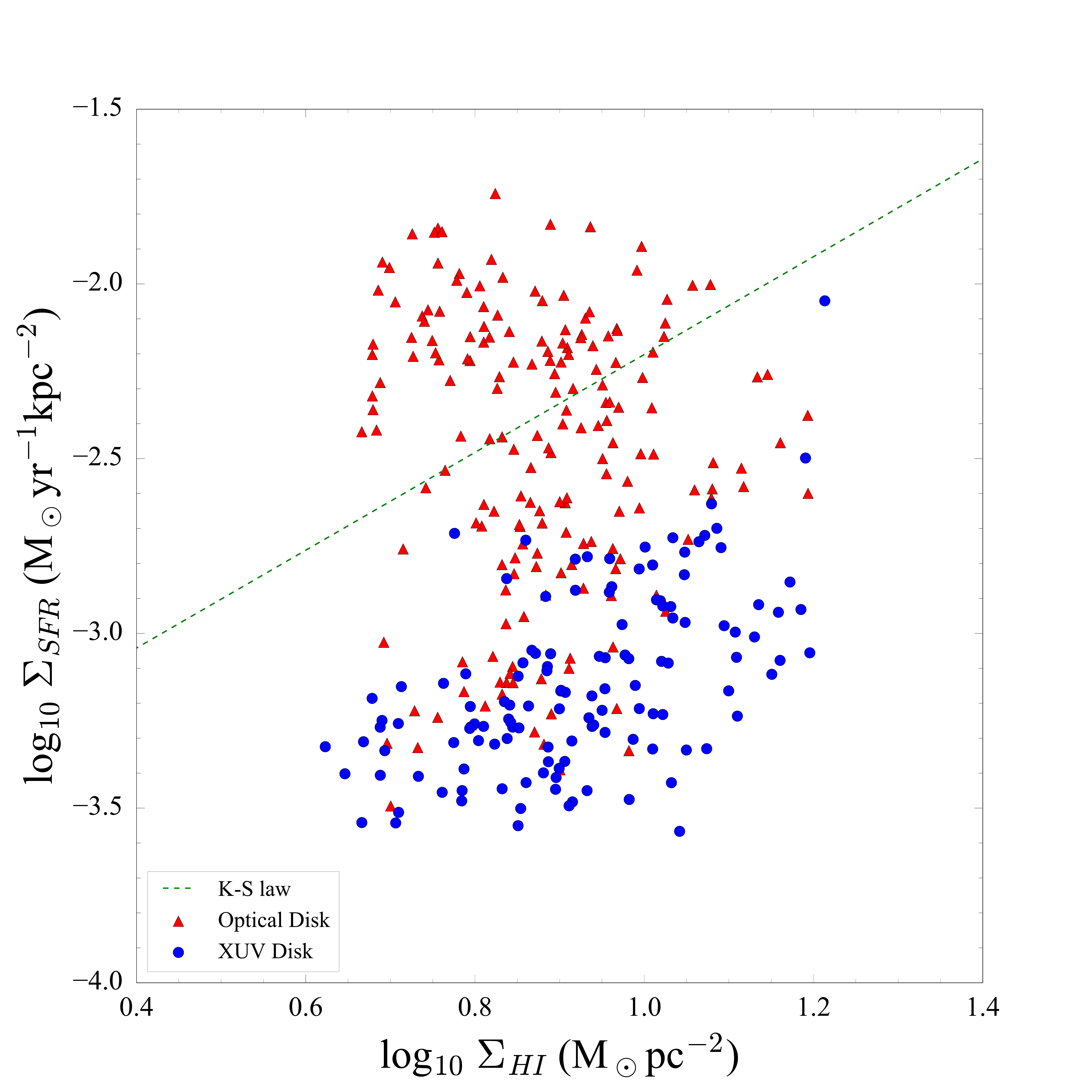}

\includegraphics[width=0.325\textwidth , trim = 0cm 0mm 0cm 1cm, clip,scale=0.045]{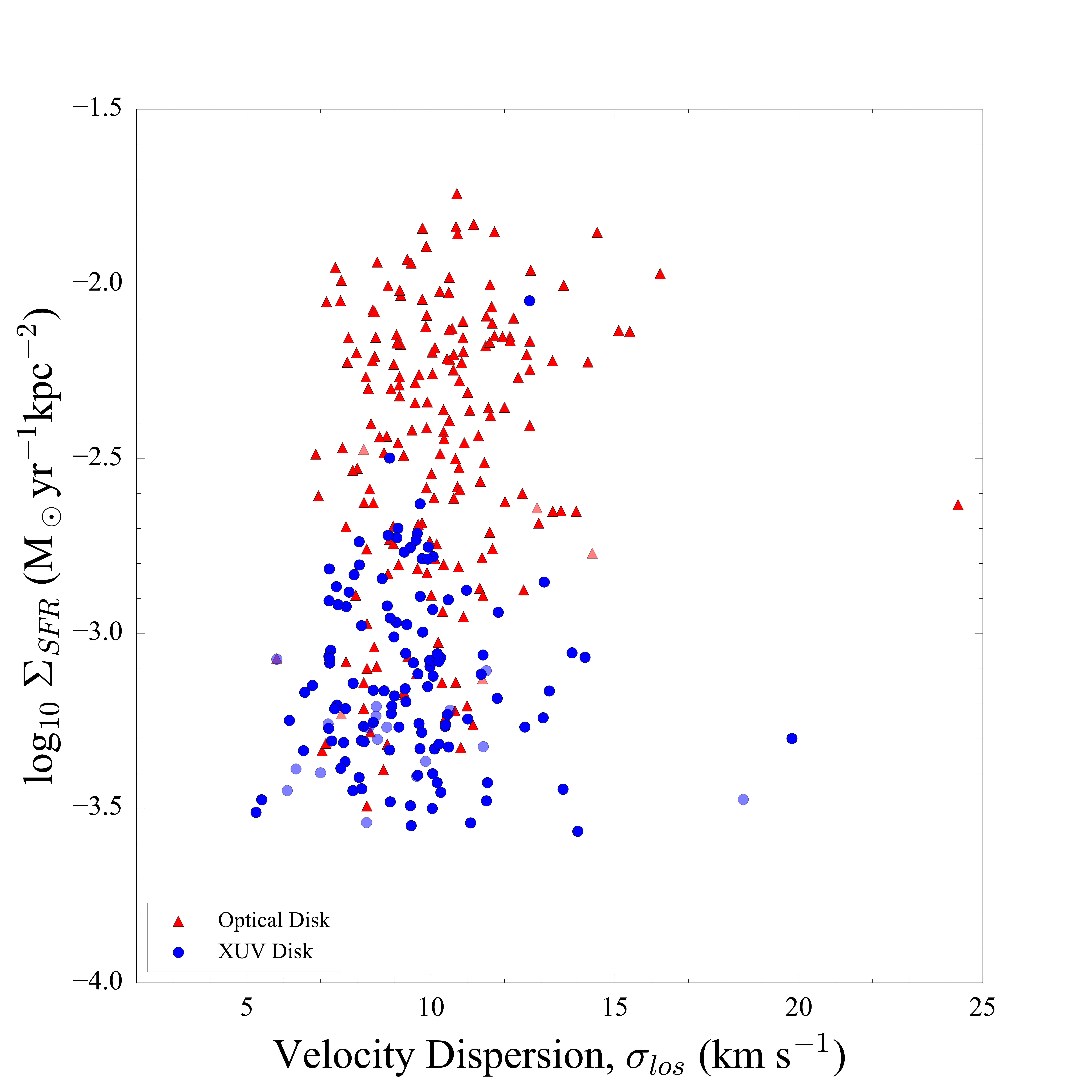}  
\includegraphics[width=0.325\textwidth , trim = 0cm 0mm 0cm 1cm, clip,scale=0.045]{BCD_snr3_disp_vs_sfr.pdf}
\includegraphics[width=0.325\textwidth , trim = 0cm 0mm 0cm 1cm, clip,scale=0.045]{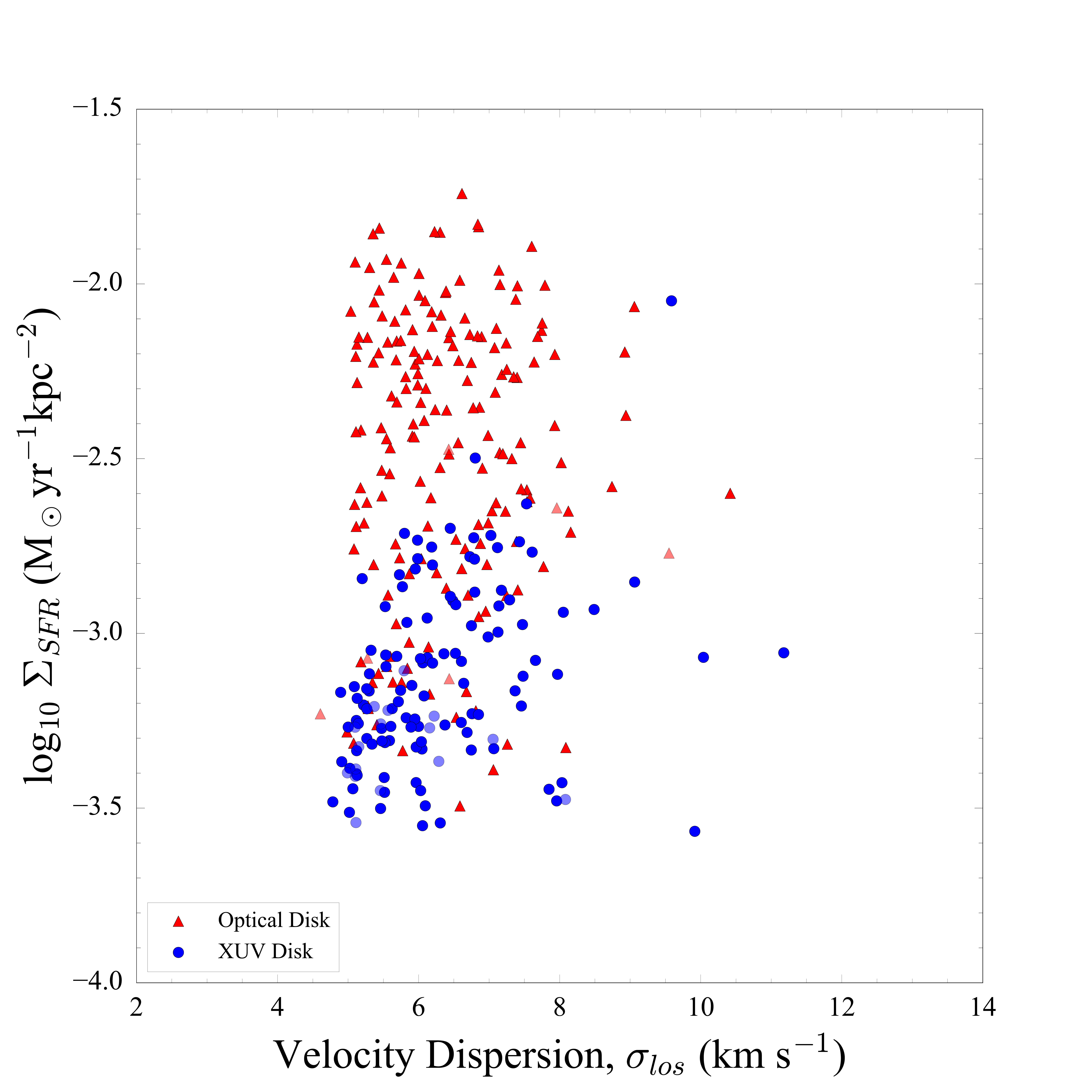}

\subfloat[2$\sigma$]{\includegraphics[width=0.33\textwidth ,  trim = 0cm 0mm 0cm 1cm, clip,scale=0.05]{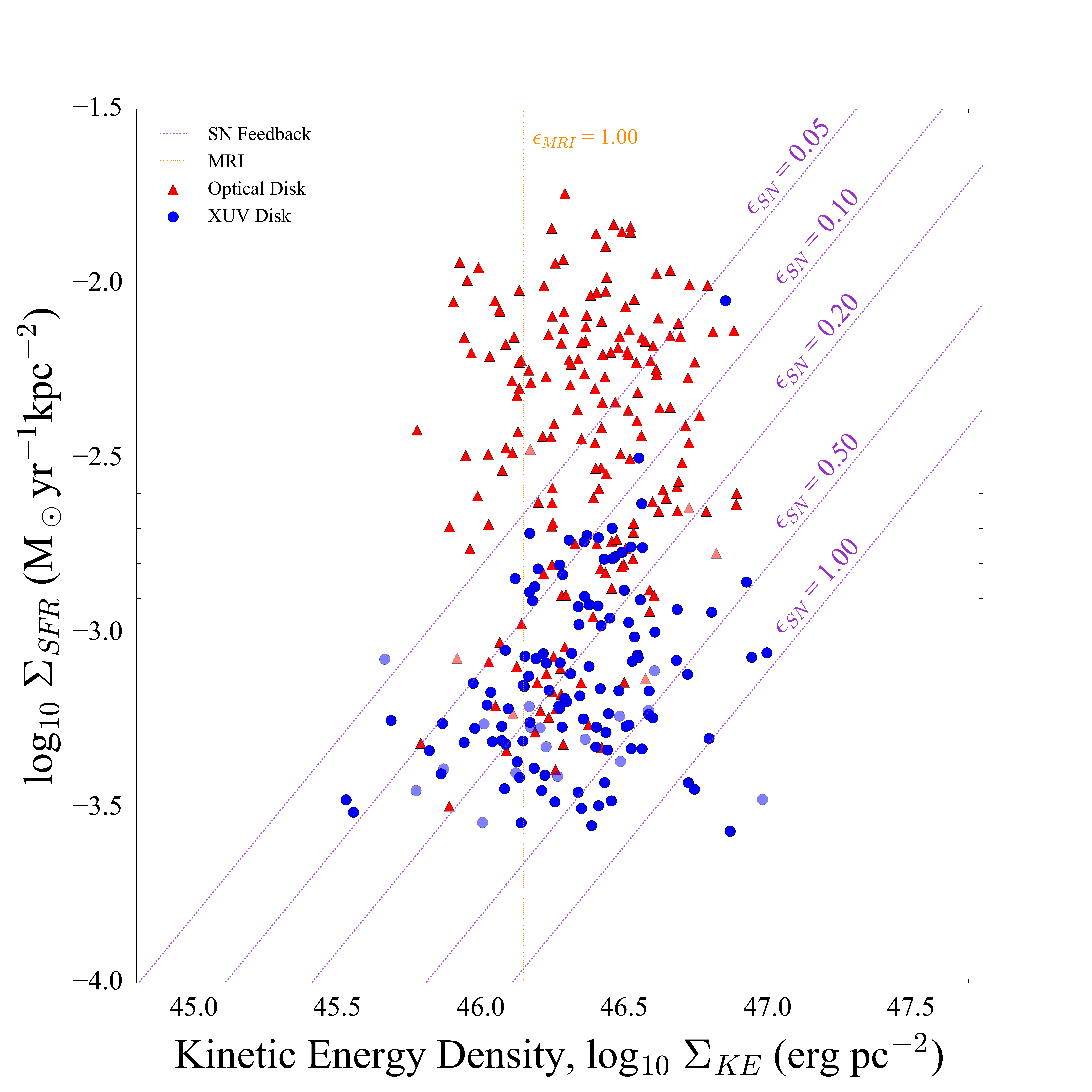} \label{fig:sd2}}
\subfloat[3$\sigma$]{\includegraphics[width=0.33\textwidth ,  trim = 0cm 0mm 0cm 1cm, clip,scale=0.05]{BCD_snr3_ke_vs_sfr.pdf}\label{fig:sd3}}
\subfloat[4$\sigma$]{\includegraphics[width=0.33\textwidth ,  trim = 0cm 0mm 0cm 1cm, clip,scale=0.05]{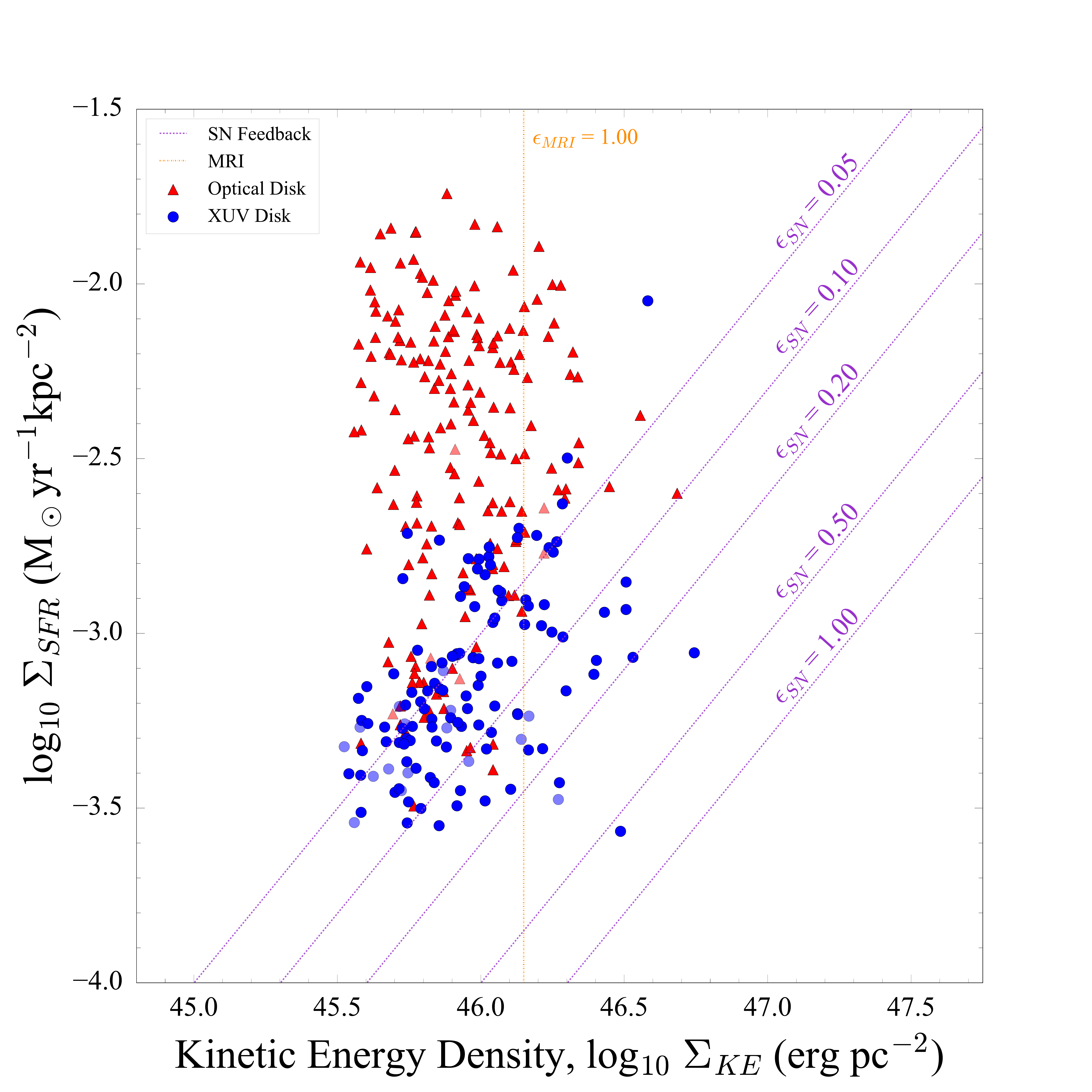} \label{fig:sd4}}

   \setcounter{figure}{14}

\caption{Correlations of \sfr\ with \mhi\ (top panel), \disp\ (middle panel), and \kesd\ (bottom panel) above (a) 2$\sigma$, (b) 3$\sigma$, and (c) 4$\sigma$. The red triangles and blue circles are regions in the optical  
and the XUV disk. Regions with area smaller than the \hi\ beam are shown in different transparency in the \sfr-\disp\ and \sfr-\kesd\ plots. The green ({\em dashed}) line in the top panel shows the Kennicutt-Schmidt law for index of 1.4 \citep{kenn98}. Models of SN energy input with different efficiency values ($\epsilon_{SN}=1$, 0.5, 0.2, 0.1, 0.05) and MRI input at maximum efficiency ($\epsilon_{MRI}=1$) from \cite{tamb09} are shown in purple and orange lines in the bottom panel. }
    \label{fig:sig}
\end{figure*}

We also study the effect of resolution by using only D-configuration VLA data. We produced the \mhi, \disp, and \kesd\ maps above 3$\sigma$ at a resolution of 59\arcsec$\times$41\arcsec. The typical size of the star forming regions
are 6\arcsec--40\arcsec(150 pc--1 kpc), hence, these maps represent semi-local average ISM properties around the star-forming regions. 
Figure \ref{fig:dconf} shows the \disp\ vs \sfr\ and \kesd\ vs \sfr\ relations for these maps. We find that the gas is kinematically spread out (8-25~\kms). In Figure \ref{fig:dconfb} and  \ref{fig:dconfc}, we find that in the optical disk, both \disp\ and \kesd\ show moderate correlation with \sfr\ with Pearson $r$ value of 0.59 and 0.65, respectively. The correlations are weak in the XUV disk with Pearson $r$ value of 0.23 and 0.34, respectively. Figure \ref{fig:dconfc} also shows that mechanisms other than SN feedback are likely dominant in the XUV disk in contributing to the turbulence of the ISM with MRI playing no role. These results are likely not representing a real scenario as feedback acts on scales much smaller than 0.5 kpc \citep{comb12}. The observations are probably caused by the \kesd\ values now representing semi-local averages around the FUV clumps at that resolution.

\begin{figure*}[!htb]
    \centering
 \setcounter{figure}{15}
\subfloat[]{\includegraphics[width=0.33\textwidth , trim = 0cm 0mm 1mm 0cm, clip,scale=0.05]{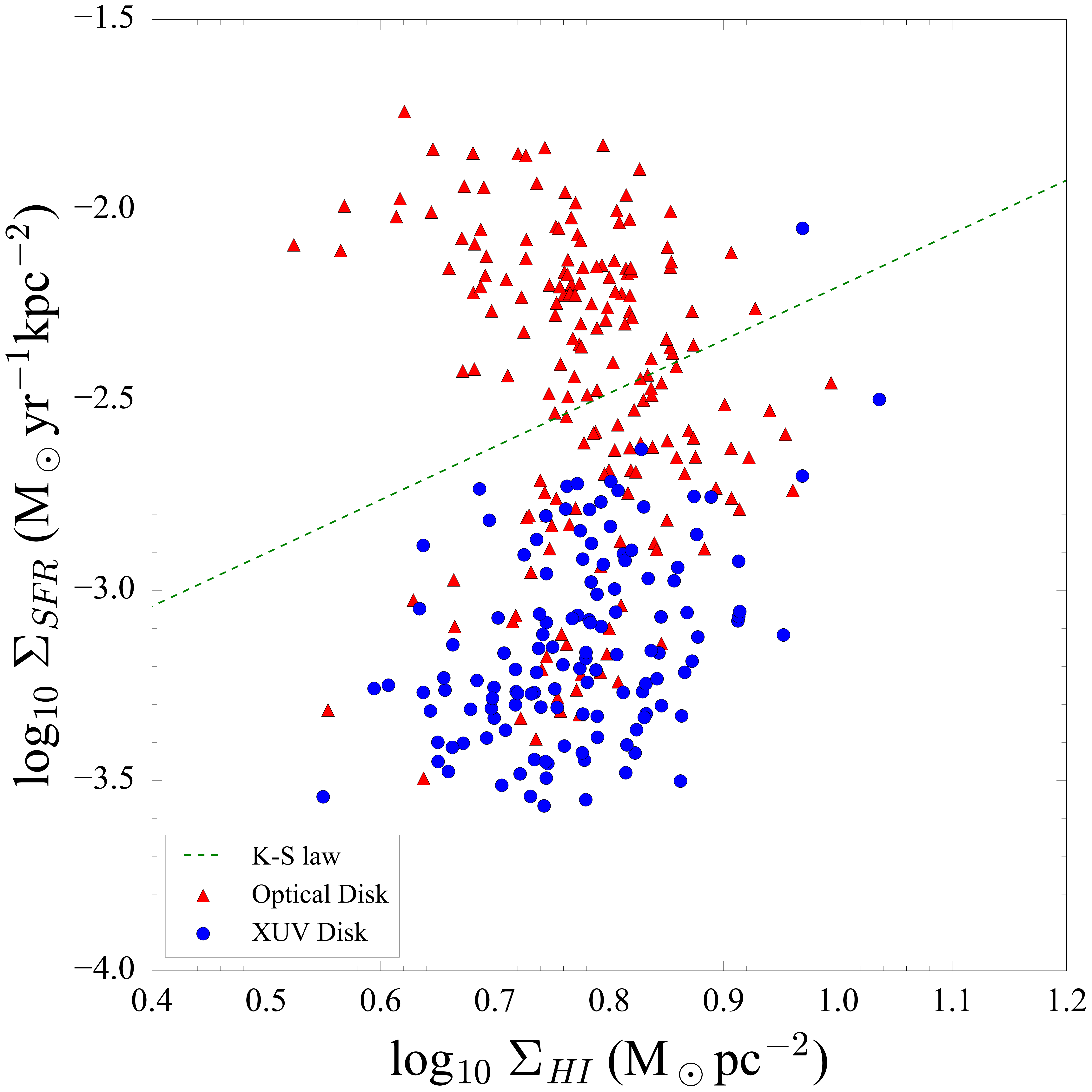} \label{fig:dconfa}}
\subfloat[]{\includegraphics[width=0.33\textwidth ,  trim = 0cm 0mm 1mm 0cm, clip,scale=0.05]{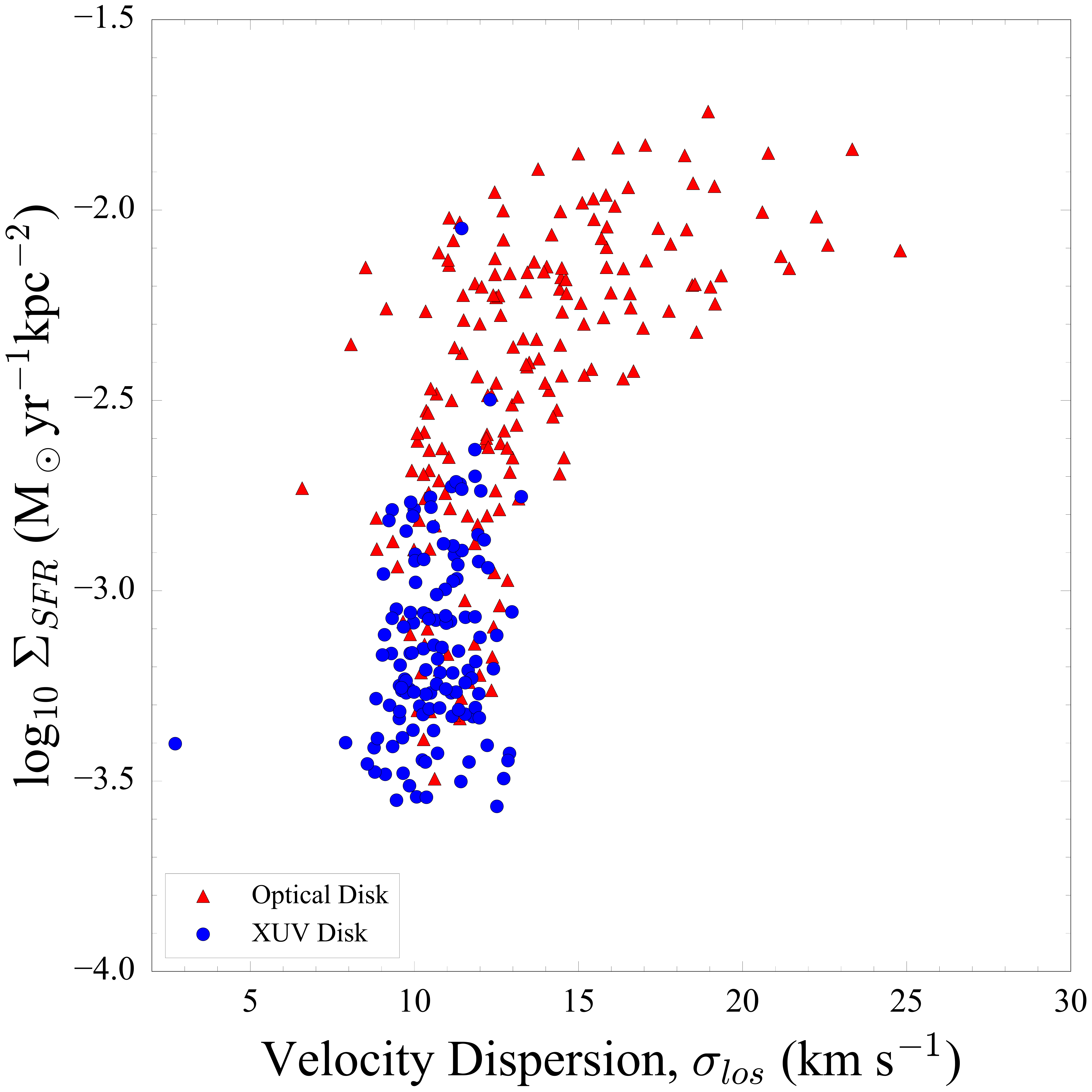}  \label{fig:dconfb}}
\subfloat[]{\includegraphics[width=0.33\textwidth , trim = 0mm 0mm 1cm 0cm, clip,scale=0.05]{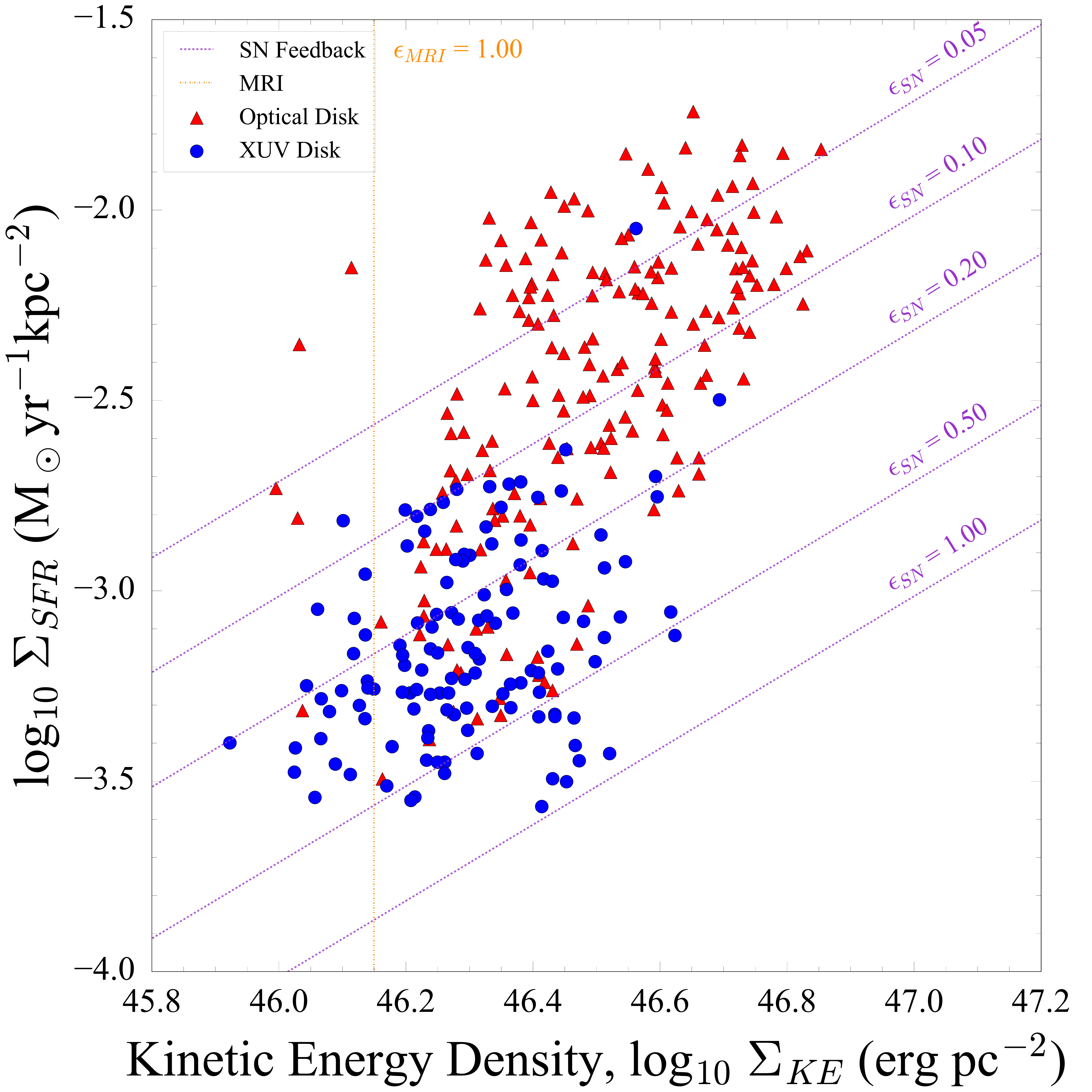} \label{fig:dconfc}}
   \setcounter{figure}{15}

\caption{Correlations between\sfr\ and (a) \mhi, (b) \disp, and (c) \kesd\ from the VLA D-configuration data. The red triangles and blue circles are regions in the optical  
and the XUV disk, respectively, with red ({\em solid}) and blue ({\em dotted}) lines representing the lines of
best fit to the corresponding points. The green ({\em dashed}) line in the top panel shows the Kennicutt-Schmidt law for index of 1.4 \citep{kenn98}. Models of SN energy input with different efficiency values ($\epsilon_{SN}=1$, 0.5, 0.2, 0.1, 0.05) and MRI input at maximum efficiency ($\epsilon_{MRI}=1$) from \cite{tamb09} are shown in purple and orange lines.  }
    \label{fig:dconf}
\end{figure*}


\bibliography{myref.bib}



\end{document}